\documentclass[prx,showpacs,twocolumn,floats,10pt,aps,citeautoscript,longbibliography,superscriptaddress]{revtex4-2}
\usepackage{dcolumn}
\usepackage{microtype} 
\usepackage{graphicx} 
\usepackage{xspace} 

\usepackage[usenames,dvipsnames]{xcolor} 

\usepackage{amsmath} 
\usepackage{amssymb}
\usepackage{bbm}  
\renewcommand{\vec}[1]{{\boldsymbol{#1}}} 
\newcommand{\mr}[1]{\ensuremath{\mathrm{#1}}}
\newcommand{\mc}[1]{\ensuremath{\mathcal{#1}}}
\allowdisplaybreaks 
\usepackage{enumitem} 

\usepackage{scalefnt} 

\usepackage[colorlinks, citecolor={blue!50!black}, urlcolor={blue!50!black}, linkcolor={red!50!black}]{hyperref}

\usepackage{scalerel,stackengine}  
\newcommand\pig[1]{\scalerel*[5pt]{\big#1}{%
\ensurestackMath{\addstackgap[1.5pt]{\big#1}}}}
\newcommand\pigl[1]{\mathopen{\pig{#1}}}
\newcommand\pigr[1]{\mathclose{\pig{#1}}}
\newcommand\Pig[1]{\scalerel*[5pt]{\Big#1}{%
\ensurestackMath{\addstackgap[1.5pt]{\Big#1}}}}
\newcommand\pigg[1]{\scalerel*[5pt]{\bigg#1}{%
\ensurestackMath{\addstackgap[1.5pt]{\bigg#1}}}}

\newcommand{\eqdisconnectedpaperI}{(31)}

\newcommand{\paperI}{Ref.~\onlinecite{Kugler2021}}

\newcommand{\Eq}[1]{Eq.~\eqref{#1}}
\newcommand{\Eqs}[1]{Eqs.~\eqref{#1}}
\newcommand{\Equ}[1]{Equation~\eqref{#1}}
\newcommand{\Equs}[1]{Equations~\eqref{#1}}
\newcommand{\Sec}[1]{Sec.~\ref{#1}}
\newcommand{\App}[1]{Appendix~\ref{#1}}
\newcommand{\Fig}[1]{Fig.~\ref{#1}}
\newcommand{\Figs}[1]{Figs.~\ref{#1}}

\def\ZF/{\mbox{ZF}}
\def\MF/{\mbox{MF}} 
\def\KF/{\mbox{KF}} 

\def\Oc{\mathcal{O}}
\def\Ac{\mathcal{A}}
\def\Bc{\mathcal{B}}
\def\Cc{\mathcal{C}}
\def\Dc{\mathcal{D}}
\def\Gc{\mathcal{G}} 
\def\Pc{\mathcal{P}}
\def\Hc{\mathcal{H}}
\def\Kc{\mathcal{K}}

\def\Tc{\mathcal{T}}

\newcommand{\mi}{\mathrm{i}} 
\newcommand{\md}{\mathrm{d}} 
\newcommand{\TO}{\mathcal{T}} 
\newcommand{\nint}{\!\int\!} 
\newcommand{\nbint}[2]{\!\int_{#1}^{#2}\!} 

\newcommand{\ve}{{\ensuremath{\varepsilon}}}

\newcommand{\eS}{S}
\newcommand{\ub}[1]{{\ensuremath{\underline{#1}}}} 
\newcommand{\ovb}[1]{{\ensuremath{\overline{#1}}}} 
\newcommand{\ovwt}[1]{{\ensuremath{\widetilde{#1}}}} 
\newcommand*{\ndots}{\kern-0.15em.\kern-0.05em.\kern-0.05em.}  
\newcommand*{\nidots}{.\kern-0.05em.\kern-0.05em.} 
\newcommand*{\ncdots}{\kern-0.15em\cdot\kern-0.2em\cdot\kern-0.2em\cdot\kern-0.15em}  
\newcommand\mydots{\makebox[1em][c]{.\hfil.\hfil.}\thinspace} 

\newcommand{\pdag}{{\protect\vphantom{dagger}}}

\newcommand{\sve}{{\underline{\ve}}}

\newcommand{\Tr}{\mathrm{Tr}}
\newcommand{\tr}{\mathrm{tr}}

\def\X{{\scriptstyle {\rm X}}}
\def\D{{\scriptstyle {\rm D}}}
\def\K{{\scriptstyle {\rm K}}} 
\def\x{{\scriptscriptstyle {\rm X}}}
\def\d{{\scriptscriptstyle {\rm D}}}
\def\k{{\scriptscriptstyle {\rm K}}}

\def\rixs{{\scriptscriptstyle {\rm RIXS}}}
\def\xas{{\scriptscriptstyle {\rm XAS}}}
\def\mim{{\scriptscriptstyle {\rm MIM}}}
\def\siam{{\scriptscriptstyle {\rm AIM}}}

\def\aaim{{\scriptscriptstyle {\rm AAIM}}}

\def\xone{\x}
\def\xtwo{\xb}
\def\xthree{\xh}
\def\xfour{\xt}

\def\eb{{\overline{e}}}

\def\sh{{\widehat{s}}}
\def\sb{{\overline{s}}}

\def\nh{{\widehat{n}}}
\def\nb{{\overline{n}}}
\def\Xt{{\widetilde{\mathrm{\scriptstyle X}}}}
\def\Xh{{\widehat{\mathrm{\scriptstyle X}}}}
\def\Xb{{\overline{\mathrm{\scriptstyle X}}}}
\def\xt{{\widetilde{\mathrm{\scriptscriptstyle X}}}}
\def\xh{{\widehat{\mathrm{\scriptscriptstyle X}}}}
\def\xb{{\overline{\mathrm{\scriptscriptstyle X}}}}

\newcommand{\uindex}[1]{{\ensuremath{\underline{#1}}}} 
\def\ui{\uindex{i}}
\def\uj{\uindex{j}}
\def\uone{\uindex{1}}
\def\utwo{\uindex{2}}
\def\uthree{\uindex{3}}

\usepackage{xparse} 

\NewDocumentCommand{\cor}{mod()}
{
	#1\IfValueTF{#2}{[#2]}{}\IfValueTF{#3}{(#3)}{}
}

\newcommand{\ik}[3]{|{#1}\rangle_{#3}^{#2}} 
\newcommand{\ib}[3]{{}_{#3\hspace{-0.1mm}}^{#2}\hspace{-0.1mm} \langle{#1}|} 
\newcommand{\ibk}[6]{{}_{#3\hspace{-0.1mm}}^{#2}\hspace{-0.1mm} \langle{#1}|{#4}\rangle_{#6}^{#5}} 
\newcommand{\ip}[4]{|{#1}\rangle_{#3}^{#2}\, {}_{#4\hspace{1.0mm}}^{#2}\hspace{-1.0mm} \langle{#1}|} 
\newcommand{\iP}[2]{\Pc_{#2}^{#1}} 
\newcommand{\io}[6]{[{#1}_{{#3}{#4}}^{#2}]_{{#5}{#6}}^{\vphantom{1}}} 
 
\newcommand{\ioc}[4]{{#1}_{{#3}{#4}}^{{#2}}} 

\newcommand{\iE}[3]{E_{#3#2}^{#1}} 

\newcommand{\iEdiff}[5]{E_{#2#3,#4#5}^{#1}} 

\newcommand{\iEdiffc}[3]{E_{#2#3}^{#1}} 

\newcommand{\id}[2]{\delta_{#1#2}} 

\newcommand{\idsuper}[2]{\delta^{#1#2}} 

\newcommand{\rhored}{\overline{\rho}\hspace{0.5pt}} 
\newcommand{\varrhored}{\overline{\varrho}\hspace{1pt}} 

\newcommand{\dtilde}{{\ovwt{d}}}
\newcommand{\dtildedag}{{\ovwt{d}}^{\hspace{0.4mm}\dagger}}

\newcommand{\disconnected}{{\mathrm{dis}}} 
\newcommand{\connected}{{\mathrm{con}}} 
\newcommand{\broad}{{\mathrm{b}}} 
\newcommand{\Lorentz}{{\scriptscriptstyle {\mathrm{L}}}} 
\newcommand{\SL}{{\scriptscriptstyle {\mathrm{SL}}}} 
\newcommand{\CL}{{\scriptscriptstyle {\mathrm{CL}}}} 
\newcommand{\Fermi}{{\scriptscriptstyle {\mathrm{F}}}} 
\newcommand{\bfactor}{b} 
\newcommand{\Tk}{{T_{\scriptscriptstyle \mathrm{K}}}} 
\newcommand{\bfactorL}{{b_{\scriptscriptstyle \mathrm{L}}}} 
\newcommand{\twopoint}{{2\mathrm{p}}} 
\newcommand{\fourpoint}{{4\mathrm{p}}} 
\newcommand{\interaction}{{\mathrm{int}}} 
\newcommand{\Gtilde}{\smash{\ovwt{G}}}
\newcommand{\threshold}{{\mathrm{th}}} 
\newcommand{\normalized}{{\mathrm{norm}}} 
\newcommand{\charge}{{\mathrm{ch}}} 
\newcommand{\Gammap}{\Gamma_{\! p}} 
\newcommand{\Deltah}{\Delta_\mr{h}}
\newcommand{\Up}{U_{\hspace{-0.8pt} p}} 
\newcommand{\omegain}{\omega_\mr{in}}
\newcommand{\omegaout}{\omega_\mr{out}}
\newcommand{\omegaloss}{\omega_\mr{loss}}

\begin{document}

\title{Computing local multipoint correlators using the numerical renormalization group}
\author{Seung-Sup B.~Lee}
\affiliation{Arnold Sommerfeld Center for Theoretical Physics, 
Center for NanoScience,\looseness=-1\,  and 
Munich Center for \\ Quantum Science and Technology,
Ludwig-Maximilians-Universit\"at M\"unchen, 80333 Munich, Germany}
\author{Fabian B.~Kugler}
\affiliation{Arnold Sommerfeld Center for Theoretical Physics, 
Center for NanoScience,\looseness=-1\,  and 
Munich Center for \\ Quantum Science and Technology,
Ludwig-Maximilians-Universit\"at M\"unchen, 80333 Munich, Germany}
\affiliation{Department of Physics and Astronomy, Rutgers University, Piscataway, New Jersey 08854, USA}
\author{Jan von Delft}
\affiliation{Arnold Sommerfeld Center for Theoretical Physics, 
Center for NanoScience,\looseness=-1\,  and 
Munich Center for \\ Quantum Science and Technology,
Ludwig-Maximilians-Universit\"at M\"unchen, 80333 Munich, Germany}

\begin{abstract}
Local  three- and four-point correlators yield important insight into strongly correlated systems and have many applications. 
However, the nonperturbative, accurate computation of multipoint correlators is challenging, particularly in the real-frequency domain for systems at low temperatures. 
In the accompanying paper, we introduce generalized spectral representations for multipoint correlators. Here, we develop a numerical renormalization group (NRG) approach, capable of efficiently evaluating these spectral representations, to compute 
local three- and four-point correlators of quantum impurity models.
The key objects in our scheme are partial spectral functions,
encoding the system's dynamical information. Their computation via NRG allows us to simultaneously resolve various multiparticle excitations
down to the lowest energies. By subsequently convolving the partial spectral functions with appropriate kernels, we obtain multipoint correlators
in the imaginary-frequency Matsubara, 
the real-frequency zero-temperature,
and the real-frequency Keldysh formalisms. 
We present exemplary results for the connected four-point correlators of the Anderson impurity model, and for resonant inelastic x-ray scattering (RIXS) spectra of related impurity models.
Our method can treat temperatures and frequencies---imaginary or real---of all magnitudes, from large to arbitrarily small ones.
\end{abstract}
\date{\today}

\maketitle

\section{Introduction}

Correlation functions beyond the standard two-point (2p) functions,
such as 3p and 4p functions, play a central role in many-body theory. 
For instance, the 4p vertex describes the effective interaction between two particles in a many-body environment and signals pairing instabilities; 
3p functions give the full detail of a particle reacting to an external perturbation
and connect to 2p correlators through various Ward identities. 
Through their physical significance and mutual relations, 
2p, 3p, and 4p functions form the basis of the microscopic Fermi-liquid theory. 

In the accompanying paper~\cite{Kugler2021}, we describe a novel approach for analyzing and computing $\ell$-point ($\ell$p) correlators. 
Its central idea is to separate system-dependent spectral information 
from formalism-dependent analytic properties.
This is achieved via the convolution of partial spectral functions (PSFs)
with various types of convolution kernels.
We consider three different frameworks, 
the imaginary-frequency Matsubara formalism (MF), 
the real-frequency zero-temperature formalism (ZF), 
and the real-frequency finite-temperature Keldysh formalism (KF).
The convolution kernels are responsible for the analytic structure of the theory; 
they depend on the choice of formalism, but are system independent. 
The PSFs encode the spectral properties of the system; 
they are independent of the choice of formalism and can be viewed as 
the central porters of information about the dynamics of the system. 
They have generalized Lehmann representations, expressed through the eigenenergies and corresponding many-body eigenstates of the system. 
Having compact support, bounded by the largest energy scale in the system,
they are particularly convenient for numerical storage. 

In \paperI, we illustrated our spectral approach for 
$\ell$p functions by showing numerical results of $4$p vertices
for selected quantum impurity models. Such functions are needed,
e.g., in diagrammatic extensions~\cite{Rohringer2018} of dynamical mean-field theory (DMFT)~\cite{Georges1996,Kotliar2006}. 
There, one starts from the self-consistent DMFT solution of an impurity model, capturing local correlations, 
and aims to include nonlocal correlations by utilizing various diagrammatic relations, 
taking local $4$p functions as input 
\cite{Rohringer2018, 
Toschi2007,Held2008,Toschi2011,Galler2017,Galler2019, 
Rubtsov2008, Brener2008, Hafermann2009a, 
Rohringer2013, 
Ayral2015,Ayral2016, 
Taranto2014,Wentzell2015,Vilardi2019}. 

The local $4$p vertices
shown in \paperI\ were computed using a generalization of the numerical renormalization group (NRG) \cite{Wilson1975,Bulla2008}. This powerful impurity solver allows one to 
directly compute real-frequency correlators, at arbitrarily low temperatures,
with exponentially fine resolution at the lowest excitation energies.
For this reason, it has been used extensively to compute $2$p correlators \cite{Bulla2008},
at high accuracy for single-orbital models \cite{Peters2006,Weichselbaum2007,Zitko2009,Lee2016,Lee2017}
and even multiorbital ones \cite{DeLeo2005,Costi2009,Moca2012,Hanl2013,Hanl2014a,Mitchell2014,Stadler2015,Stadler2016,Lee2018,Stadler2019,Kugler2019,Kugler2020}. 
Importantly, these features also offer new opportunities for computing local $4$p correlators, compared to other well-established impurity solvers, such as exact diagonalization (ED)~\cite{Antipov2015}
and quantum Monte Carlo (QMC)~\cite{Bauer2011,Gaenko2017,Gull2011,Hafermann2013,Shinaoka2017a,Huang2015,
Parcollet2015,Seth2016,Haule2007,
Parragh2012,Wallerberger2019,Gunacker2015,Gunacker2016,Kaufmann2017} methods.

The present paper offers a detailed discussion of how NRG can be utilized to compute 3p and 4p correlators for quantum impurity models. 
To this end, we generalize the full-density-matrix (fdm) NRG approach~\cite{Peters2006,Weichselbaum2007}, 
devised for $2$p correlators, to the 3p and 4p cases.
To tackle the novel dependence of $\ell$p PSFs on $\ell \!-\! 1$ frequencies,
we introduce an additional, iterative scheme to finely resolve regimes involving frequencies 
of different magnitudes, $|\omega_i| \! \ll \! |\omega_j|$ ($1 \! \leq \! i, j \! < \! \ell$).
The bulk of this paper is devoted to a detailed description of this approach.

We have performed numerous benchmark checks of our new method, focusing on the computation of the local connected 4p correlator and corresponding
4p vertex of various impurity models. 
We obtained excellent agreement with (i) analytical predictions for the 
power-law behavior of the ZF vertex for x-ray absorption,
(ii) QMC results for the MF vertex of the Anderson impurity model (AIM) at intermediate temperatures,
(iii) exact results for the KF vertex of the AIM with infinitely strong interactions (the Hubbard atom),
(iv) perturbative results for the KF vertex of the AIM with weak interactions,
and (v) perturbative results for the ZF and KF connected correlator for the AIM with weak interactions. The numerical results for checks (i)--(iv) were  discussed in \paperI.
Here, we present the underlying PSFs and the results for check (v).
The success of all these tests establishes the reliability of our method, and its ability to treat temperatures and frequencies---imaginary or real---of all magnitudes, including very small ones. For example, for the AIM, our approach can reach temperatures much lower than the Kondo temperature and accurately resolve correspondingly low frequencies.

In this paper, we also include an application of great physical interest, 
the computation of resonant inelastic x-ray scattering (RIXS) spectra.
RIXS is a powerful experimental technique for probing various excitations in solids over a wide energy range~\cite{Kotani2001,Ament2011}. However, 
the effects of many-body correlations on RIXS spectra are poorly understood, even for simple models. 
Numerical calculations of RIXS spectra are typically based either on ED of small systems~\cite{Tsutsui2003,Ishii2005,Chen2010,Stavitski2010,Kourtis2012,Haverkort2012,Wohlfeld2013,
Haverkort2014,Green2016,Jia2016,Hariki2018,Wang2019:RIXS} or on a Bethe--Salpeter approach building on \textit{ab initio} calculations
using density functional theory~\cite{Vinson2011,Gilmore2015,Gilmore2021}.
These methods have limited ability for capturing
strong-correlation phenomena characterized by low energies and long length scales, such as the Kondo effect, or the emergence of a small quasiparticle coherence energy scale in many correlated metals. 

Our method is ideally suited for overcoming this limitation. We demonstrate this with proof-of-principle calculations of the RIXS spectra of two minimal models: the Mahan impurity model (MIM) involving a free conduction band interacting with a core hole, 
used to describe x-ray absorption in metals in a seminal work by Mahan \cite{Mahan1967},
and an augmented AIM (AAIM), 
involving the AIM and a core hole. We elucidate how the RIXS spectra of these models are affected by Anderson orthogonality and the Kondo effect, respectively.
Note that the scope of the AIM goes beyond impurity models, as
DMFT and its diagrammatic extensions describe lattice systems 
by the AIM with a self-consistently determined bath.

As an overview, we here summarize the workflow of 
our NRG method for the case of local $4$p correlators.
First, one constructs a Wilson chain by discretizing the impurity model and obtains a complete basis of (approximate) energy eigenstates of the 
entire chain; \Sec{sec:WilsonChains} outlines this step.
Then, one computes $4$p PSFs in a recursive way, described in \Sec{sec:PartialSpectralFunctions}, such that contributions to $4$p PSFs involving frequencies of
widely different magnitudes can be obtained
by invoking the routines for $3$p and $2$p PSFs.
To enable such a recursive approach, 
we rephrase the established scheme for computing $2$p PSFs,
introducing additional notational and diagrammatic conventions, as elaborated in \Sec{sec:Operators}.

By convolving the $4$p PSFs with the kernels given in \Sec{subsec:SummarySpectralRepresentations}, one obtains the full $4$p correlators in the MF, ZF, and KF.
For the real-frequency ZF or KF correlators, it is necessary to broaden the $4$p PSFs, which are discrete due to the discretization in the beginning. Further, to describe genuine two-particle properties, such as $4$p vertices, one needs to extract the connected part of $4$p correlators. This can be numerically challenging.
In \Sec{sec:from-PSF-to-G}, we describe strategies for improving the accuracy of these steps.

Our result for connected 4p correlators
are presented in Sec.~\ref{sec:Results-Connected}, and
for RIXS spectra in Sec.~\ref{sec:RIXS}.
Finally, Sec.~\ref{sec:SummaryOutlook} offers a summary and an outlook to 
applications opened up by our NRG approach to $3$p and $4$p correlators.

\section{Spectral representations}
\label{subsec:SummarySpectralRepresentations}

We begin with a summary of the spectral representations of $\ell$p correlators derived in \paperI, defining $\ell$p PSFs and their convolution kernels.
To describe key ideas and introduce notation, we will first focus on the \ZF/, stating analogous \MF/ and \KF/ results thereafter.

\subsection{Zero-temperature formalism}
\label{sec:ZF-correlators}

In the \ZF/ time domain, an $\ell$p correlator is defined by a time-ordered product, 
$\Gc(\vec{t}) = (-\mi)^{\ell-1} \langle \TO \prod_{i=1}^\ell \Oc^i(t_i) \rangle$, 
with $\vec{t}= (t_1, \, \ndots \, , t_\ell)$. It involves 
a tuple $\vec{\Oc} = (\Oc^1, \, \ndots \, , \Oc^\ell)$
of operators, each time-evolving as $\Oc^i(t) = e^{\mi \mathcal{H}t} \Oc^i e^{-\mi \mathcal{H}t}$, and $\langle ... \rangle = {\rm Tr} [ \varrho \, ... ]$
denotes thermal averaging, with density matrix 
$\varrho = e^{- \beta \Hc}/Z$ and inverse temperature $\beta = 1/T$.
The time-ordered product can be expressed as a sum over permutations yielding
all possible operator orderings,
\begin{align}
\label{eq:Gc_t_KS_ZF} 
\Gc(\vec{t}) = 
\sum_p \vec{\zeta}^p \Kc(\vec{t}_p)
\langle \prod_{i=1}^\ell \Oc^{\ovb{i}}(t_{\ovb{i}}) \rangle , 
\end{align}
where the kernel 
$\Kc(\vec{t}_p)
= \prod_{i=1}^{\ell-1} \big[- \mi \theta(t_{\ovb{i}}-t_{\ovb{i+1}}) 
\bigl]$ is nonzero only if the permuted
times $\vec{t}_p = (t_{\ovb{1}}, \, \ndots, t_{\ovb{\ell}})$ 
satisfy $t_{\ovb{i}} > t_{\ovb{i+1}}$. 
Here, $p(1  2 \, \ndots  \ell) = (\ovb{1}  \, \ovb{2} \, \ndots  \ovb{\ell} ) $ [or $p=(\ovb{1} \, \ovb{2} \, \ndots  \ovb{\ell} )$ for short] denotes
the permutation replacing $i$ by $p(i) = \ovb{i}$, and $\sum_p$ includes 
all such permutations. For example, if $p=(312)$, then $(t_1,t_2,t_3)_p = (t_3,t_1,t_2)$. 
The sign factor $\vec{\zeta}^p$ equals $-1$ if $\vec{\Oc}_p = (\Oc^\ovb{1}, \, \ndots , \Oc^{\ovb{\ell}})$ differs from $\vec{\Oc}$ by an odd number of
transpositions of fermionic operators; otherwise $\vec{\zeta}^p = +1$.

The Fourier transform of $\Gc(\vec{t})$, as defined in \paperI, is  
\begin{align}
\Gc(\vec{\omega})
& =
\nint \md^\ell t \, e^{\mi \vec{\omega} \cdot \vec{t}} 
\Gc(\vec{t})
=
2\pi \delta(\omega_{1\cdots \ell}) \, G(\vec{\omega}) ,
\label{eq:ZF_frequencydomain}
\end{align}
with $\vec{\omega} = (\omega_1, \, \ndots \, , \omega_\ell)$.
Using the shorthand notation $\omega_{1\cdots i} = \sum_{j=1}^{i} \omega_j$,
the $\delta$ function on the right expresses energy
conservation, $\omega_{1 \cdots \ell} = 0$, following from time-translational invariance for $\Gc(\vec{t})$. 
Because of the multiplicative structure of
Eq.~\eqref{eq:Gc_t_KS_ZF}, $G(\vec{\omega})$
can be expressed as an $(\ell \!-\! 1)$-fold convolution of 
the Fourier transforms of $\Kc$ and the operator product, resulting
in an expression of the form
\begin{align}
G (\vec{\omega})
= \sum_p \vec{\zeta}^p
\nint \md^{\ell-1} \omega_p' \,
K( \vec{\omega}_p - \vec{\omega}_p' )
S[\vec{\Oc}_p](\vec{\omega}_p') 
.
\label{eq:G_w_ZF}
\end{align}
Here, $G$, $K$, and $S$ each have only $\ell\!-\!1$ independent arguments,
with $\vec{\omega}$ and $\vec{\omega}'_p = (\omega'_{\ovb{1}}, \ndots, \omega'_{\ovb{\ell}})$ understood
to obey energy conservation, $\omega_{1 \cdots \ell} = 0$
and $\omega'_{\ovb{1} \cdots \ovb{\ell}} = 0$. 
The \ZF/ convolution kernel $K$ can be chosen as
\begin{equation}
K(\vec{\omega}_p - \vec{\omega}_p') = 
\prod_{i = 2}^{\ell}  \xi_{\ovb{i}\cdots\ovb{\ell}}^{-1} 
, \quad
\xi_\ovb{i} = -\omega_\ovb{i} + \omega'_\ovb{i} + \mi \gamma_\ovb{i} ,
\label{eq:K_w_ZF}
\end{equation}
with the shorthand 
$\xi_{\ovb{i}\cdots\ovb{j}} = \sum_{m = i}^{j} \xi_\ovb{m}$.
This corresponds to Eq.~(54a) of Ref.~\onlinecite{Kugler2021}.
For numerical calculations, the imaginary parts have small but finite values $\gamma_i \!>\! 0$.
The PSFs $S$ have a
Lehmann representation:
\begin{flalign}
& S [\vec{\Oc}_p] (\vec{\omega}'_p)
= 
\! \sum_{\ub{1}, \nidots, \ub{\ell}} \!
\rho_{\ub{1}} \prod_{i=1}^{\ell-1} \! 
\Big[ O^{\ovb{i}}_{\ub{i} \, \ub{i+1}} 
\,
\delta(\omega'_{\ovb{1}\cdots \ovb{i}} 
\! - \! E_{ \ub{i+1} \, \ub{1} } )
\Big] O^{\ovb{\ell}}_{\ub{\ell}\ub{1}} . 
\hspace{-1cm} & 
\label{eq:PartialSpectralFinalExpressionNRG}
\end{flalign}
Here, each underlined summation index $\ui$ enumerates 
a complete set of  many-body eigenstates $|\ub{i}\rangle$ 
of the Hamiltonian $\Hc$, 
with eigenenergies $E_\ui$, transition energies $E_{\uj\ui} = E_\uj - E_\ui$,
and matrix elements $O_{\ui\uj} = \langle \ui|\Oc|\uj\rangle$, 
$\rho_{\ub{1}} = e^{-\beta E_{\ub{1}}}/Z$.
(We use calligraphic symbols for operators, roman ones for matrix elements.)
Equations \eqref{eq:G_w_ZF}--\eqref{eq:PartialSpectralFinalExpressionNRG}
give the spectral representation for \ZF/ $\ell$p correlators.

\subsection{Matsubara formalism}
\label{sec:MF-correlators}

In the \MF/, operators time-evolve 
as $\Oc^i(\tau) = e^{\mathcal{H} \tau} \Oc^i e^{-\mathcal{H}\tau}$,
and $\ell$p correlators are defined as $\Gc(\vec{\tau}) =
(-1)^{\ell-1} \langle \TO \prod_{i=1}^\ell \Oc^i(\tau_i) \rangle$. 
Operators are time-ordered on the imaginary-time interval $\tau_i \in (0, \beta)$.
The Fourier transform of $\Gc(\vec{\tau})$ takes the form 
\begin{align}
\Gc(\mi \vec{\omega}) 
= 
\nbint{0}{\beta} \md^\ell \tau \, 
e^{\mi \vec{\omega} \cdot \vec{\tau}} \Gc(\vec{\tau})
= 
\beta \delta_{\omega_{1 \cdots \ell},0}
\, G(\mi \vec{\omega}) ,  
\label{eq:Gc_iw_IF}
\end{align}
with $\vec{\omega}$ denoting a set of discrete
Matsubara frequencies, and the Kronecker symbol
enforcing energy conservation. 
A permutation expansion for $\Gc(\tau)$ analogous
to \Eq{eq:Gc_t_KS_ZF} leads to the spectral representation, 
\begin{align}
G(\mi\vec{\omega})
& =
\sum_p \vec{\zeta}^p
\nint \md^{\ell-1} \omega_p' \,
K(\mi \vec{\omega}_p - \vec{\omega}_p') \, 
S[\vec{\Oc}_p](\vec{\omega}_p')
,
\label{eq:G_iw_KS_IF} 
\end{align}
with real frequencies $\vec{\omega}'_p$, and $\omega_{1 \cdots \ell} = 0$ and  $\omega'_{\ovb{1} \cdots \ovb{\ell}} = 0$
understood. The PSFs $S$ are again given by \Eq{eq:PartialSpectralFinalExpressionNRG}. The \MF/ kernel, expressed through
$\vec{\Omega}_p = \mi \vec{\omega}_p - \vec{\omega}_p' = (\Omega_\ovb{1}, \, \ndots, \, \Omega_\ovb{\ell})$ and $\Omega_{\ovb{1}\cdots \ovb{i}} = \sum_{j = 1}^{i} \Omega_\ovb{j}$, reads 
\begin{align}
K(\vec{\Omega}_p) =
\begin{cases}
\displaystyle
\prod_{i=1}^{\ell-1}
\Omega_{\ovb{1}\cdots\ovb{i}}^{-1}
, \hspace{2.6cm} \text{if} \quad 
\prod_{i=1}^{\ell-1}
\Omega_{\ovb{1}\cdots\ovb{i}}
\neq 0 
, 
\\[-3mm]
\\
\displaystyle
- \frac{1}{2} 
\Big[
\beta + 
\sum^{\ell-1}\limits_{\substack{i=1 \\ i \neq j}}
\Omega_{\ovb{1}\cdots\ovb{i}}^{-1}
\Big]
\prod^{\ell-1}\limits_{\substack{i=1 \\ i \neq j}}
\Omega_{\ovb{1}\cdots\ovb{i}}^{-1} \, 
, \! \quad
\textrm{if} \quad  \Omega_{\ovb{1}\cdots\ovb{j}} = 0
. \vspace{-4mm} \phantom{.}
\end{cases}
\label{eq:K_Omega_IF}
\end{align}
The first line gives the regular part of the kernel,
applicable if all denominators are nonzero. If this is
not the case, anomalous contributions arise. The second
line gives their form under the assumption that at most 
one denominator vanishes, say, $\Omega_{\ovb{1} \cdots \ovb{j}}$, 
for some $j < \ell$. This includes the case of
arbitrary $2$p correlators, $3$p correlators with only one bosonic operator, as well as $4$p correlators of fermionic operators, such that
$\omega_{\ovb{1} \cdots \ovb{i}}$, with $i < \ell$, produces at most one bosonic frequency.
Equations \eqref{eq:PartialSpectralFinalExpressionNRG}, 
\eqref{eq:G_iw_KS_IF}, and \eqref{eq:K_Omega_IF} 
give the spectral representation for \MF/ $\ell$p correlators.

\subsection{Keldysh formalism}
\label{sec:KF-correlators}

\KF/ correlators in the contour basis, $\Gc^{\vec{c}}$, are
defined as 
$\Gc^{\vec{c}}(\vec{t}) = (-\mi)^{\ell-1}
\langle \TO_{\rm c} \prod_{i=1}^\ell \Oc^i(t_i^{c_i}) \rangle $,
where $t_i^{c_i}$ are real times and $\TO_{\rm c}$ denotes contour ordering. They carry a tuple of contour indices
$\vec{c} \!=\! c_1 \ncdots \, c_\ell$, with 
$c_i = -$ or $+$ if operator  $\Oc^i$
resides on the forward or backward branch of the Keldysh contour, respectively.
In this work, we will treat
\KF/ correlators in the Keldysh basis, $\Gc^{\vec{k}} $, carrying  
a tuple $\vec{k} = k_1 \ncdots \,  k_\ell$ of Keldysh indices 
$k_i \in \{1,2\}$. The correlator $\Gc^{\vec{k}} $ 
is obtained from $\Gc^{\vec{c}} $ 
by applying a linear transformation $D$ to
each contour index,  
$\Gc^{\vec{k}} (\vec{t})
= \! \sum_{c_1 \nidots c_\ell}
\prod_{i=1}^\ell  [ D^{k_i ,  c_i} ]
\Gc^{\vec{c}} (\vec{t})$, with 
$D^{k_i, \mp} = (\pm 1)^{k_i} / \sqrt{2}$.

The Fourier transform of $\Gc^{\vec{k}}$ is 
defined as in \Eq{eq:ZF_frequencydomain}. The resulting
$G^{\vec{k}}(\vec{\omega})$ has the spectral representation,
\begin{align}
G^{\vec{k}} (\vec{\omega})
& =
\frac{2}{2^{\ell/2}}
\sum_p \vec{\zeta}^p \!
\nint \md^{\ell-1} \omega_p' \,  
K^{\vec{k}_p}( \vec{\omega}_p - \vec{\omega}_p' )
S[\vec{\Oc}_p](\vec{\omega}_p')
,  
\label{eq:G_w_KF}
\end{align}
involving real frequencies $\vec{\omega}$ and $\vec{\omega}'_p$, 
with $\omega_{1 \cdots \ell} = 0$ and  $\omega'_{\ovb{1} \cdots \ovb{\ell}} = 0$
understood. The PSFs $S$ are yet again given by \Eq{eq:PartialSpectralFinalExpressionNRG}. 
The \KF/ kernel can be expressed as
\begin{equation}
K^{\vec{k}_p} \!
= \!
\sum_{\lambda = 1}^{\ell}
(-1)^{k_{\ovb{1}\cdots \ovb{\lambda-1}}} 
\frac{ 1 + (-1)^{k_\ovb{\lambda} }}{2}
\bigg[
\prod_{i = 1}^{\lambda-1} \xi_{\ovb{1}\cdots\ovb{i}}^{-1}
\prod_{j = \lambda+1}^{\ell} \! \xi_{\ovb{j}\cdots\ovb{\ell}}^{-1}
\bigg],
\label{eq:K_w_KF}
\end{equation}
with $\xi_{\ovb{i}}$ from \Eq{eq:K_w_ZF}.
This corresponds to Eqs.~(63) and (52) of \paperI.
Equations \eqref{eq:PartialSpectralFinalExpressionNRG}, 
\eqref{eq:G_w_KF}, and \eqref{eq:K_w_KF}
give the spectral representation for \KF/ $\ell$p correlators needed here,
concluding  our summary of the results derived in \paperI.

\subsection{Structure of partial spectral functions}
\label{sec:PSF-structure}

A very attractive feature of the above spectral representations
is that all three formalisms, \ZF/, \MF/, and \KF/, 
contain the \textit{same}
PSFs, given by \Eq{eq:PartialSpectralFinalExpressionNRG}.
These PSFs are thus the central porters of dynamical information.
The main goal of this paper is to describe a numerical 
algorithm for computing them.  We conclude this section with some
general remarks on their structure.

First, for a given permutation $p$, the kernels $K$ and the PSFs $S$ all depend on 
the $\ell \!-\! 1$ integration variables  $(\omega'_{\ovb{1}}, \, \ndots \, , \omega'_{\ovb{\ell-1}})$ only in the combinations $\ve_i = \omega'_{\ovb{1} \cdots \ovb{i}}$, for $1 \! \le \! i \! <  \! \ell$.
For numerical purposes, it is convenient to use these combinations as independent integration variables. 
We collect them in an $(\ell \!-\! 1)$-tuple,
$\vec{\ve} =  (\ve_1, \, \ndots \, , \ve_{\ell-1})$
and express the PSFs through this tuple, defining
\begin{align}
\label{eq:define-tilde-S-early}
\eS_p(\vec{\ve})=
S[\vec{\Oc}_p](\vec{\omega}'_p)|_{\omega'_{\ovb{1} \cdots \ovb{i}} = \ve_i}. 
\end{align}
We will henceforth display the subscript on $\eS_p(\vec{\ve})$ only 
in formulas involving a permutation sum, such as \Eq{eq:G_w_KF}; elsewhere the subscript $p$ will be suppressed, it being understood that $\eS(\vec{\ve})$ refers to a specified permutation.

When computing $\eS(\vec{\ve})$, the spectral resolution can be controlled individually for each $\ve_i$. 
This is crucial for NRG computations, where the spectral resolution attainable for each $\ve_i$ is not uniform but proportional to 
$|\ve_i|$.
For a given permutation $p$, the index contraction pattern of $\eS(\vec{\ve})$, as given by \Eq{eq:PartialSpectralFinalExpressionNRG} with
$\ve_i = \omega'_{\ovb{1} \cdots \ovb{i}}$ there, 
can be depicted diagrammatically as follows (for $\ell = 4$): 

\noindent
{\centering
\includegraphics[width=0.96\linewidth]{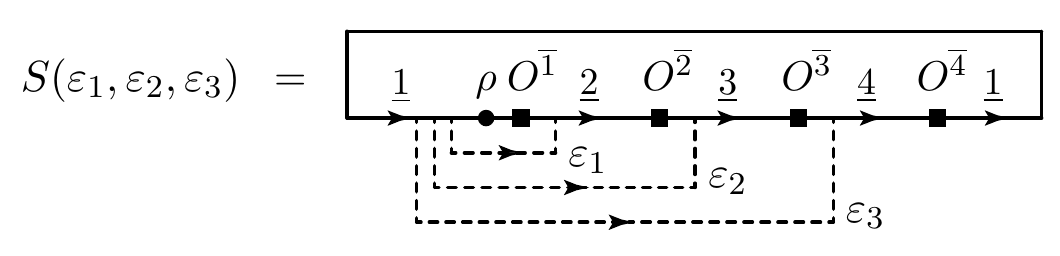}
\par}

\noindent
Here, the  matrix
elements  $O^{\ovb{i}}_{\ub{i}\ub{i+1}}$ and $\rho_{\ub{1}}$
are represented by black squares and a dot, respectively.
An arrow $\ub{i}$ on a solid line connecting two such symbols denotes a sum over $\ub{i}$. It points away from the second (ket) index and toward the first (bra) index of the corresponding operator matrix elements.
Each $\delta$ function is represented by a dashed line, whose arrow
points from the incoming index $\ub{1}$ to the outgoing
index $\ub{i +1}$ in the corresponding condition
$\ve_i  = E_{\ub{i+1}} - E_{\ub{1}}$. 
It will be convenient to express $\eS$ in a form reflecting the
nested structure of the dashed lines:
\begin{align}
\label{eq:4pt_tilde_spectral_rep_compact}
\cor{\eS}
(\vec{\ve}) 
= 
\mathrm{Tr} 
\Big[ \Bigl( \hspace{-0.45mm}\pigl( \hspace{-0.35mm}\bigl(\hspace{-0.2mm} 
(\varrho \, \Oc^{\ovb{1}})^{\ve_1} \Oc^{\ovb{2}}\bigr)^{\ve_2} 
\, \ndots\Oc^{\ovb{\ell-1}}  \pigr)^{\ve_{\ell-1}} \Bigr) 
\Oc^{\ovb{\ell}} \Bigr] . 
\end{align}
Here, $\Oc^{\ve}$ is defined as an operator with matrix elements 
\begin{align}
\label{eq:AomegaMatrixElementsIntroversion1}
[O^{\ve}]_{\ui \uj} = (O_{\ui \uj})^{\ve} = O_{\ui \uj} \delta(\ve - E_{\uj\ui})  
,
\end{align}
with $(\;_{\ui \uj} )^{\ve}$ a shorthand notation for multiplication by a $\delta$ function. Similarly, the matrix elements of 
$\bigl((\varrho \, \Oc^{\ovb{1}})^{\ve_1} \Oc^{\ovb{2}}\bigr)^{\ve_2} $ 
involve a combination of $\delta$ functions with nested arguments:
\begin{align}
\label{eq:NestedMatrixElements}
 \bigl((\rho \, O^{\ovb{1}})^{\ve_1} O^{\ovb{2}}\bigr)^{\ve_2}_{\uone \uthree}  & = \sum_\utwo 
\bigl((\rho_\uone 
O^{\ovb{1}}_{\uone \utwo})^{\ve_1} 
O^{\ovb{2}}_{\utwo \uthree}\bigr)^{\ve_2} 
	\\ \nonumber
& = \sum_\utwo 
\rho_\uone O^{\ovb{1}}_{\uone \utwo} 
\delta(\ve_1 - E_{\utwo \uone})  
O^{\ovb{2}}_{\utwo \uthree}
\delta(\ve_2 - E_{\uthree \uone})  
.
\end{align} 
This nesting complicates the numerical computation of $\eS$ using NRG,
as will be elaborated in the next sections.

\section{Wilson chains}
\label{sec:WilsonChains}

In this section, we summarize the key ingredients of
Wilson's NRG approach for solving quantum impurity models.
We begin with the construction of Wilson shells, sets of approximate energy eigenstates resolving successively lower-energy parts of the spectrum with ever finer resolution. We then review how complete sets of eigenstates for the entire system can be constructed from the Wilson shell eigenstates. Finally, we give a brief preview of how these complete sets of states can be used in Lehmann representations to compute a $2$p PSF. This sets the stage for  a more detailed discussion of the computation of PSFs for $\ell=2,3$, and $4$ in subsequent sections.

\subsection{Wilson shells}
\label{sec:WilsonShells}

A quantum impurity model describes a discrete degree of freedom coupled
to a continuous bath of excitations.
In the NRG approach devised by 
Wilson \cite{Wilson1975}, the bath is discretized on a logarithmic grid characterized by a discretization parameter $\Lambda$ $(>\!1)$.
The discretized bath levels have energies typically scaling as $\sim \pm \Lambda^{-k} D$,
with integer $k\ge 0$ and $D$ the half-bandwidth (taken as unit of energy).
Then, the model is mapped onto a semi-infinite ``Wilson chain,'' where the impurity is represented by the leftmost site, having index $-1$, and the bath by the sites $0, 1, \ndots \, $. Site $n$ has on-site energy $\epsilon_n$ and couples to site $n \!-\! 1$ via hopping amplitude $t_n$.
The exponential decay of $t_n \sim \Lambda^{-n/2}$ and the even stronger decay of $\epsilon_n$ ensures energy-scale separation. 
In practice, one studies finite chains, choosing the largest site index $N$ sufficiently large to resolve all relevant energy scales of the problem, including the temperature (requiring $\Lambda^{-N/2} \ll T)$. 

The Hilbert space of such a ``length-$N$'' chain is spanned by the local basis states $\{ |\boldsymbol{\sigma}_N\rangle\} = 
\{|\sigma_{N}\rangle \otimes \dots \otimes |\sigma_{-1}\rangle\}$,
with local dimension $d$ per site for sites $n\ge 0$. (Following
Ref.~\onlinecite{Weichselbaum2007}, we build the direct product
basis from right to left, as if filling the chain in the order $c^\dagger_{N} \, \ndots \, c^\dagger_{-1}|0\rangle$, where for brevity we displayed only site indices.)
Exploiting energy-scale separation,  the chain can be diagonalized iteratively,
starting with the impurity and increasing the subchain length $n$, 
one site at a time. This amounts to resolving ever smaller energy scales, resulting in a \textit{top-down} (high- to low-energy) refinement strategy. However, the subchain Hilbert space dimension, $\sim\! d^n$, increases exponentially. Hence, a truncation scheme is needed, in which some states are \textit{kept} ($\K$) and others \textit{discarded} ($\D$). 

Let $n_0$ be the last site for which 
the Hamiltonian of a length-$n_0$ subchain $\Hc^{n_0}$ can 
be diagonalized exactly without truncation. The set of 
eigenstates of $\Hc^{n_0}$ is known as \textit{Wilson shell} $n_0$. Its lowest-lying levels have characteristic level spacing $\sim\! \Lambda^{-n_0/2}$, as
that is the smallest energy scale in 
$\Hc^{n_0}$. Now, one partitions this shell into two sectors, containing low-lying kept and high-lying discarded states, respectively, $\{\ik{s}{n_0}{\x}\}$, $\X \!\in\! \{\K, \D\}$, with eigenenergies $\iE{n_0}{s}{\x}$. Then, one proceeds iteratively as follows. 

For any $n \!>\! n_0$, suppose that $\Hc^{n-1}$ has been diagonalized, yielding the eigenstates $\{\ik{s}{n-1}{\x}\}$, $\X \!\in\! \{\K, \D\}$ of shell $n\!-\!1$, with eigenenergies $\iE{n-1}{s}{\x}$.
Now, add site $n$ and diagonalize $\Hc^{n}$ 
on the truncated space spanned by the direct product of the new site and the kept sector of shell $n\!-\!1$, 
$\{|\sigma_n\rangle \!\otimes\! \ik{s}{n-1}{\k}\}$. (The neglect of the high-lying discarded sector during this step is justified by energy-scale separation \cite{Wilson1975}.) The resulting eigenstates form  \textit{shell $n$}. It has low-lying level spacings
$\sim\! \Lambda^{-n/2}$; hence, it provides a refined description of the $\K$ sector of shell $n\!-\!1$, which had larger spacings $\sim\! \Lambda^{-(n-1)/2}$. If $n \!<\! N$, partition the eigenstates of shell $n$ again into low-lying kept and high-lying discarded sectors, $\{\ik{s}{n}{\x}\}$,  with eigenenergies $\iE{n}{s}{\x}$, concluding the $n$th iteration step.
At the last iteration $n \!=\! N$, one declares all 
eigenstates of $\Hc^N$ as discarded, since there is no next iteration in need of kept states.

The eigenstates obtained in step $n$ can be written as linear combinations of the form
\cite{Weichselbaum2007}
\vspace{-2mm}
\begin{subequations}
\label{eq:MPSiteration}
\begin{align} 
\label{eq:MPSiteration-a}
\ik{s'}{n}{\x} & = |\sigma_n\rangle \otimes \ik{s}{n-1}{\k} 
\io{M}{\sigma_n}{\k}{\x}{s}{s'} \, ,  \; \;  \; \; 
\raisebox{-3.5mm}{\includegraphics[width=0.15\linewidth]{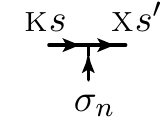}} 
\\
\label{eq:MPSiteration-b}
\ib{s'}{n}{\x} & = \ib{s}{n-1}{\hspace{3.5mm} \k} \otimes \langle \sigma_n| 
\io{M}{\sigma_n \dagger}{\x}{\k}{s'}{s}   \, , 
\;\; \;  \raisebox{-3.5mm}{\includegraphics[width=0.15\linewidth]{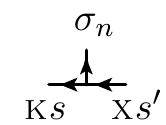}} 
\end{align}
\end{subequations}%
\vspace{-3mm}

\noindent 
where sums over the repeated indices $s$, $\sigma_n$ are implied.
The coefficients are arranged as matrices, $\ioc{M}{\sigma_n}{\k}{\x}$, with elements labeled by $ss'$.
The matrices 
$\ioc{M}{\sigma_n}{}{}$ and $\ioc{M}{\sigma_n \dagger}{}{}$ 
appearing in the 
ket~\eqref{eq:MPSiteration-a} and bra~\eqref{eq:MPSiteration-b}
can be graphically represented by three-leg vertices, 
with incoming or outgoing arrows associated with the kets on 
the right or left of \Eq{eq:MPSiteration-a}, respectively,
and all arrows inverted for the bra version,  Eq.~\eqref{eq:MPSiteration-b}.
Equations~\eqref{eq:MPSiteration} also apply for the first few iterations up to $n_0$, just without truncation. (For the impurity site, $\ioc{M}{\sigma_{-1}}{\phantom{\k}}{\k}$ is a vector, not a matrix.) The eigenstates of $\Hc^n$ thus obtained are matrix product states (MPSs)~\cite{Weichselbaum2012b,Schollwoeck2011} of the form 
\begin{subequations}
\label{subeq:MPSdefinitionKetBra}
\begin{align}
\label{eq:MPSdefinitionKet}
\ik{s}{n}{\x} & = |\sigma_{n}, \ndots \, , \sigma_{-1}\rangle 
\bigr[\ioc{M}{\sigma_{-1}}{\phantom{\k}}{\k}\ioc{M}{\sigma_{0}}{\k}{\k}  
\, \ndots  
\ioc{M}{\sigma_{n-1}}{\k}{\k}
\ioc{M}{\sigma_{n}}{\k}{\x}
\bigl]_s  
\\ 
\nonumber
& = \raisebox{-4mm}{\includegraphics[width=0.32\linewidth]{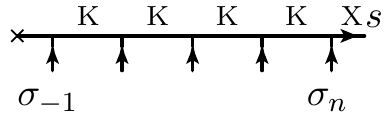}}\;\; , 
\\
\label{eq:MPSdefinitionBra}
\ib{s}{n}{\x} & = \langle \sigma_{-1}, \ndots \, , \sigma_{n}|
\bigr[\ioc{M}{\sigma_{n}\dagger}{\x}{\k}
\ioc{M}{\sigma_{n-1}\dagger}{\k}{\k} \!\!
\ndots  
\ioc{M}{\sigma_{0}\dagger}{\k}{\k}  
\ioc{M}{\sigma_{-1} \dagger}{\k}{\phantom{\k}}
\bigl]_s
\\ 
\nonumber
& = \raisebox{-4mm}{\includegraphics[width=0.32\linewidth]{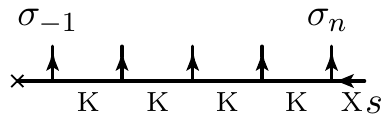}} \; \; . 
\end{align}
\end{subequations}
They form an orthonormal set, 
\begin{align}
\ibk{s}{n}{\x}{\sb}{n}{\xb} 
\label{eq:MPSorthogonality}
= \raisebox{-6.6mm}{\includegraphics[width=0.43\linewidth]{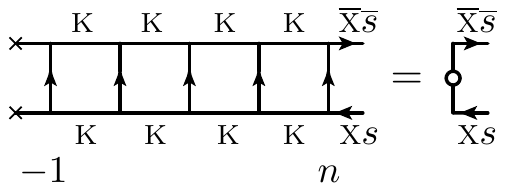}} 
= \id{\x}{\xb} \delta_{s\sb} 
.  
\end{align}
In our MPS diagrams, sums over $s$ or $\sigma$ indices,
indicated by bonds between vertices, will usually not be displayed but understood to be
implicit. Vertices on right- or left-pointing lines for kets or bras represent
$M$'s or $M^\dagger$'s, respectively, lines connecting vertices denote index
contractions, and the empty circle indicates an identity matrix. We will henceforth suppress arrows on vertical lines representing  local state spaces.

\subsection{Anders--Schiller basis for Wilson chain} 
\label{sec:ASBasis}

Following Anders and Schiller \cite{Anders2005}, we next construct states defined not on subchains but the full length-$N$ chain. For
this purpose, we change our perspective: whereas, above, $\Hc^n$ 
was defined on a length-$n$ chain, we henceforth define it 
as $\Hc^n \!=\! \Hc^N|_{t_{\nb > n} = \epsilon_{\nb > n} = 0}$, living 
on the full length-$N$ chain just as $\Hc^N$, but with 
couplings $t_{\nb} \!=\! 0$ and on-site energies $\epsilon_{\nb} \!=\! 0$ turned off for all $\nb \!>\! n$. 
A top-down refinement of the chain, 
starting from $\Hc^{n_0}$, then successively replaces
 $\Hc^{n-1}$ by $\Hc^{n}$, ``switching on'' 
the parameters $t_n$, $\epsilon_n$.
By diagonalizing each $\Hc^n$ and combining
its eigenstates with the discarded states of previous iterations,
one obtains a sequence of sets of basis states, each spanning the Hilbert space of the full chain, but with ever finer energy resolution for low-lying energies.

For any subchain ending at site $n$ ($\ge n_0$), we denote the set of states spanning its environment (the rest of the chain) by
$\{\ik{e}{n}{} = |\sigma_N, \dots, \sigma_{n+1}\rangle \}$.
Then, the states 
\begin{align}
\label{eq:defineAndersSchillerStates}
\ik{se}{n}{\x}  = \ik{e}{n}{} \otimes \ik{s}{n}{\x}
= \raisebox{-4mm}{\includegraphics[width=0.46\linewidth]{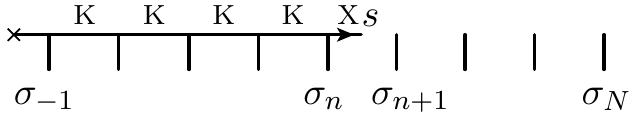}} 
\end{align}
resolve the spectrum of $\Hc^n$, with low-lying level spacings $\sim\! \Lambda^{-n/2}$, but carry an additional
degeneracy $d^{N-n}$ due to the environmental degrees of freedom.
For shell $n_0$, the set of all states $\{\ik{se}{n_0}{\x}\}$
forms a complete basis for the full length-$N$ chain, albeit with
a coarse energy resolution and very high degeneracy. For any later shell $n$, 
the union of all its states with the discarded
states from previous iterations, 
$\cup_{n_0 \leq \nb < n} \{\ik{se}{\nb}{\d}\} \cup \{\ik{se}{n}{\x} \}$,
likewise forms a complete basis for the length-$N$ chain. 
[Completeness is proven below; see Eq.~\eqref{eq:refineidentity}.]
With increasing $n$, the low-energy spectral resolution 
becomes exponentially finer and the 
environmental degeneracy exponentially smaller. 
For the last shell at site $N$, all states are defined
as $\D$ and there are no environmental states and no environmental degeneracy. The set $\cup_{n_0 \leq n \le N} \{\ik{se}{n}{\d}\}$, 
comprising the discarded states of all shells, is again 
complete on the full length-$N$ chain and
known as Anders--Schiller (AS) basis. 

Evoking energy-scale separation, we now adopt the \textit{NRG approximation}. It states 
that, when acting on states from shell $n$, the Hamiltonian of the full chain, $\Hc = \Hc^N$, can be approximated by that of the first
$n$ sites, $\Hc^n$, yielding the eigenvalue $\iE{n}{s}{\x}$:
\begin{align}
\label{eq:NRG-approximation}
\Hc \ik{se}{n}{\x} \simeq \iE{n}{s}{\x} \ik{se}{n}{\x}.
\end{align}
The AS basis, collecting all discarded states, thus forms a \textit{complete} set of approximate
\textit{eigenstates} of the full $\Hc$, with spectrum $\{\iE{n}{s}{\d}\}$
(and degeneracy $d^{N-n}$ for each $\iE{n}{s}{\d}$).

\subsection{Spectral functions from NRG: Brief preview}
\label{sec:previewl=2}

Here, we briefly preview the computation 
of $2$p PSF by NRG, to set the stage for the formal developments
of subsequent sections. 
Since the AS basis
is complete and equipped with eigenenergy information (albeit approximate), it can in principle be used to compute spectral functions  via Lehmann representations \cite{Peters2006,Weichselbaum2007},
with the identification $|\ui\rangle = \ik{se}{n}{\d}$.
For $\ell = 2$, e.g., \Eq{eq:4pt_tilde_spectral_rep_compact}
yields 
\begin{align}
\label{eq:l=2LehmanExample}
& \eS[\Ac, \Bc](\ve)  
= \sum_{\uone \utwo} \bigl(\hspace{-0.2mm} (\varrho \Ac)_{\uone \utwo}\bigr){}^\ve
\Bc_{\utwo \uone} 
\\ 
\nonumber
& \qquad \simeq  
\sum_{nse,\nb \sb \eb} 
\ib{se}{n}{\d} \varrho \Ac \ik{\sb\eb}{\nb}{\d} \, 
\ib{\sb\eb}{\nb}{\d} 
\Bc \ik{se}{n}{\d}  
\,
\delta(\ve \!-\! \iE{\nb}{\sb}{\d} \!+\! \iE{n}{s}{\d} )
\end{align}
with approximate eigenenergies $\iE{\nb}{\sb}{\d}$, $\iE{n}{s}{\d}$.
Since the energy resolution of the AS basis becomes
finer with increasing $n$, the spectral resolution attainable 
becomes exponentially fine with decreasing energy 
$\ve$. This fact makes NRG a singularly powerful tool
for studying the low-energy dynamics of quantum impurity models.

Yet, the Lehmann representation \eqref{eq:l=2LehmanExample} is not the final expression used for NRG calculations.
The reason is that the evaluation of \textit{shell-off-diagonal} $(n \!\neq\! \nb)$ contributions to the double sum is computationally demanding, without improving the accuracy of the results.
A shell-off-diagonal contribution, say with transition energy
$\ve \!=\! \iE{\nb}{\sb}{\d} \!-\! \iE{n}{s}{\d}$ for $n \!<\! \nb$, 
involves the difference between two energy values with different resolutions, $\sim\! \Lambda^{-\nb/2}$ and $\sim\! \Lambda^{-n/2}$.
The frequency resolution of such off-diagonal contributions is dominated by the coarser $\Lambda^{-n/2}$ of the earlier shell.
The better resolution of the later shell thus is futile, yielding no added benefit---the later shell is \textit{overrefined}.

We thus arrive at a central principle for the NRG computation of spectral functions: 
avoid shell-off-diagonal contributions and find shell-diagonal representations.
Reserving a detailed discussion for later sections,
we here just state the main idea: the $\K$ sector of a given shell $n$ 
may be viewed as a coarse-grained description of 
all later shells $\nb > n$ (after all, the latter are 
obtained by refining the former). Thus, the
overrefined off-diagonal contributions to Lehmann representations can be coarse grained \textit{bottom-up} (low- to high-energy) by replacing
$\sum_{\nb > n,\sb\eb} \ik{\sb\eb}{\nb}{\d}  \ib{\sb\eb}{\nb}{\d}$ by $ \sum_{\sb\eb} \ik{\sb\eb}{n}{\k}
\ib{\sb\eb}{n}{\k}$, i.e., the projector onto all shells later than $n$
by the projector onto the $\K\K$ sector of shell $n$.
As shown in Refs.~\onlinecite{Peters2006,Weichselbaum2007} and
recapitulated in Sec.~\ref{sec:l=2PartialSpectralFunctions}, this leads to a coarse-grained version of \Eq{eq:l=2LehmanExample}
with just a single sum over shell-diagonal contributions: 
\begin{align}
\nonumber
\eS[\Ac, \Bc](\ve)  
& \simeq  \! \sum_{nse,\sb\eb} \sum_{\x\xb}^{\neq \k\k}
\ib{se}{n}{\x} \varrho \Ac  \ik{\sb\eb}{n}{\xb} \, 
 \ib{\sb\eb}{n}{\xb} 
 \Bc \ik{se}{n}{\x}  
 \\
 \label{eq:l=2LehmanExample-diagonal}
& \hspace{2cm}
\times \delta(\ve - \iE{n}{\sb}{\xb} + \iE{n}{s}{\x} ).
\end{align}
This is the final expression used in NRG calculations.
Here, each transition energy  
$\iE{n}{s}{\x} \!-\! \iE{n}{\sb}{\xb}$ involves two energies from the same shell $n$, computed
with the same accuracy $\sim\! \Lambda^{-n/2}$. 
For each $n$, the sum over sectors includes only  $\D\D$, $\D\K$, and $\K\D$ matrix
elements; $\K\K$ contributions are excluded, since these are represented, in refined fashion, by later shells. 
Finally, we note that the frequency dependence of the spectral function is resolved with a resolution comparable
to $| \ve| $, furnished by those shells in the expansion 
for which $ \Lambda^{-n/2}$ matches $|\ve|$. 

The above ideas  form the basis of the fdm-NRG approach
of Ref.~\onlinecite{Weichselbaum2007} for computing  $2$p PSFs.
Our purpose here is to  generalize it to $3$p PSFs and 
$4$p PSFs. To this end, we will need 
various formal properties of Wilson shell projectors, 
expressed in a compact and economical notation.
The next few sections thus develop a formalism for expanding 
operator products in terms of operator projections to specific shells.

In the material that follows, the expansions of nonlocal operators (\Sec{subsubsec:NonLocalOperators}) and their products (\Sec{sec:OperatorProducts}), the operator slicing scheme needed for such expansions (\Sec{eq:BinningSlicing}), and the subsequent methods for computing multipoint PSFs and correlators (Secs.~\ref{sec:l=3PartialSpectralFunctions}--\ref{sec:from-PSF-to-G}) are novel methodological developments from this paper.
To explain the new notions efficiently, below we will reformulate some well-established ideas and methods, such as the expansion of local operators and the binning~\cite{Weichselbaum2007,Peters2006,Weichselbaum2012b}, in terms of our formalism.

\section{Operator expansions}
\label{sec:Operators}

This section describes how to expand
various types of operators and operator products along a Wilson chain.
We begin by defining projectors onto specific Wilson shells and discussing
their properties. We then consider the expansions
of local operators, acting nontrivially only on sites $-1$ to $n_0$.
Thereafter, we turn to their time- or frequency-dependent versions,
which are nonlocal, in that they depend on the Hamiltonian of the 
full chain. Next, we discuss the representation
of the density operator on the Wilson chain. 
Finally, we consider expansions of products of two or more operators. 

\subsection{Shell projectors}
\label{sec:ShellProjectors}

We begin by noting that AS states defined on shells $n$
and $\nb$ have overlaps of the following forms [cf.\ \Eq{subeq:MPSdefinitionKetBra}]:
\newline
\includegraphics[width=\linewidth]{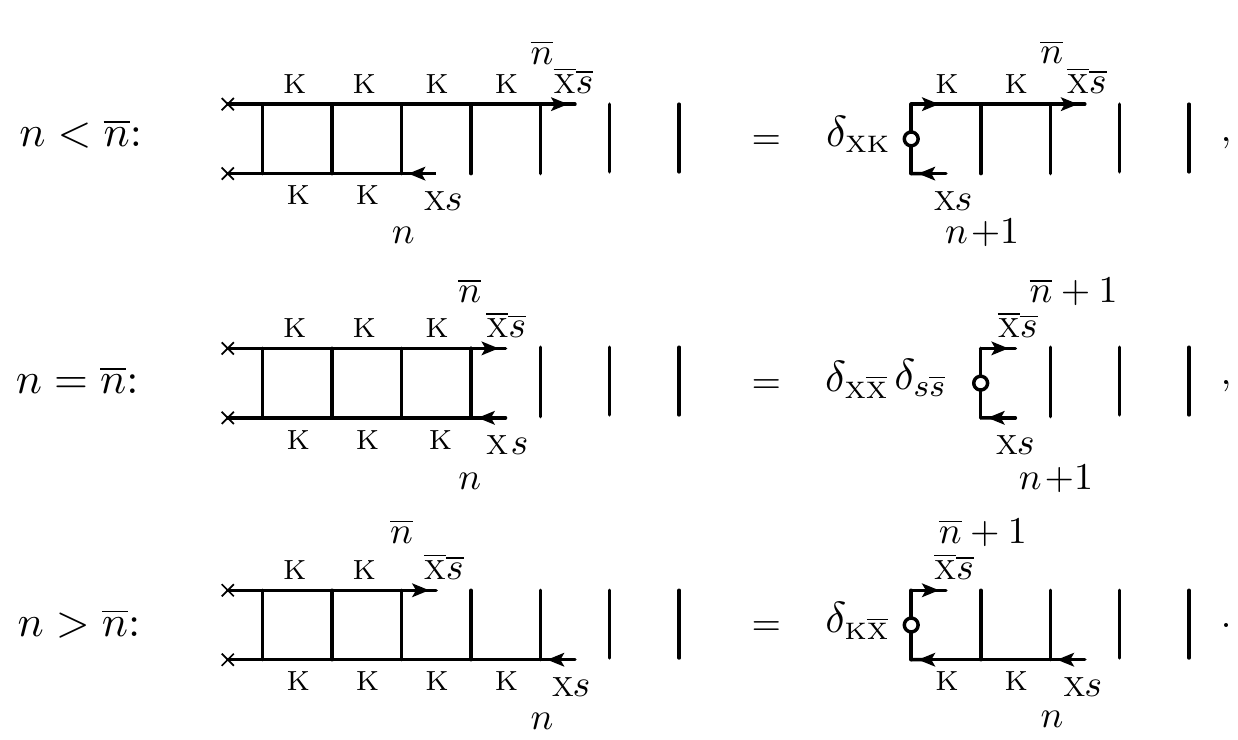}

\vspace{-5mm}
\begin{alignat}{2} 
\label{eq:braketoverlaps} 
\ibk{se}{n}{\x}{\sb\eb}{\nb}{\xb}  
=  
\begin{cases}
\id{\x}{\k} \, 
\bigl[
\ioc{M}{\sigma_{n+1}}{\k}{\k} \mydots 
\ioc{M}{\sigma_\nb}{\k}{\xb}\bigr]_{s \sb } \delta_{e \eb} 
&   \textrm{if} \;\; n < \nb  ,
\\[0.5mm]
\id{\x}{\xb} \delta_{s \sb} \delta_{e \eb} 
&   \textrm{if} \;\; n = \nb  ,
\\[1mm]
\bigl[\ioc{M}{\sigma_{n}\dagger}{\x}{\k}
\mydots \ioc{M}{\sigma_{\nb + 1}\dagger}{\k}{\k}\bigr]_{s \sb}  
 \, \id{\k}{\xb} \delta_{e \eb} \!\!
&   \textrm{if} \;\; n > \nb  . 
\end{cases}
\end{alignat}
(Here, $\delta_{e \eb} $
refers to the environmental modes of sites $> \max(n , \nb)$.)
In words, $n \!\neq\! \nb$ overlaps vanish unless the ``earlier'' sector is $\K$, 
and $n \!=\! \nb$ overlaps yield identity  [by \Eq{eq:MPSorthogonality}]. 
These properties will be used repeatedly.

We next introduce shell projectors as convenient tools for working
with the AS basis. For any  $n \!\geq\! n_0$, let
\begin{align}
\label{eq:projectorX}
\iP{n}{\x} = \sum_{se} \ip{se}{n}{\x}{\x}  = 
\raisebox{-4.4mm}{\includegraphics[width=0.28\linewidth]{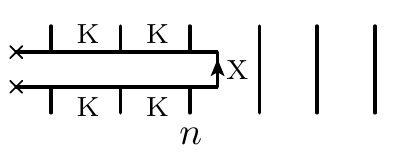}} 
\end{align} 
denote the projector onto the $\X$ sector (with $\X \!\in\! \{\K,\D\}$)
of shell $n$. 
Since the $\K$ sector of shell $n$ spans
both the $\K$ and $\D$ sectors of shell $n \!+\! 1$, their
projectors satisfy a \textit{refinement identity}:
\begin{align} 
\label{eq:refinekeptIterate}
\iP{n}{\k} = \sum_{\x = \k, \d} \iP{n+1}{\x} . 
\end{align}
 
\vspace{-3mm} \noindent
{\centering
\includegraphics[width=0.8\linewidth]{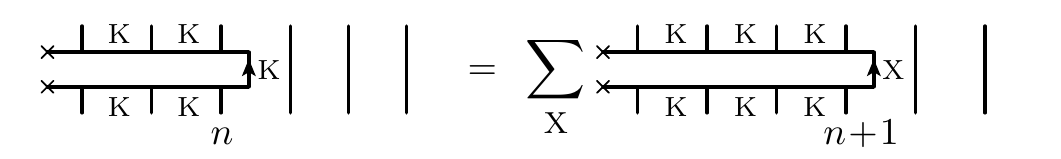} 
\par}
\noindent
This can be used to iteratively refine the description of the $\K$ sector of shell $n$ in terms of the $\K$ and $\D$ sectors
of subsequent shells up to any shell $\nh$, with $n \!<\! \nh \!\le\! N$:
\begin{align} 
\label{eq:refinekept}
\iP{n}{\k} = \sum_{\nb = n + 1}^\nh \iP{\nb}{\d} \, + \, \iP{\nh}{\k} . 
\end{align}
If $\nh \!=\! N$, the last term is absent, since we define all states of shell $N$ as $\D$, so that it has no $\K$ sector: 
\begin{align} 
\label{eq:refinekepttoN}
\iP{n}{\k} \, = \sum_{\nb = n + 1}^N \iP{\nb}{\d}  \, 
= \!\! \raisebox{-5.5mm}{\includegraphics[width=0.44\linewidth]{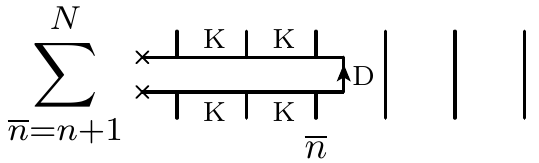}}  
\end{align}
This identity reflects the fact that the $\K$ sector of shell $n$, spanned by the basis $\{\ik{se}{n}{\k}\}$, 
is also spanned by  $\cup_{\nb > n} \{\ik{\sb \eb}{\nb}{\d}\}$, providing a basis for the union of the $\D$ sectors 
of all later shells. The latter basis has a finer energy resolution than the former. Thus, \Eq{eq:refinekepttoN} can be used in two ways: 
top down, where the right-hand side refines the left, 
or bottom up, where the left-hand side coarse grains the right.
Depending on context, we will adopt either one or the other point of view below.

Now, consider
the identity operator on the full $N$-site chain, expressed 
via  the $\K$ and $\D$ projectors for site $n_0$:
\begin{align}
\label{eq:resolve-identity-n0}
\mathbbm{1}^N 
 = \sum_{\boldsymbol{\sigma}_N} |\boldsymbol{\sigma}_N\rangle \langle 
\boldsymbol{\sigma}_N| 
& = \sum_{\x} \iP{n_0}{\x}  
. 
\end{align}

\vspace{-3mm} 
\noindent
{\centering
\includegraphics[width=0.82\linewidth]{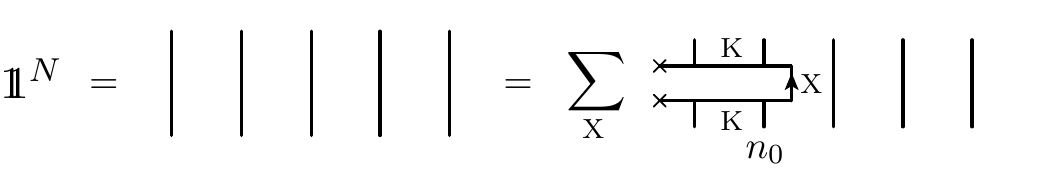}  
\par}
\noindent 
This resolves the identity using 
eigenstates of $\Hc^{n_0}$, but since $n_0$ typically is small, 
the energy resolution is poor.
To improve the resolution, we may refine the $\K$ sector of shell $n_0$
via \Eq{eq:refinekeptIterate},  obtaining a sequence of equivalent resolutions of the identity, for any $\nh$ from $n_0$ to $N$:
\begin{align}
\mathbbm{1}^N 
\label{eq:refineidentity}
& 
 = \sum_{n = n_0}^\nh \iP{n}{\d} \, + \, \iP{\nh}{\k} 
= \sum_{n = n_0}^N \iP{n}{\d} 
= \raisebox{-4mm}{\includegraphics[width=0.35\linewidth]{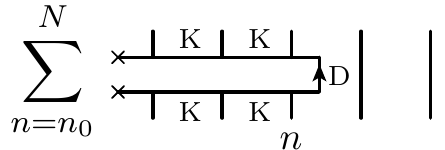}} .  
\end{align}
\noindent 
This proves completeness of the AS basis, and of
its less finely resolved analogues, mentioned after \Eq{eq:defineAndersSchillerStates}. Importantly, any of these
bases may thus be used for evaluating the PSFs,
as indicated in Sec.~\ref{sec:previewl=2}.

We henceforth view sites $n_0$ to $N$ as 
default summation range for $n$, 
writing 
$\sum_n$,  $\sum_{n\le \nh}$, and $\sum_{n > \nb}$
for
$\sum_{n = n_0}^N$, $\sum_{n = n_0}^\nh$, and $\sum_{n = \nb+1}^N$,
respectively.

We will often encounter products of projectors. Such products
can be simplified using the following identity,
\begin{align}
\label{eq:projector-properties}
\iP{n}{\x} \iP{\nb}{\xb}
= 
\idsuper{n <}{\nb} \id{\x}{\k} \iP{\nb}{\xb}
+ 
\idsuper{n}{\nb} \id{\x}{\xb} \iP{n}{\x} 
+
\idsuper{n >}{\nb} \iP{n}{\x} \id{\k}{\xb} 
, 
\end{align}
obtained via \Eq{eq:braketoverlaps}.
The $\delta$ symbols indicate that
the first, second, and third terms contribute only for $n \!<\! \nb$, 
$n \!=\! \nb$, and $n \!>\! \nb$, respectively:

\centerline{\includegraphics[width=\linewidth]{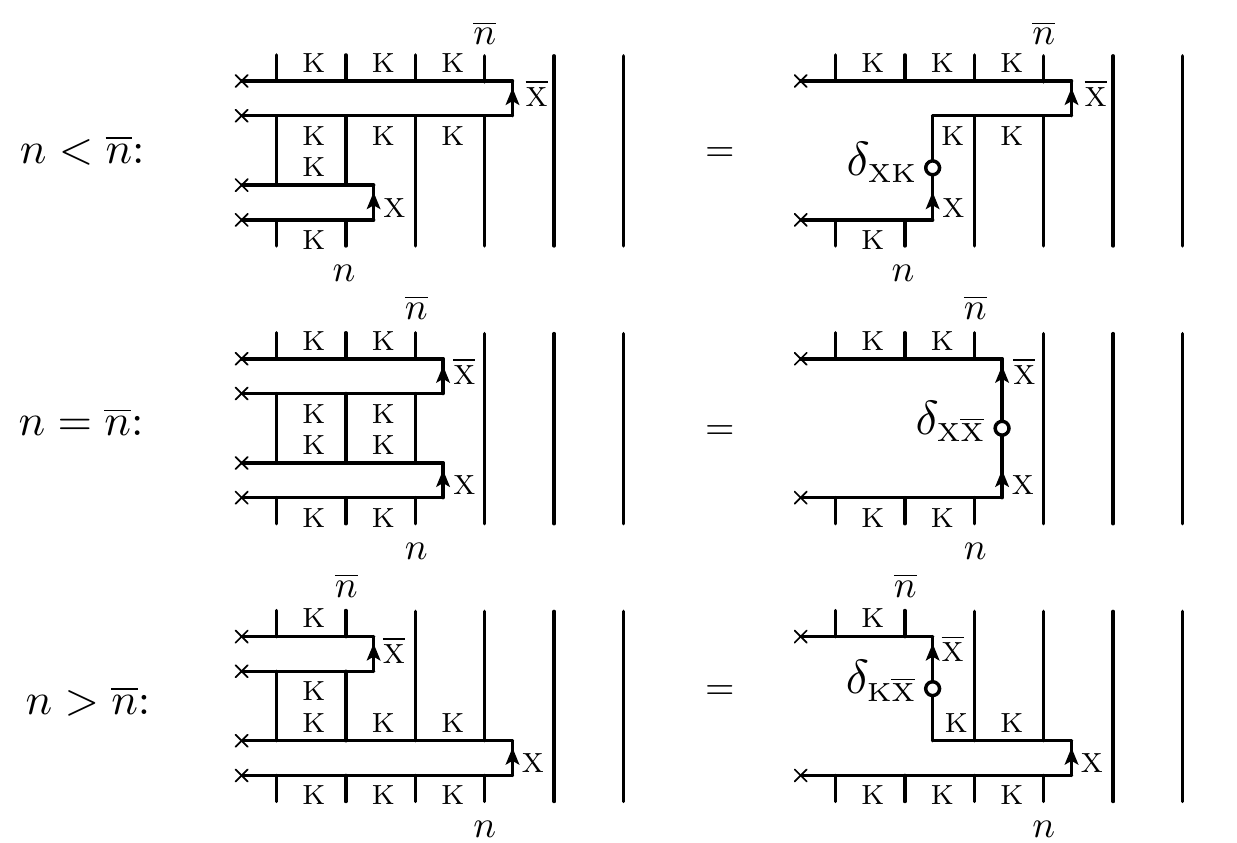}}

The projector product identity \eqref{eq:projector-properties} implies the following rules: a projector product 
\begin{itemize}[topsep=2pt,itemsep=0pt,parsep=0pt,partopsep=0pt,labelindent=-4mm,itemindent=!,leftmargin=!]
\item
with $n \!\neq\! \nb$  vanishes unless the ``earlier'' sector is $\K$, and then it equals  the projector of the later shell;  
\item
with $n \!=\! \nb$  is sector diagonal; 
\item
with $\K\D$ or $\D\K$ vanishes unless $\K$ 
is ``earlier'' than $\D$; 
\item
with $\D\D$  is site diagonal.
\end{itemize}
These rules will be used extensively in later sections.

\subsection{Expanding local operators}
\label{subsubsec:LocalOperators}

A local operator, acting nontrivially only on sites $-1$ to $n_0$,
can be represented exactly in the basis of shell $n_0$,
\begin{align}
\Oc 
& = 
\!\!\!\! 
\sum_{\boldsymbol{\sigma}_{n_0},\overline{\boldsymbol{\sigma}}_{\!n_0},e} \!\!\!
|\boldsymbol{\sigma}_{n_0}\rangle 
|e\rangle^{\! n_0}
O_{\boldsymbol{\sigma}_{n_0}\overline{\boldsymbol{\sigma}}_{\! n_0}}
\langle\overline{\boldsymbol{\sigma}}_{\! n_0}| 
{}^{n_0 \!}\langle e|
= \sum_{\x\xb} \iP{n_0}{\x} \Oc \iP{n_0}{\xb}  . 
\end{align}
{\centering
\includegraphics[width=0.8\linewidth]{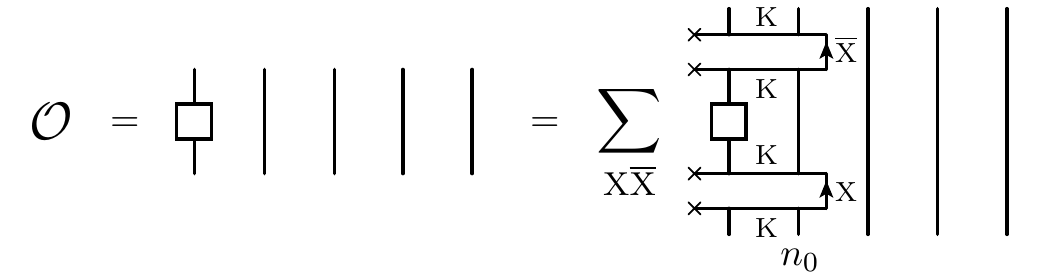} 
\par}
\noindent 
The operator $\Oc$ can be expanded in various
ways along the Wilson chain. Such expansions
involve operator projections to specified combinations
of shells and sectors:
\begin{align}  
\ioc{\Oc}{n\nb}{\x}{\xb}   
& = \iP{n}{\x} \Oc \iP{\nb}{\xb}  
= 
\sum_{se,\sb \eb}\ik{se}{n}{\x} \, \io{O}{n\nb}{\x}{\xb}{se,}{\sb\eb} \, 
\ib{\sb \eb}{\nb}{\xb} 
.
\end{align} 
One obvious expansion
strategy sandwiches $\Oc$ between 
two identities, resolved through the AS basis
using \Eq{eq:refineidentity}:
\begin{align}
\label{eq:OExpansionASbasis}
\Oc = \mathbbm{1}^N \Oc \, \mathbbm{1}^N =
\sum_{n\nb} \iP{n}{\d} \Oc \iP{\nb}{\d}
= \sum_{n\nb} 
\ioc{\Oc}{n\nb}{\d}{\d}
. 
\end{align}
\includegraphics[width=\linewidth]{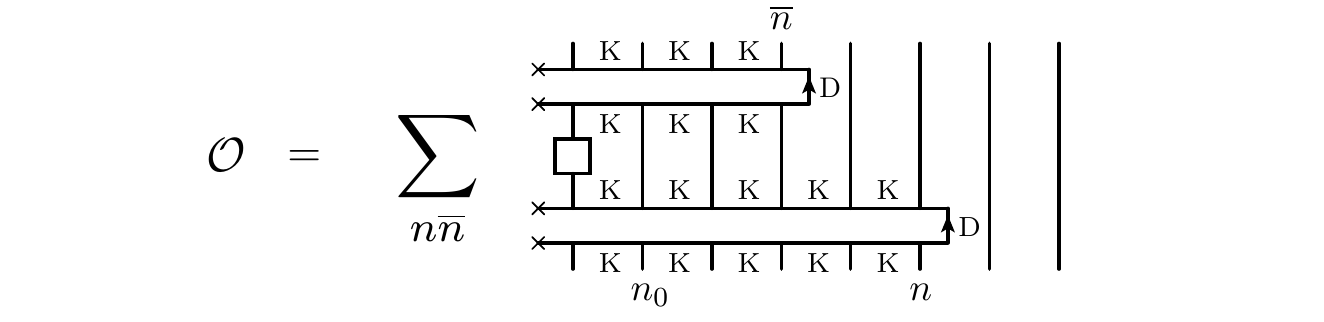}
This expansion involves shell-off-diagonal terms. However, for reasons
explained in Sec.~\ref{sec:previewl=2}, we prefer an expansion that involves only shell-diagonal terms.

One way to obtain a shell-diagonal expansion is bottom-up coarse graining.
We replace $\sum_{\nb > n}\iP{\nb}{\d}$ in \Eq{eq:OExpansionASbasis} by $\iP{n}{\k}$, using \Eq{eq:refinekepttoN}:
\begin{align}
\label{eq:OExpansion-shell-diagonal}
\Oc & 
= \sum_{n} \left[ \iP{n}{\d} \Oc \iP{n}{\d}
+ \iP{n}{\d} \Oc \iP{n}{\k} + \iP{n}{\k} \Oc \iP{n}{\d} \right]
 = \sum_n \sum_{\x\xb}^{\neq \k\k} \ioc{\Oc}{nn}{\x}{\xb} 
 . 
\end{align}
The expansions \eqref{eq:OExpansionASbasis}
and \eqref{eq:OExpansion-shell-diagonal} are equivalent
if $\Oc$ is local, as assumed here, but not for nonlocal 
ones, such as $\Oc(t)$ or 
$\Oc^\ve$. This will be elaborated in the next section.

When working with shell-diagonal projections, we will use the
shorthand notation $\ioc{\Oc}{n}{\x}{\xb}  = \ioc{\Oc}{nn}{\x}{\xb}$
and write
$\Oc^n = \sum_{\x\xb} \ioc{\Oc}{n}{\x}{\xb}$
for the sum over all sectors. The
matrix elements $\io{O}{nn}{\x}{\xb}{se,}{\sb\eb}$ 
then have the form 
$\delta_{e\eb} \io{O}{n}{\x}{\xb}{s}{\sb} $, 
where the elements $\io{O}{n}{\x}{\xb}{s}{\sb} 
= \ib{se}{n}{\x} \Oc \ik{\sb e}{n}{\xb}$
are known at site $n$, and for later sites can 
be computed recursively: 
\begin{align}
\label{eq:OperatorEarlyToLateProjection}
\io{O}{n}{\x}{\xb}{s}{\sb} 
& = 
\sum_{s' \sb'}
\io{M}{\sigma_{n} \dagger}{\x}{\k}{s}{s'}
\io{O}{n-1}{\k}{\k}{s'}{\sb'} 
\io{M}{\sigma_{n}}{\k}{\x}{\sb'}{\sb} 
. 
\end{align}
These relations are graphically represented as\\
\includegraphics[width=\linewidth]{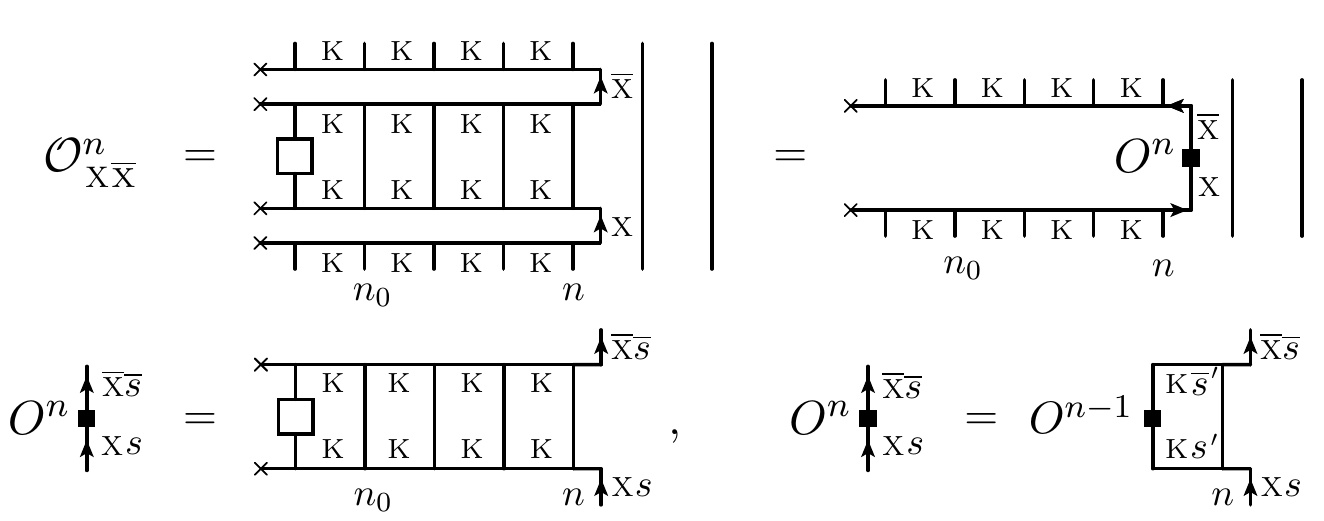}
\noindent 
To be consistent with \Eq{eq:defineAndersSchillerStates}, we depict the first (second) index of $\io{O}{n}{\x}{\xb}{s}{\sb}$ by
an incoming (outgoing) leg.   

We now discuss a more economical, top-down approach, leading directly to the shell-diagonal expansion \eqref{eq:OExpansion-shell-diagonal} without a shell-off-diagonal detour. Starting from 
shell $n_0$, we iteratively refine
only the $\K\K$ sector of each successive shell.  
For a given shell $n$   $(\ge \! n_0)$, we may 
use \Eq{eq:refinekept} to  refine its $\K\K$ sector in terms of shell $n \!+\! 1$, 
\begin{align}
\label{eq:OperatorRefinementIterationIdentity}
\ioc{\Oc}{n}{\k}{\k} = \ioc{\Oc}{n+1}{}{} 
= \sum_{\x\xb} \ioc{\Oc}{n+1}{\x}{\xb} .
\end{align}
\includegraphics[width=\linewidth]{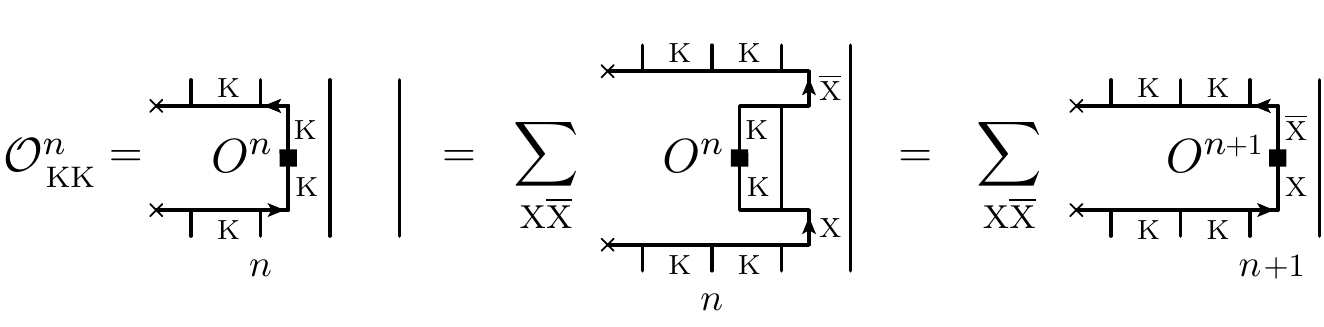}
\noindent 
Iterating up to any shell $\nh$ larger than $n$, we obtain 
\begin{subequations}
\label{subeq:Operator-expansion-n-to-nhat}
\begin{align}
\label{eq:Operator-expansion-n-to-nhat-KKseparate}
\ioc{\Oc}{n}{\k}{\k} 
= \sum_{\nb > n}^{\nh} 
\sum_{\x\xb}^{\neq \k\k}
\ioc{\Oc}{\nb}{\x}{\xb} \, + \, \ioc{\Oc}{\nh}{\k}{\k} 
.
\end{align}
By construction, all terms generated in this manner are shell diagonal,
and each sum on $\X\Xb$ always excludes the case $\K\K$,
except for the site $\nh$ where the expansion terminates.  If it
terminates at $\nh=N$, where there is only a $\D\D$ sector,
there is no $\K\K$ contribution at all:
\begin{align}
\label{eq:Operator-expansion-n-to-nhat}
\ioc{\Oc}{n}{\k}{\k} 
= \sum_{\nb > n}
\sum_{\x\xb}^{\neq \k\k}
\ioc{\Oc}{\nb}{\x}{\xb} 
.
\end{align}
\end{subequations}%
Just as \Eq{eq:refinekepttoN}, the identities \eqref{subeq:Operator-expansion-n-to-nhat} can be used either top down,
with the right side refining the left, or bottom up, with the
left side coarse graining the right.

Starting from site $n_0$, we can go top down to
construct a sequence of equivalent Wilson-chain expansions 
for $\Oc$,
\begin{align}
\Oc 
& = \sum_{n \le \nh} \sum_{\x\xb}^{\neq \k\k}
\ioc{\Oc}{n}{\x}{\xb} 
\, + \, \ioc{\Oc}{\nh}{\k}{\k}
\label{eq:Operator-expansion-n0-to-N}
= \sum_n \sum_{\x\xb}^{\neq \k\k}  \ioc{\Oc}{n}{\x}{\xb} 
, 
\end{align}
where $\nh$ may take any value from $n_0$ to $N$. For the latter choice, shown on the right,
\Eq{eq:Operator-expansion-n0-to-N} matches \Eq{eq:OExpansion-shell-diagonal}.
Thus, the top-down and bottom-up approaches for expanding operators
are equivalent.

The shell-diagonal expansion of local operators is exact, 
as follows directly from the completeness of the AS basis. 
For nonlocal operators, the expansion becomes approximate, 
but it is a reasonable approximation given logarithmic energy resolution $\sim\! \Lambda^{-n/2}$.
In the next section, we will demonstrate the expansion of nonlocal operators in the top-down approach.

\subsection{Expanding nonlocal operators}
\label{subsubsec:NonLocalOperators}

Next, we consider time-dependent operators and their Fourier transforms, needed for the computation of spectral functions. 
Even if $\Oc$ is local, $\Oc(t) = e^{\mi \Hc t} \Oc e^{-\mi \Hc t}$ is not, 
since its definition involves the full Hamiltonian acting on all sites, $\Hc^N$. 
The same is true for its Fourier transform 
$\Oc^\ve = \int \frac{\md t}{2\pi} e^{\mi \ve t} \Oc(t)$, 
with matrix elements given by \Eq{eq:AomegaMatrixElementsIntroversion1},
needed in \Eq{eq:4pt_tilde_spectral_rep_compact} for PSFs.
Hence, such operators must \textit{a priori} be expressed through 
Wilson-chain expansions involving the \textit{entire} chain.
For reasons explained in \Sec{sec:previewl=2} above, this should be done in a manner leading to a shell-diagonal expansion.

We derive the
shell-diagonal expansion of frequency-dependent nonlocal operator $\Oc^\ve$ in the top-down way.
[The expansion of time-dependent operators $\Oc (t)$ can be obtained by Fourier transforming the expansion of $\Oc^\ve$.]
Hence, we start from a coarse-grained representation of $\Oc^\ve$, say
$\sum_{\x\xb} (\ioc{\Oc}{n_0}{\x}{\xb})^\ve$, where the Hamiltonian
is represented by $\Hc^{n_0}$. We then 
refine it by successively turning on the couplings $t_n$ and
energies $\epsilon_n$
along the chain, i.e., replacing $\Hc^{n-1}$ by $\Hc^n$.
This amounts to repeating the iterative
refinement of the $\K\K$ sector of shell $n$ 
via \Eqs{eq:OperatorRefinementIterationIdentity} and
\eqref{eq:Operator-expansion-n-to-nhat-KKseparate}, but now with added
frequency labels $\ve$: 
\begin{flalign}
\label{eq:Operator-Omega-expansion-n-to-N}
&
(\ioc{\Oc}{n}{\k}{\k})^\ve 
\mapsto  \! \sum_{\x\xb} (\ioc{\Oc}{n+1}{\x}{\xb} )^\ve
\mapsto \! \sum_{\nb > n}^{\nh} \! 
\sum_{\x\xb}^{\neq \k\k}
(\ioc{\Oc}{\nb}{\x}{\xb})^\ve 
+ (\ioc{\Oc}{\nh}{\k}{\k})^\ve.
\hspace{-1cm}
&
\end{flalign}
This expansion is approximate as the frequency-dependent matrix elements 
employ the NRG approximation:
\begin{subequations}%
\begin{align}
(\io{O}{n}{\x}{\xb}{s}{\sb})^\ve 
& = \io{O}{n}{\x}{\xb}{s}{\sb} 
\, \delta(\ve - \iEdiff{n}{\xb}{\x}{\sb}{s}) 
, 
\label{eq:Oomega} \\
\iEdiff{n}{\xb}{\x}{\sb}{s} 
& = \iE{n}{\sb}{\xb}  - \iE{n}{s}{\x} 
.  
\end{align}
\end{subequations}

Diagrammatically, we represent the $\delta$ function in \Eq{eq:Oomega} 
by a dashed line bracketing the symbol for 
the matrix element. The arrow on the dashed line points from the incoming to the outgoing
energies in the corresponding condition
$\ve  = \iE{n}{\sb}{\xb} - \iE{n}{s}{\x}$:\\
\includegraphics[width=\linewidth]{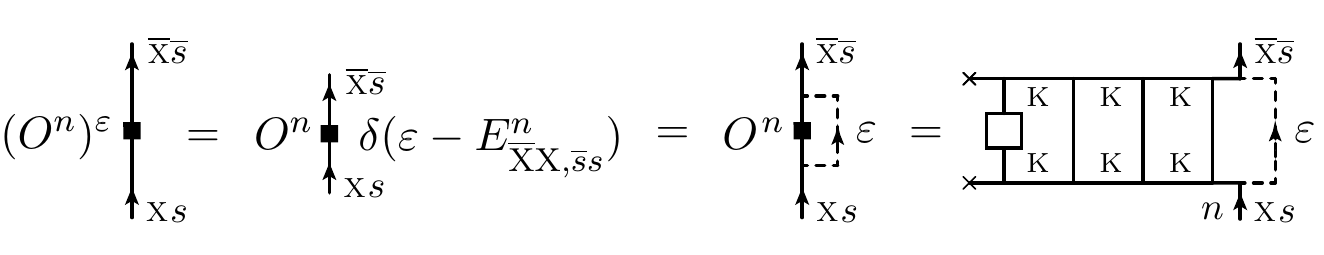}
For a given $\ve$, nonzero  matrix elements
will come mainly from those shells whose low-lying level spacing
$\sim \Lambda^{-n/2}$  is of order $\ve$. However, energy differences between high-lying levels
of earlier shells will also contribute a bit, since the level spacing
within a shell rapidly decreases with increasing energy.

In \Eq{eq:Operator-Omega-expansion-n-to-N}, the $\K\K$ projector on the left of each $\mapsto$ 
can be viewed in two ways: as a coarse-grained placeholder,
to be replaced by the more refined expression on its right once
the need for further refinement arises, or, if there is no such need, 
as the final expression terminating the expansion.
We will henceforth adopt this refinement strategy
throughout, but for simplicity use the ``='' notation
for both local and nonlocal operators.
Since the frequency dependence of $\Oc^\ve$ 
enters simply via a factor multiplying its matrix elements, 
we will often drop the superscript, writing just $\Oc$.
Where needed, the frequency dependence can easily be restored.

\subsection{Density matrix}
We now discuss the thermal density matrix $\varrho$, 
a nonlocal operator,
needed for thermal averages. 
In the AS basis,
\begin{align}
\label{eq:FullDensityMatrix}
\varrho & = \sum_n \ioc{\varrho}{n}{\d}{\d} , \quad 
\ioc{\varrho}{n}{\d}{\d}  =  
\sum_{s\sb e}
\ik{se}{n}{\d}  \io{\rho}{n}{\d}{\d}{s}{\sb} \ib{\sb e}{n}{\d} 
.
\end{align}
Further employing the NRG approximation,
$\varrho$ is fully diagonal, where the weight contributed by each discarded state of shell $n$ is given by a Boltzmann factor:
\begin{align}
\label{eq:DensityMatrixDiscardedWeight}
 \io{\rho}{n}{\d}{\d}{s}{\sb}  & \simeq \delta_{s\sb} 
e^{-\beta \iE{n}{s}{\d}} / Z 
,  \quad 
Z = \Tr [e^{-\beta \Hc^N}] 
. 
\end{align}

{\centering
\includegraphics[width=0.8\linewidth]{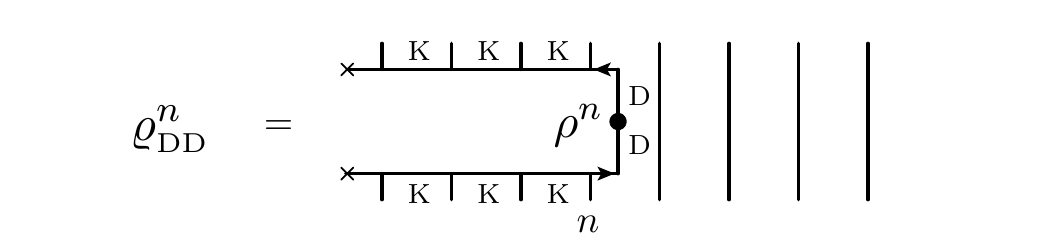}
\par} 
\noindent
The sector projections of $\varrho$ for shell $n$, defined as
$\ioc{\varrho}{n}{\x}{\xb} = \iP{n}{\x} \varrho \iP{n}{\xb}$
and found via \Eq{eq:projector-properties},
are given by 
\begin{align}
\label{eq:DensityMatrixShellProjection}
\ioc{\varrho}{n}{\d}{\d} , 
\qquad 
\ioc{\varrho}{n}{\k}{\k} = \sum_{\nb > n} \ioc{\varrho}{\nb}{\d}{\d} 
,
\qquad
\ioc{\varrho}{n}{\k}{\d} = \ioc{\varrho}{n}{\d}{\k} = 0 
.
\end{align}
Thus, $\ioc{\varrho}{n}{\x}{\xb}$ is sector diagonal.
Note that $\ioc{\varrho}{n}{\k}{\k}$ encompasses a sum over
all later shells; this sum provides a refinement of shell
$n$ in the spirit of \Eq{eq:Operator-expansion-n-to-nhat}. Hence, 
for any $\nh$ from $n_0$ to $N$, 
we can express the density matrix as
[cf.~\Eq{eq:Operator-expansion-n0-to-N}]
\begin{align}
\label{eq:DensityMatrixUpToNhat}
\varrho = \sum_{n \le \nh} \ioc{\varrho}{n}{\d}{\d}
\, + \, \ioc{\varrho}{\nh}{\k}{\k} 
.
\end{align}

Next, we consider the reduced density matrix, 
$\varrhored^n_{\x\xb}$, for a subchain of length $n$, obtained from  $\varrho^n_{\x\xb}$ in bottom-up fashion
by tracing out the environmental states of all later sites, starting
from $N$ and working back to $n \!+\! 1$:
\begin{align}
\label{eq:DefineReducedDensityMatrix}
\ioc{\varrhored}{n}{\x}{\xb} = 
\underset{\text{sites\,} > n}{\Tr} [\ioc{\varrho}{n}{\x}{\xb}]
= \sum_{e} 
\ib{e}{n}{} \ioc{\varrho}{n}{\x}{\xb} \ik{e}{n}{} 
.
\end{align}
\Equ{eq:DensityMatrixShellProjection} implies $\ioc{\varrhored}{n}{\k}{\d} = \ioc{\varrhored}{n}{\d}{\k} = 0$. We find the matrix
elements of $\ioc{\varrhored}{n}{\d}{\d}$ and 
$\ioc{\varrhored}{n}{\k}{\k}$ using \Eqs{eq:DensityMatrixShellProjection} and \eqref{eq:DefineReducedDensityMatrix} and the following diagrams, 
which depict them as circled dots:
 
{\centering
\includegraphics[width=0.85\linewidth]{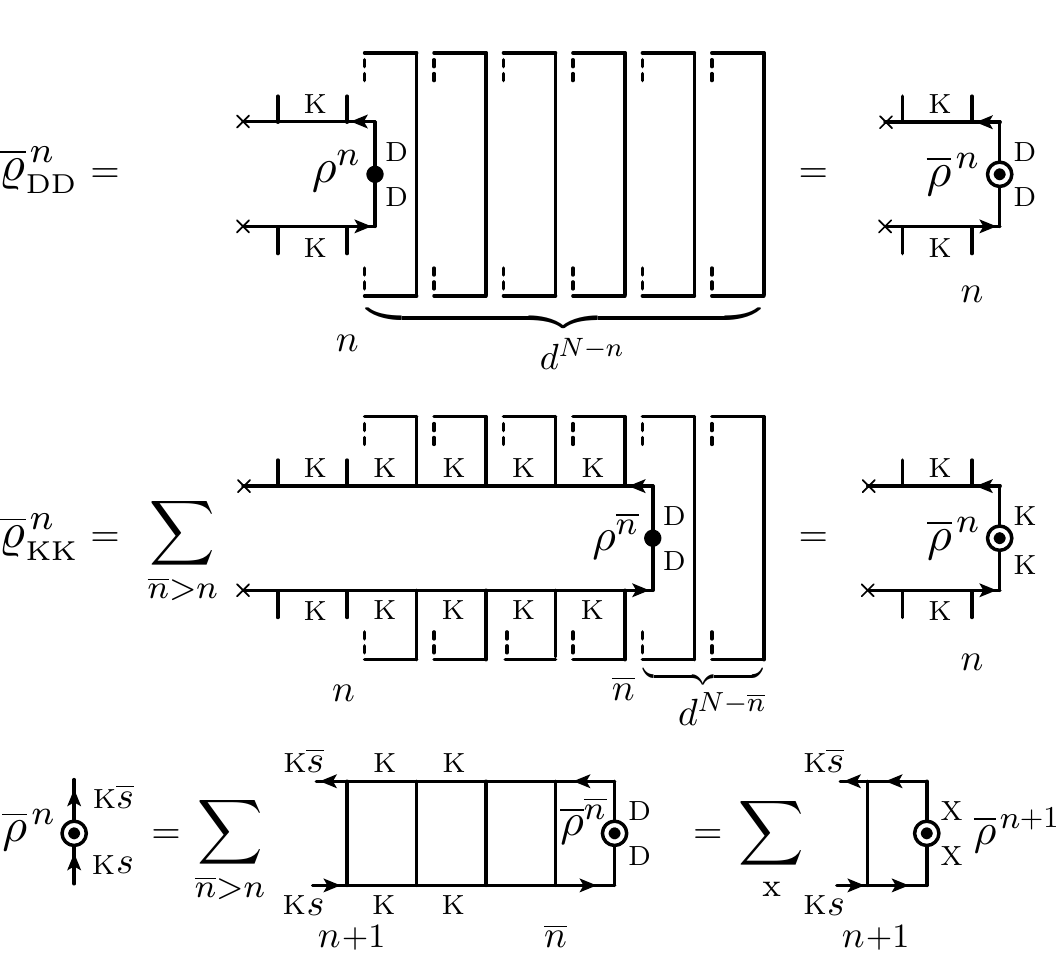}
\par}
\noindent 
The vertically aligned dashed lines, understood to be connected, represent the trace over sites $>n$. 
For $\ioc{\varrhored}{n}{\d}{\d}$, this trace yields
a degeneracy factor $d^{N-n}$:
\begin{subequations}
\begin{align}
\label{eq:ReducedDensityMatrixDD}
\io{\rhored}{n}{\d}{\d}{s}{\sb} & = \io{\rho}{n}{\d}{\d}{s}{\sb} d^{N-n} 
= w_n \delta_{s\sb} 
e^{-\beta \iE{n}{s}{\d}} / Z_n^\d 
.
\end{align}
On the right, 
$Z_n^\d = \sum_{s} e^{-\beta \iE{n}{s}{\d}} $ is the partition
function of the $\D$ sector of shell $n$ (without its environmental states),
and  $w_n = d^{N-n} Z_n^\d/Z$ the relative weight contributed by that sector
to the total partition function. Finally, for 
$\ioc{\varrhored}{n}{\k}{\k}$, the sum over
$\nb \!>\! n$  in \Eq{eq:DensityMatrixShellProjection} yields 
\begin{align*}
\io{\rhored}{n}{\k}{\k}{s}{\sb} & = 
{\displaystyle \sum_{\nb > n}}
\bigl[\ioc{M}{\sigma_{n+1}}{\k}{\k} \!\! 
 \ndots \, \ioc{M}{\sigma_{\nb}}{\k}{\d} 
 \, \ioc{\rhored}{\nb}{\d}{\d} 
  \ioc{M}{\sigma_{\nb} \dagger}{\d}{\k}
\, \ndots \ioc{M}{\sigma_{n + 1} \dagger}{\k}{\k}\bigr]_{ s \sb} 
.
\end{align*}
Starting at $n  =  N - 1$, it can be computed recursively via a backward sweep 
(see above diagram, last line, right-hand part):
\begin{align}
\label{eq:ReducedDensityMatrixKK}
\io{\rhored}{n}{\k}{\k}{s}{\sb} & = 
\sum_\x \bigl[\ioc{M}{\sigma_{n+1}}{\k}{\x} 
 \; \ioc{\rhored}{n+1}{\x}{\x} 
 \, \ioc{M}{\sigma_{n + 1} \dagger}{\x}{\k}\bigr]_{ s\sb } 
 . 
\end{align} 
\end{subequations} 

The reduced density matrix $\ioc{\varrhored}{n}{\k}{\k}$ residing in the $\K$ sector is used when the density matrix contracts with the shell-diagonal expansion of operators [cf.~Eqs.~\eqref{eq:OExpansion-shell-diagonal} and \eqref{eq:Operator-Omega-expansion-n-to-N}].
If the projection of an operator onto a $\K$ sector does not need further refinement (e.g., since the operator is local), the density matrix to contract with the projection does not need to resolve later shells.
Then, the reduced density matrix instead of the original $\varrho$ can be used for the contraction through the $\K$ sector.
We will see an example of this in the next section.

\subsection{Thermal averages}
\label{sec:LocalExpectationValues}

The expectation value $\langle\Oc \rangle = \Tr [  \varrho \, \Oc]$
of a local operator, nontrivial only on sites $\le n_0$,
can be computed as follows: 
\begin{align}
\label{eq:LocalExpectationChain}
\langle \Oc \rangle 
& = \sum_{\x} 
\underset{\text{all sites}}{\Tr}  
[\ioc{\varrho}{n_0}{\x}{\x} \ioc{\Oc}{n_0}{\x}{\x} ] 
= 
\sum_{\x}
\underset{\text{sites} \le n_0}{\Tr}  
[ \ioc{\varrhored}{n_0}{\x}{\x}  \ioc{\Oc}{n_0}{\x}{\x}] . 
\end{align}%
In the first step, we expressed $\varrho \, \Oc$  through their sector projectors for iteration $n_0$,
recalling that $\varrho^{n_0}$ is sector diagonal and the trace cyclic. 
Since $\Oc$ is local, it acts as the identity operator on the Hilbert space of all sites $> n_0$.
In the second step, we exploited this to trace out these sites, reducing $\ioc{\varrho}{n_0}{\x}{\x}$ to 
$\ioc{\varrhored}{n_0}{\x}{\x}$.
Performing the remaining trace over the Hilbert space of sites $\le n_0$
yields a simple trace over matrix elements (denoted by $\mathrm{tr}$) of shell $n_0$,
\begin{flalign}
\label{eq:LocalExpectationShellnZero}
\langle \Oc \rangle
& = 
\sum_{\x} \underset{\text{shell\;} n_0}{\tr}  
[ \ioc{\rhored}{n_0}{\x}{\x} \ioc{O}{n_0}{\x}{\x} ] 
= 
\sum_{\x} 
\sum_{s \sb} \io{\rhored}{n_0}{\x}{\x}{s}{\sb}  \io{O}{n_0}{\x}{\x}{\sb}{s} 
, 
\hspace{-0.5cm}
&
\end{flalign}
with the eigenstates of shell $n_0$ made explicit on the right.
We will mostly use the more compact trace-of-matrix-product
notation of the middle expression, suppressing $s$ indices.
The diagram depicts the expressions on the right of 
\Eq{eq:LocalExpectationChain} and \Eq{eq:LocalExpectationShellnZero},
respectively: 
\newline
\includegraphics[width=\linewidth]{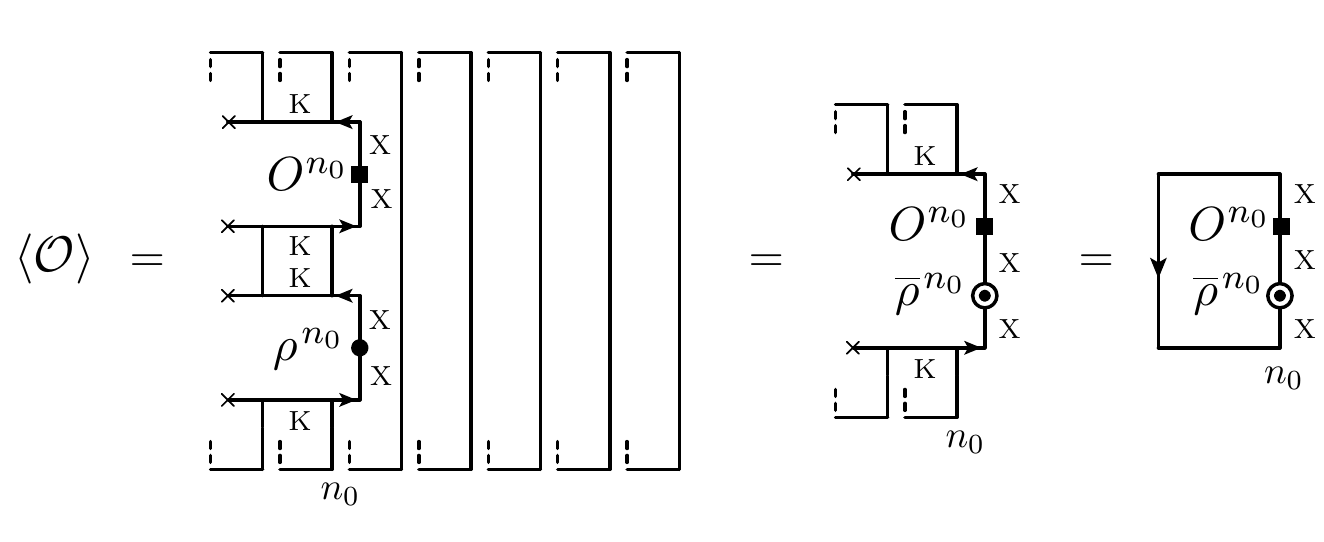} 

For future reference, we note that the above strategy 
also works for thermal averages
of shell-$n$ projections:
\begin{align}
\label{eq:LocalExpectationShelln}  
\langle \ioc{\Oc}{n}{\x}{\xb} \rangle = 
\underset{\text{all sites}}{\Tr}  
[\ioc{\varrho}{n}{\x}{\x} \ioc{\Oc}{n}{\x}{\xb} ]
= \id{\x}{\xb} \underset{\text{shell\;}n}{\tr}  
[ \ioc{\rhored}{n}{\x}{\x} \ioc{O}{n}{\x}{\x} ] 
. 
\end{align}
For the first step, we used \Eq{eq:DensityMatrixUpToNhat}, 
$\varrho = \sum_{\nb \le n} \ioc{\varrho}{\nb}{\d}{\d}
+ \ioc{\varrho}{n}{\k}{\k}$, and noted that
$\ioc{\Oc}{n}{}{}$ is orthogonal to the $\D$ sectors
from all earlier shells [cf. \Eq{eq:projector-properties}]. 
The second step mimics \Eq{eq:LocalExpectationShellnZero},
with $\id{\x}{\xb}$ from the cyclicity of the trace.

Importantly, the above strategy
also works for  averages of shell-diagonal products, 
$\langle \Ac^n \Bc^n \rangle$, $\langle \Ac^n \Bc^n \Cc^n \rangle$,
or $\langle \Ac^n \Bc^n \Cc^n \Dc^n \rangle$. Such 
products, just like $\Oc^n$, are orthogonal to all earlier $\D$ sectors 
and trivial on all later shells. Hence, \Eq{eq:LocalExpectationShelln}
applies again, with $O^n_{\x \x}$ on the right replaced by a corresponding product of
blocks of matrix elements, e.g., $\sum_{\x\xb}
\ioc{\rhored}{n}{\x}{\x} \ioc{A}{n}{\x}{\xb}\ioc{B}{n}{\xb}{\x}$. 
We will exploit this fact when computing spectral functions below.

\subsection{Expanding operator products}
\label{sec:OperatorProducts}

The computation of $\ell$p PSFs
inevitably involves products of frequency-dependent operators, 
e.g., $\bigl(\hspace{-0.2mm} ((\varrho\Ac)^{\ve_1} \Bc)^{\ve_2} \Cc\bigr)^{\ve_3} \Dc$ for $\ell \!=\! 4$. These require Wilson-chain expansions. Moreover, the latter should be shell diagonal,
to avoid overrefinement and to facilitate computing expectation values via \Eq{eq:LocalExpectationShelln}.
We now explain how such expansions can be obtained. 

One approach starts by individually expressing
each operator in the product in terms
of its full-chain expansion \eqref{eq:Operator-expansion-n0-to-N}
from site $n_0$ all the way to site $N$. However, the resulting
expansion is not shell diagonal.  It can be brought into shell-diagonal form by bottom-up coarse graining the $\D$ sectors of
shell-off-diagonal terms, but this requires tedious rearrangements, discussed in \App{sec:OPEAppendix}.

A simpler, top-down approach is to iteratively refine entire products 
rather than individual operators, in a manner that retains
shell diagonality throughout.
For brevity, we use notation appropriate for local operators. [For nonlocal ones, frequency labels should be added and
the ``='' signs in \Eqs{subeq:OperatorRefinementsIterate} and
\eqref{subeq:OperatorRefinements} below read as
``$\mapsto$'' refinements, as discussed at the end
of Sec.~\ref{subsubsec:NonLocalOperators}.]
Starting
from  $n_0$, we refine, for each shell $n$, only the all-$\K$ sector of the product, representing it by a sum over all sectors of shell $n\!+\!1$. 
Using iteration steps analogous to \Eq{eq:OperatorRefinementIterationIdentity}, 
\begin{subequations}
\label{subeq:OperatorRefinementsIterate}
\begin{align}
\label{eq:OperatorDoubleProductRefinementIterate}
\ioc{\Ac}{n}{\k}{\k} 
\ioc{\Bc}{n}{\k}{\k} 
 & = \Ac^{n+1} \Bc^{n+1}
 , 
\\
\label{eq:OperatorTripleProductRefinementIterate}
\ioc{\Ac}{n}{\k}{\k} 
\ioc{\Bc}{n}{\k}{\k} 
\ioc{\Cc}{n}{\k}{\k} 
 & = \Ac^{n+1} \Bc^{n+1} \Cc^{n+1} 
 , 
\\
\label{eq:OperatorQuadrupleProductRefinementIterate}
\ioc{\Ac}{n}{\k}{\k} 
\ioc{\Bc}{n}{\k}{\k} 
\ioc{\Cc}{n}{\k}{\k} 
\ioc{\Dc}{n}{\k}{\k} 
 & = \Ac^{n+1} \Bc^{n+1} \Cc^{n+1} \Dc^{n+1} 
 , 
\end{align}
\end{subequations}
one finds the following generalizations of 
expansion \eqref{eq:Operator-expansion-n0-to-N} for a single operator:
\begin{subequations}
\label{subeq:OperatorRefinements}
\begin{align}
\label{eq:OperatorDoubleProductRefinement}
\Ac \Bc & = \sum_n \sum_{\x \xb \xh}^{\neq \k\k\k} 
\ioc{\Ac}{n}{\x}{\xb} \ioc{\Bc}{n}{\xb}{\xh} 
, 
\\
\label{eq:OperatorTripleProductRefinement}
\Ac \Bc \Cc & = \sum_n \sum_{\x\xb \xh \xt}^{\neq \k\k\k\k}
\ioc{\Ac}{n}{\x}{\xb} \ioc{\Bc}{n}{\xb}{\xh} \ioc{\Cc}{n}{\xh}{\xt} 
,
\\
\label{eq:OperatorQuadrupleProductRefinement}
\Ac \Bc \Cc \Dc & = \sum_n \sum_{\x\xb \xh \xt \x'}^{\neq \k\k\k\k\k}
\ioc{\Ac}{n}{\x}{\xb} \ioc{\Bc}{n}{\xb}{\xh} 
\ioc{\Cc}{n}{\xh}{\xt} \ioc{\Dc}{n}{\xt}{\x'} 
.
\end{align}
\end{subequations}
We depict shell-diagonal operator products as follows:\\
\includegraphics[width=\linewidth]{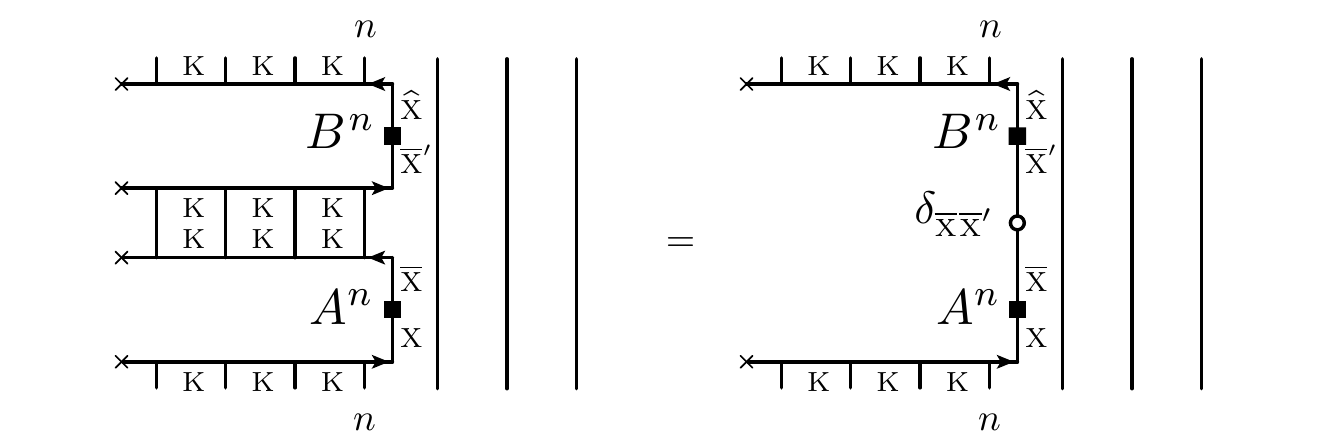}

As an important example, consider the composite
operator $\varrho \Ac  = \sum_{\x \xb} \ioc{(\varrho \Ac )}{n_0}{\x}{\xb}$.
Since $\varrho$ is sector diagonal, the Wilson-chain 
expansion \eqref{eq:OperatorDoubleProductRefinement} of 
$\varrho \Ac $ reduces to 
\begin{align}
\label{eq:ShellProjectionAnrhon}
\varrho \Ac  = \sum_n \sum_{\x \xb}^{\neq \k\k} 
\ioc{(\varrho \Ac)}{n}{\x}{\xb} , \qquad
\ioc{(\varrho \Ac )}{n}{\x}{\xb} = 
\ioc{\varrho}{n}{\x}{\x} \ioc{\Ac}{n}{\x}{\xb} 
.
\end{align}
Operator products as in \Eqs{subeq:OperatorRefinements} always appear in thermal averages,
such that the trace equates the first and last sector index. For $\ell \!=\! 2$, e.g., \Eq{eq:OperatorDoubleProductRefinement} and \eqref{eq:LocalExpectationShelln} yield
\begin{align}
\langle \Ac \Bc \rangle = 
\sum_n \sum_{\x \xb}^{\neq \k\k} 
\underset{\text{shell\;}n}{\tr}  
[ \ioc{\rhored}{n}{\x}{\x} \ioc{A}{n}{\x}{\xb} \ioc{B}{n}{\xb}{\x} ] 
,
\end{align}
where $\K\K$ contributions are 
excluded. However, for products of $\ell \!\geq\! 3$ operators,
the expansions obtained via \Eqs{subeq:OperatorRefinements}
do contain terms involving $\K\K$ sectors. In the top-down 
approach, this is an inevitable consequence of insisting on shell diagonality. 
(In the bottom-up approach, such $\K\K$ contributions are not present \textit{a priori}. 
Yet, they do arise once one coarse grains in order to avoid
shell-off-diagonal terms; see \App{sec:OPEAppendix}.)
If the corresponding operators are frequency dependent,
$(\ioc{\Oc}{n}{\k}{\k})^\ve$, their matrix elements 
[cf.\ \Eq{eq:Oomega}] involve $\delta$ functions
of the form $\delta(\ve - \iEdiffc{n}{\k}{\k})$,
containing $\K\K$ transition energies. These 
require further refinement in case $\ve$ is much
smaller than the characteristic scale of shell $n$.
Strategies for achieving this will 
be discussed in detail in Secs.~\ref{sec:l=3PartialSpectralFunctions} and \ref{sec:l=4PartialSpectralFunctions}.

In \Eqs{subeq:OperatorRefinements},
each product is shell diagonal; hence, the energy differences in 
the $\delta$ functions of a given product are all computed with the same accuracy. To be specific, the 
matrix elements of $(\Ac^n)^{\ve_1} \Bc^n$ or $((\Ac^n)^{\ve_1} \Bc^n)^{\ve_2}$ are given by  [cf. \Eq{eq:NestedMatrixElements}]
\begin{flalign}
\label{eq:NestedMatrixElementsWilsonChain}
&\bigl[(\ioc{A}{n}{\x}{\xb}{}{})^{\ve_1} \! 
\ioc{B}{n}{\xb}{\xh}\bigr]_{s \sh}
= 
\sum_{\sb} 
\io{A}{n}{\x}{\xb}{s}{\sb} \,  
\delta(\ve_1 \!-\! \iEdiff{n}{\xb}{\x}{\sb}{s})
\io{B}{n}{\xb}{\xh}{\sb}{\sh} , \hspace{-1cm}&
\\[1mm] 
\nonumber
&\bigl(\bigl[(\ioc{A}{n}{\x}{\xb}{}{})^{\ve_1}
\ioc{B}{n}{\xb}{\xh}\bigr]_{s \sh}\bigr)^{\ve_2}
\\ \nonumber
& \qquad =
\sum_{\sb} 
\io{A}{n}{\x}{\xb}{s}{\sb} \, 
\delta(\ve_1 \!-\! \iEdiff{n}{\xb}{\x}{\sb}{s})
\io{B}{n}{\xb}{\xh}{\sb}{\sh} \, 
\delta(\ve_2 \!-\! \iEdiff{n}{\xh}{\x}{\sh}{s}) 
\\
\label{eq:NestedMatrixElementsWilsonChainl=3}
& \qquad =
\sum_{\sb} 
\bigl(\io{A}{n}{\x}{\xb}{s}{\sb}\bigr)^{\ve_1}
\io{B}{n}{\xb}{\xh}{\sb}{\sh} \, 
\delta(\ve_2 \!-\! \iEdiff{n}{\xh}{\x}{\sh}{s}) 
, 
& 
\end{flalign}
and analogously for higher products. 
We depict the resulting contraction patterns as follows:\\
\includegraphics[width=\linewidth]{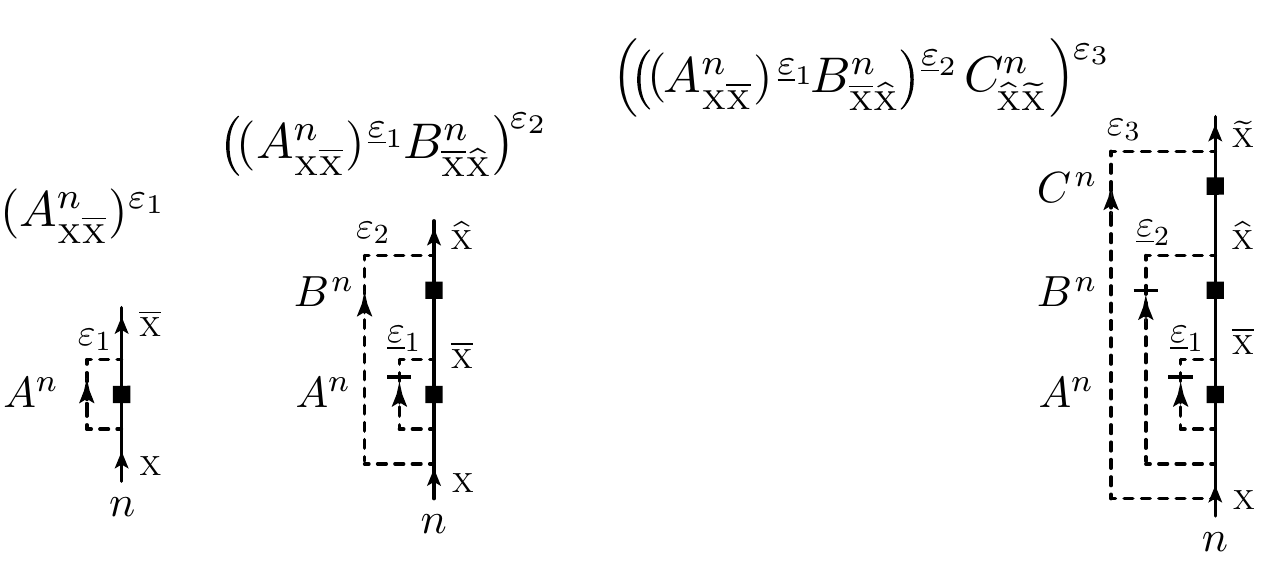}
The underlined $\sve$ frequencies and
slashes through some dashed lines indicate ``slicing'', a 
numerical strategy for dealing with products of $\delta$ functions 
discussed in  Sec.~\ref{eq:BinningSlicing}.

\section{Partial spectral functions}
\label{sec:PartialSpectralFunctions}

We now have all ingredients needed for computing PSFs.
We start with the case $\ell \!=\! 2$, recapitulating
the fdm-NRG approach of Ref.~\onlinecite{Weichselbaum2007}. We then
discuss some numerical techniques used to deal with
the $\delta$ functions arising in Lehmann representations. 
Finally, we consider the cases $\ell = 3$ and $\ell = 4$.
We will denote $2$p PSFs by 
$\eS(\ve)$,  $3$p PSFs by 
$\eS(\vec{\ve})$ with $\vec{\ve} \!=\! (\ve_1, \ve_2)$,
and $4$p PSFs by $\eS(\vec{\ve})$
with $\vec{\ve} \!=\! (\ve_1, \ve_2, \ve_3)$,
using the same symbols $\eS$ throughout.
The number of independent frequency arguments, $\ell \!-\! 1$, will always be clear from the context.

\subsection{Partial spectral functions: \texorpdfstring{$\ell=2$}{l=2}}
\label{sec:l=2PartialSpectralFunctions}

We begin with the case $\ell = 2$.
By \Eq{eq:4pt_tilde_spectral_rep_compact}, the $2$p PSF of 
$\Ac$ and $\Bc$ is given by 
\begin{subequations}
\label{subeq:ABomegaCompactl=2} 
\begin{align}
\label{eq:ABomegaCompactl=2}
\cor{\eS}[\Ac,\Bc](\ve) 
& =  \Tr [(\varrho \Ac)^{\ve} \Bc ]
\\
& = \Tr [\varrho \Ac \Bc^{-\ve}  ] 
.
\label{eq:ABomegaCompactl=2Alternatives}
\end{align}
\end{subequations}
We denote it by $\eS(\ve)$ for short.
The above two forms of writing the trace are equivalent
and can be used interchangeably.
They implement the $\delta$ functions occurring in \Eq{eq:PartialSpectralFinalExpressionNRG}
in different ways while exploiting
the fact that $\varrho$ is diagonal in the energy eigenbasis. Analogous equivalent forms also 
exist for the shell-diagonal traces encountered below. In the present
section, we focus on the first form; in subsequent sections,
we will refer to both.

We start from \Eq{eq:ABomegaCompactl=2}. To finely
resolve the $\ve$ dependence 
of the product $(\varrho \Ac)^{\ve} \Bc$,  we use \Eq{eq:OperatorDoubleProductRefinement} to express it
as a sum over all shells $\ge n_0$, obtaining 
the expansion
\begin{align}
\label{eq:sumnABomegaCompactl=2}
\cor{\eS} (\ve) 
& = \sum_{n}
\hspace{-0.5mm} \sum_{\x \xb}^{\neq \k\k} \hspace{-0.5mm} 
\ioc{\eS}{n}{\x}{\xb}[\Ac ,\Bc ] (\ve)  
, 
\end{align}
where each $\eS^n$ is a trace over a shell-diagonal product,
\begin{align}
\label{eq:ProjectorProduct2}
\ioc{\eS}{n}{\xone \xtwo}{}[\Ac,\Bc] (\ve)& 
\hspace{-0.3mm}=\hspace{-0.4mm} 
\Tr [
(\hspace{-0.2mm}
\ioc{(\varrho \Ac)}{n}{\xone}{\xtwo} )^{\ve}
\ioc{\Bc }{n}{\xtwo}{\xone} ] 
.
\end{align}
The following diagram schematically
depicts the iterative refinement of $\K\K$ sectors
(red squares) leading to \Eq{eq:sumnABomegaCompactl=2}.

\noindent \includegraphics[width=0.98\linewidth]{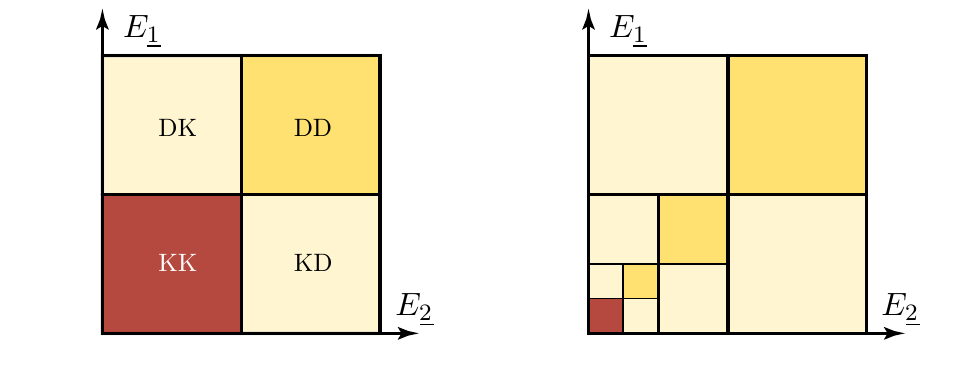}

\noindent%
The traces in \Eq{eq:ProjectorProduct2}
can be computed using 
\Eq{eq:LocalExpectationShelln}, 
replacing an operator trace over the entire Wilson chain 
by a shell-$n$ trace over products of shell-$n$ matrices,
and $\varrho$ by the reduced density matrix 
$\rhored^n$ of shell $n$: 
\begin{align}
\label{eq:ABomegaFinalShelln}
\ioc{\eS}{n}{\x}{\xb}[\Ac,\Bc](\ve) & = 
\underset{\text{shell\;}n}{\tr}  
[ ( \hspace{-0.2mm}\ioc{(\rhored A)}{n}{\x}{\xb} )^{\ve} 
\ioc{B}{n}{\xb}{\x} ] 
.
\end{align}
The depiction of \Eq{eq:ABomegaFinalShelln} mimics 
the diagram for \Eq{eq:LocalExpectationShellnZero}:
\\
\includegraphics[width=\linewidth]{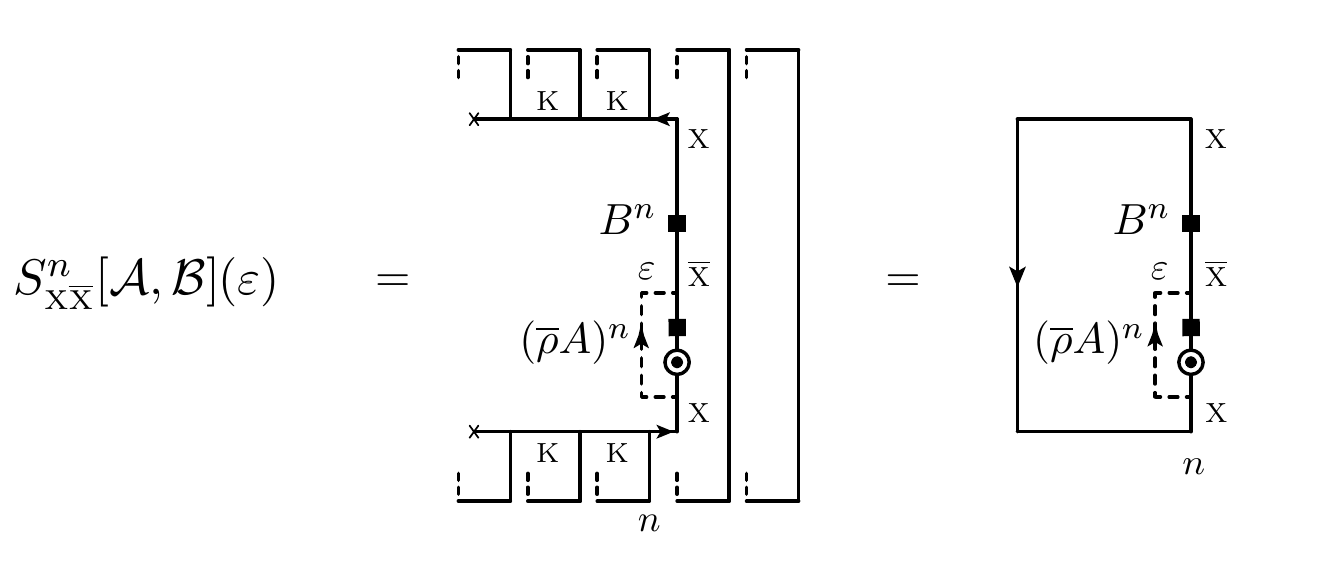} 
Combining  \Eqs{eq:sumnABomegaCompactl=2}--\eqref{eq:ABomegaFinalShelln} and making the matrix
trace over shell $n$ from the latter explicit, we obtain our final formula
for $2$p PSFs:
\begin{align}
\label{eq:l=2ABFinalResult}
\cor{\eS}(\ve) 
= \sum_n   \sum_{\x \xb }^{\neq \k \k} \! 
\sum_{s \sb}
\io{(\rhored  A)}{n}{\x}{\xb}{s}{\sb} 
\, \delta(\ve \!-\! \iEdiff{n}{\xb}{\x}{\sb}{s})
\io{B}{n}{\xb}{\x}{\sb}{s} 
. 
\end{align}
This formula, restating \Eq{eq:l=2LehmanExample-diagonal} in more
compact notation, reproduces an expression first given in Ref.~\onlinecite{Weichselbaum2007}. 

In numerical practice, the Wilson chain is diagonalized in a forward sweep, yielding all coefficients $\io{M}{n}{\k}{\x}{s}{s'}$ and energies $\iE{n}{\x}{s}$ from site $n= -1$ to $N$. Subsequently, the matrix elements of $\ioc{\rhored}{n}{\x}{\x}$ are computed recursively during a backward sweep from site $N$ to $n_0$. To evaluate PSFs, the matrix elements $\ioc{A}{n}{\x}{\xb}$, $\ioc{B}{n}{\xb}{\x}$ 
and $\ioc{(\rhored A)}{n}{\x}{\xb}$ 
are evaluated in a second forward sweep up to site $N$. The $\delta$
function is dealt with by ``binning'', a technique described  next.

\subsection{Binning and slicing}
\label{eq:BinningSlicing}

\begin{figure}
\includegraphics[width=0.98\linewidth]{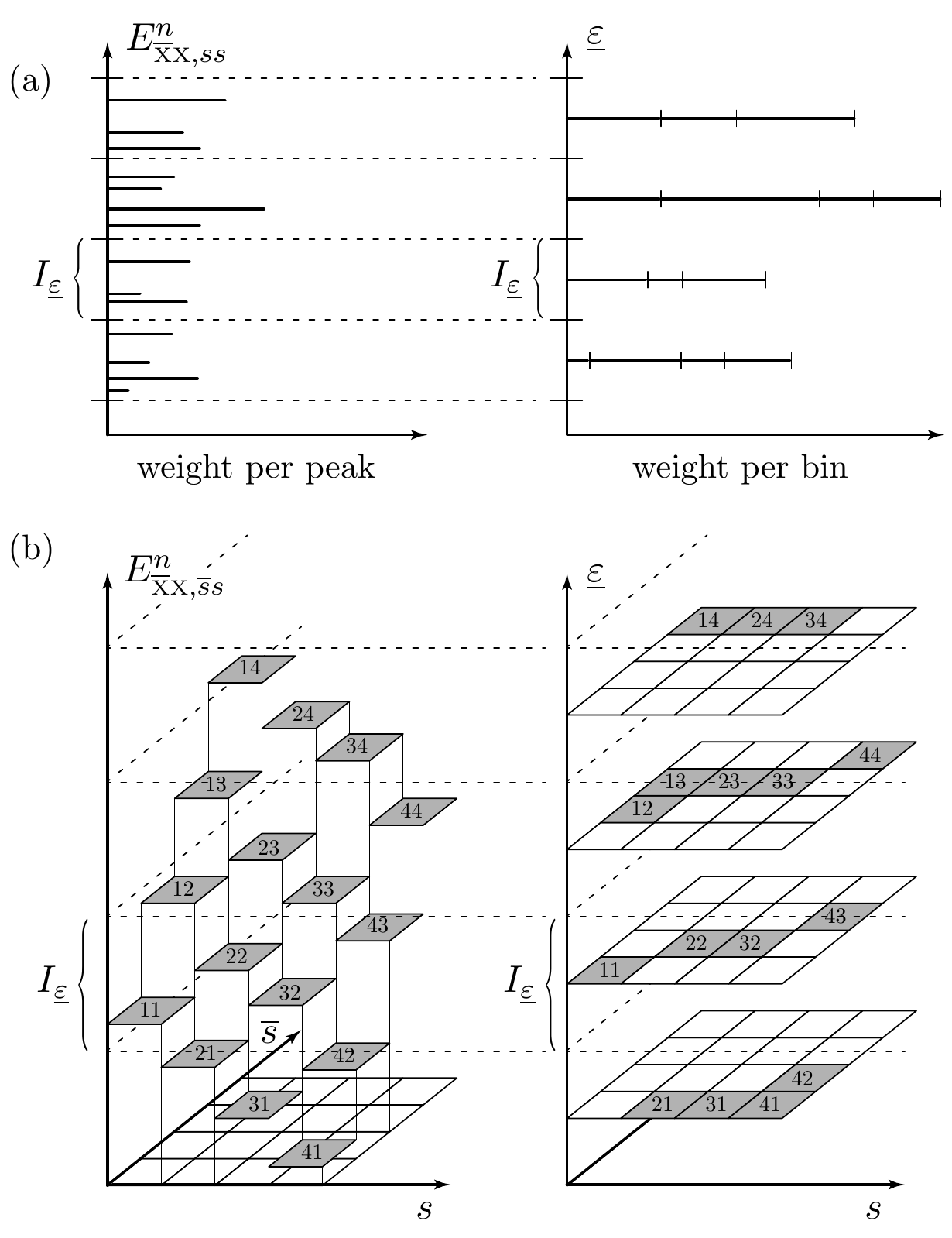}
\caption{\label{fig:BinningSlicing}
(a) Binning. Left: Each line represents a $ \delta$ peak;  
its length represents the associated weight, its vertical position the 
transition energy  $\iEdiff{n}{\xb}{\x}{\sb}{s}$. Right:
Peaks lying in the same interval
$I_\sve$ are treated as having the same energy difference $\sve$;
thus, their weights are added. 
(b) Slicing. Left: A gray shape labeled $s\sb$ represents the matrix element 
$\protect\io{O}{n}{\x}{\xb}{s}{\sb}$; 
its horizontal placement 
reflects the  values of the indices $s\sb$, the vertical position
the transition energy $\iEdiff{n}{\xb}{\x}{\sb}{s}$. Right:
Matrix elements with energy differences lying in the same interval
$I_\sve$ are treated as all having the same energy difference $\sve$ and are gathered into a matrix with elements 
$(\protect\io{O}{n}{\x}{\xb}{s}{\sb})^\sve$.
}
\end{figure}

We briefly interrupt our formal development with
an interlude on numerical matters. 
As illustrated by \Eq{eq:l=2ABFinalResult}, the Lehmann representations of PSFs involve sums over very many discrete $\delta$ functions, originating from the frequency-dependent matrix
elements in \Eq{eq:Oomega}. To obtain smooth 
functions, the discrete $\delta$ peaks have to be broadened 
at the end of the computation through convolution with a suitable
broadening kernel,
as further explained in Sec.~\ref{sec:broadening} and \App{app:logGaussian}.
Therefore, it is not necessary to keep track of the precise position, $\iEdiff{n}{\xb}{\x}{\sb}{s}$, of each peak.
(Storing all pairs of spectral weights and peak positions becomes intractable when large numbers of states are kept during the iterative diagonalization.) Instead, we may adopt a \textit{binning} strategy: 
we partition the $\ve$ axis into narrow intervals such that each interval $I_\sve$ is centered around $\sve$.
Then, for all 
spectral peaks with positions $\iEdiff{n}{\xb}{\x}{\sb}{s}$ lying within $I_\sve$, we adjust their positions to $\sve$, i.e., replace $\delta (\ve - \iEdiff{n}{\xb}{\x}{\sb}{s})$ by $\delta (\ve - \sve)$.
This discretization of frequency variables enables us to store the PSF as a histogram array;
each of its elements is the sum of the spectral weights whose transition energies fall into $I_\sve$ and are then adjusted to $\sve$.
This binning strategy is graphically explained in Fig.~\ref{fig:BinningSlicing}(a).

Because of the logarithmic bath discretization
underlying the Wilson chain, 
the $\sve$ grid should have a logarithmic structure, too, 
fine enough to resolve separate shells. 
Enumerating grid points $\sve[m]$ by an index $m$,
we choose $\sve[\pm m] = \pm 10^{(|m|-1)/n_\mr{dec}} \ve_\mr{min}$ for $m\!\neq\!0$ and $\sve[m\!=\!0] = 0$,
where $n_\mr{dec}$ is the number of grid points per decade and $\ve_\mr{min} = \sve[m\!=\!1]$ is the smallest energy resolved on the grid.
We limit the grid size by taking $|\sve[m]| \leq \ve_\mr{max}$, with $\ve_\mr{max}$ larger than all other energy scales in the system. 
The grid point at $\sve = 0$ collects spectral weights sitting at truly zero frequency (up to numerical precision) rather than small but finite frequencies. It is needed for the anomalous MF kernels in \Eq{eq:K_Omega_IF} and 
for constructing the PSFs of disconnected correlators (see \Sec{sec:ConnectedPart}).
When the grid is plotted on a logarithmic axis, 
the spacing between grid points on this axis, 
$\Delta_{\sve} = (\ln 10)/n_\mr{dec}$, 
should be much narrower than that between the discretized bath energy levels, $\Delta_\mr{LD} = \ln \Lambda$. 
In this work we use $\Lambda = 4$ and $n_\mr{dec} = 32$, yielding $\Delta_{\sve} = 0.072$ and $\Delta_\mr{LD} = 1.39$.

Whereas $2$p PSFs have only one frequency argument, the spectral contributions to $\ell$p PSFs for $\ell \!>\! 2$ have multiple frequency positions, $\ve_1, \ve_2, \ndots$.
Generally, the frequency values associated with the same spectral contribution can largely differ in size, e.g., $|\ve_1| \gg |\ve_2|$.
In such a case, the value of $\ve_1$ is determined at an earlier shell,
that of $\ve_2$ at a later shell. (See Secs.~\ref{sec:l=3PartialSpectralFunctions} and \ref{sec:l=4PartialSpectralFunctions} for details.)
Thus, information on large frequencies associated with early shells
has to be recorded while information on smaller frequencies from later
shells is still being evaluated.
For this, we use a \textit{slicing} strategy.

For a given shell $n$ and sector $\X\Xb$, the frequency-dependent matrix elements $(\io{O}{n}{\x}{\xb}{s}{\sb})^\ve$
constitute a three-dimensional object labeled by two discrete indices 
$s\sb$ and a continuous index $\ve$. Upon discretizing the latter, i.e., replacing $\ve$ by $\sve$, 
this object should be replaced by a rank-three tensor with discrete indices $s , \sb , \sve$, defined as 
\begin{align}
(\io{O}{n}{\x}{\xb}{s}{\sb})^\sve
= 
\begin{cases}
\io{O}{n}{\x}{\xb}{s}{\sb}  & \textrm{if} \quad 
\iEdiff{n}{\xb}{\x}{\sb}{s} 
\in I_\sve 
, 
\\
0 & \textrm{otherwise} 
.
\end{cases}
\end{align}%
Each $\sve$ value defines a two-dimensional slice through this
tensor, containing nonzero entries only if 
$\iEdiff{n}{\xb}{\x}{\sb}{s}$ 
lies within the bin $I_\sve$. Thus, slicing 
involves no physically motivated approximations---it is merely
a convenient way of dealing with a three-dimensional data structure.

The sliced tensor, depicted schematically in Fig.~\ref{fig:BinningSlicing}(b), has to be stored explicitly.  By keeping the $\sve$ leg open while computing tensor contractions involving $s\sb$, this task can be performed in parallel for all values of $\sve$.

To indicate diagrammatically that a frequency $\sve$ is treated by slicing, 
we will use a perpendicular slash, ``slicing'' through  the associated dashed line, as illustrated in the diagram after \Eq{eq:NestedMatrixElementsWilsonChainl=3}.
Dashed lines without slashes signify binning. Whether 
a $\delta$ function requires binning or slicing depends on the context; we will explain this in detail in the subsequent two sections.

\subsection{Partial spectral functions: \texorpdfstring{$\ell=3$}{l=3}}
\label{sec:l=3PartialSpectralFunctions}

We next turn to $3$p PSFs, $\eS [\Ac,\Bc,\Cc](\vec{\ve})$, 
with two frequency arguments, $\vec{\ve} = (\ve_1, \ve_2)$.
Their computation starts along the same
lines as that for $\ell=2$, with a top-down shell-diagonal
expansion of the threefold operator product using 
\Eq{eq:OperatorTripleProductRefinement}. However, extra
refinement efforts are required  if the two
frequencies differ significantly,  $|\ve_1| \ll |\ve_2|$ 
or $|\ve_2| \ll |\ve_1|$. We will express the 
terms describing these cases through 
$2$p PSFs of a single operator and a composite operator, 
the latter incorporating the dependence on the larger
frequency via slicing. A high resolution for the smaller
frequency can then be achieved by a further top-down expansion of the $2$p PSFs. 
This strategy, not needed for $\ell = 2$, is the main new ingredient for the NRG computation of higher-order PSFs. 

We again begin from \Eq{eq:4pt_tilde_spectral_rep_compact},
\begin{subequations}
\label{eq:ABomegaCompactl=3}
\begin{align}
\cor{\eS}[\Ac,\Bc,\Cc](\vec{\ve}) 
\label{eq:ABomegaCompactl=3a}
& = 
\Tr [(\varrho \Ac^{\ve_1} \Bc)^{\ve_2} \Cc]  
\\
\label{eq:ABomegaCompactl=3b}
& = \Tr [(\varrho \Ac)^{\ve_1} \Bc \, \Cc^{-\ve_2}]  
,
\end{align}
\end{subequations}
or $\eS(\vec{\ve})$ for short. 
The above two ways of writing the trace are equivalent
and will be used interchangeably.
Expanding \Eq{eq:ABomegaCompactl=3b} via \Eqs{eq:OperatorTripleProductRefinement}, we obtain 
\begin{align}
\label{eq:l=3PSFsumovern}
\cor{\eS} (\vec{\ve})
& =  \sum_n \sum_{\x \xb \xh}^{\neq \k\k\k} 
\ioc{\eS}{n}{\x \xb}{\xh}
[\Ac,\Bc,\Cc](\vec{\ve}) 
, 
\end{align}
where each $\eS^n$ denotes a shell-diagonal trace: 
\begin{align}
\ioc{\eS}{n}{\x \xb}{\xh}
[\Ac,\Bc,\Cc](\vec{\ve}) 
=
\Tr[
(\ioc{ (\varrho \Ac)}{n}{\xone}{\xtwo})^{\ve_1} 
\ioc{\Bc}{n}{\xtwo}{\xthree} \, 
(\ioc{\Cc}{n}{\xthree}{\xone})^{\! -\ve_2} ] 
. 
\label{eq:SnProjectorProduct3}
\end{align}
The following diagram
depicts the iterative refinement of $\K\K\K$ sectors
(red cubes) leading to \Eq{eq:l=3PSFsumovern}
(the $E_{\underline{1}}$-$E_{\underline{2}}$ 
plane lies at the front, and the $E_{\underline{3}}$ axis points away from the reader):

\noindent \includegraphics[width=0.98\linewidth]{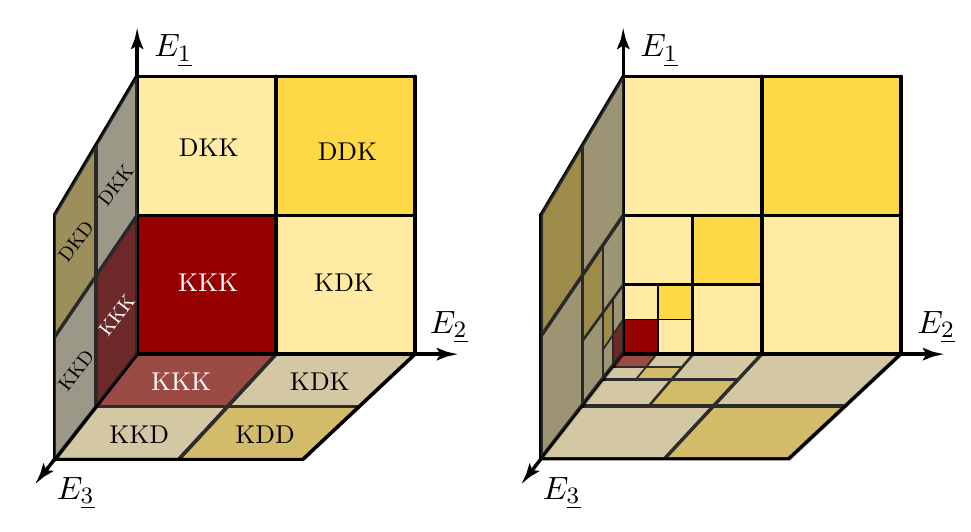}

Out of the seven contributions from the sum over sectors
in \Eq{eq:l=3PSFsumovern},
five, namely $\D\Xb\Xh$ and $\K\D\D$,  do not involve products of $\K\K$ sectors of the two frequency-dependent operators $(\varrho \Ac)^{\ve_1}$ and $\Cc^{-\ve_2}$. For instance, the $\D\D\K$ term 
$(\ioc{(\varrho \Ac)}{n}{\d}{\d})^{\ve_1} \ioc{\Bc}{n}{\d}{\k} (\ioc{\Cc}{n}{\k}{\d})^{-\ve_2}$ does not involve any $\K\K$ sectors; the $\D\K\K$ term 
$(\ioc{(\varrho \Ac)}{n}{\d}{\k})^{\ve_1} \ioc{\Bc}{n}{\k}{\k} (\ioc{\Cc}{n}{\k}{\d})^{-\ve_2}$ does, but only for the
frequency-independent $\Bc$. 
These five contributions can be computed using \Eq{eq:LocalExpectationShelln}, binning both $\ve_1$ and $\ve_2$ dependencies.
They
are adequately refined---further refinement, yielding
shell-off-diagonal energy differences, would be futile.

By contrast, the remaining two terms, $\K\K\D$ and $\K\D\K$,
do need further refinement.
Fortunately, 
this can be achieved by expressing them through \textit{two-point} PSF, albeit involving composite operators---this simplification is one of the crucial ingredients of
our strategy.
Consider, e.g., 
\begin{subequations}
\label{subeq:TnKKD}
\begin{align}
\label{eq:TnKKDa}
\ioc{\eS}{n}{\k\k\d}{} [\Ac,\Bc,\Cc](\vec{\ve}) 
= 
\Tr[
(\ioc{(\varrho \Ac)}{n}{\k}{\k})^{\ve_1} 
\ioc{\Bc}{n}{\k}{\d} 
(\ioc{\Cc}{n}{\d}{\k})^{-\ve_2} ] 
. 
\end{align}
The $\ve_1$ dependence enters via a $\K\K$ sector 
and should be further refined. By contrast, 
the $\ve_2$ dependence, entering via a $\D\K$ sector,
need not be refined further and can be sliced. We indicate this
by replacing $\ve_2$ by $\sve_2$.
We thus view $\ioc{\Bc}{n}{\k}{\d} (\ioc{\Cc}{n}{\d}{\k})^{-\sve_2}$ as the $\K\K$ sector of a composite operator with a sliced,
not direct, frequency dependence. Evoking \Eq{eq:ABomegaCompactl=2}, we can then read \Eq{eq:TnKKDa} as the
$\ve_1$-dependent $2$p PSF of a single and a composite operator:
\begin{align}
\label{eq:TnKKDb}
\ioc{\eS}{n}{\k\k\d}{} [\Ac,\Bc,\Cc](\vec{\ve}) 
\simeq
\eS [\ioc{\Ac}{n}{\k}{\k}, \ioc{\Bc}{n}{\k}{\d} 
(\ioc{\Cc}{n}{\d}{\k})^{-\sve_2} ](\ve_1) 
,
\end{align}
\end{subequations}
where we used $(\varrho\Ac)^n_{\k\k} = \varrho(\Ac)^n_{\k\k}$.
The ``$\simeq$'' sign signifies that the right-hand
side involves the numerical approximation of slicing, which
limits the resolution attainable for $\sve_2$ to $\simeq \Lambda^{-n/2}$. Note that slicing
is needed only for $\K\D$ and $\D\K$ sectors, but not for the 
$\D\D$ sector, since the latter is excluded during further refinement. 
This is numerically convenient, since slicing the $\D\D$ sector 
would require much more memory than for the $\K\D$ and $\D\K$ sectors. 

The $\K\D\K$ term can be treated
similarly. Starting from \Eq{eq:ABomegaCompactl=3a}, we express it as
\begin{subequations}
\label{eq:TnKDK}
\begin{align}
\label{eq:TnKDKa}
\ioc{\eS}{n}{\k\d\k}{}
[\Ac,\Bc,\Cc](\vec{\ve}) 
= 
\Tr \big[
\bigl(\ioc{\varrho}{n}{\k}{\k} 
( \ioc{\Ac}{n}{\k}{\d})^{\ve_1} 
\ioc{\Bc}{\,n}{\d}{\k} \bigr)^{\ve_2} \ioc{\Cc}{n}{\k}{\k} \big] 
.
\end{align}
Since $\ve_1$ enters through a $\K\D$ sector we slice
it as $\sve_1$, viewing $\bigl( (\ioc{\Ac}{n}{\k}{\d})^{\sve_1} 
\ioc{\Bc}{\,n}{\d}{\k} \bigr)^{\ve_2}$ as a composite operator.
Its dependence on $\ve_2$ enters through $\K\K$ elements;
hence, we refine it by identifying a $2$p PSF via 
\Eq{eq:ABomegaCompactl=2}:
\begin{align}
\ioc{\eS}{n}{\k\d\k}{}
[\Ac,\Bc,\Cc](\vec{\ve}) 
\label{eq:TnKDKb}
\simeq 
\eS [(\ioc{\Ac}{n}{\k}{\d})^{\sve_1} 
\ioc{\Bc}{\,n}{\d}{\k} , \ioc{\Cc}{n}{\k}{\k} ] (\ve_2) 
.
\end{align} 
\end{subequations}
We note that there also are other, equivalent ways of associating $\ve_1$ and $\ve_2$ with operators.
For example, if $\ioc{\eS}{n}{\k\d\k}{}$ is expressed via \Eq{eq:ABomegaCompactl=3b}, then the representation \eqref{eq:TnKDKb} would be obtained via \Eq{eq:ABomegaCompactl=2Alternatives}.
When implementing this scheme numerically, it is convenient to prioritize associating the original frequencies $\ve_i$ with operators rather than their sign-flipped 
versions $-\ve_i$, unless the sign flip is inevitable, e.g., as in \Eq{eq:TnKKDb}. This prioritization yields a better organization when the routine for $3$p PSFs is invoked recursively, as needed in the next section.

Since the $2$p PSFs \eqref{eq:TnKKDb} and \eqref{eq:TnKDKb}
are built from two $\K\K$ operators,
they can be refined top down
to improve the $\ve_1$ or $\ve_2$ resolution, 
respectively.
Iterative use of \Eq{eq:OperatorDoubleProductRefinementIterate} 
yields expansions similar to \Eq{eq:sumnABomegaCompactl=2}:
\begin{subequations}
\label{subeq:TnKKDKDKc} 
\begin{flalign}
\label{eq:TnKKDc} 
\ioc{\eS}{n}{\k\k\d}{} 
(\ve_1, \sve_2)
&  = \!
\sum_{\nb > n} \!\! \sum_{\,\,\xone \xtwo}^{\neq \k\k} 
\! \ioc{\eS}{\nb}{\xone \xtwo}{}
[\ioc{\Ac}{n}{\k}{\k}, \hspace{-0.3mm} \ioc{\Bc}{n}{\k}{\d} 
(\ioc{\Cc}{n}{\d}{\k})^{-\sve_2} ](\ve_1)  
,
\hspace{-0.5cm} & 
\\ 
\label{eq:TnKDKc}
\ioc{\eS}{n}{\k\d\k}{} 
(\sve_1, \ve_2)
& = \!
\sum_{\nb > n} \!\! \sum_{\,\,\xone \xtwo}^{\neq \k\k} 
\! \ioc{\eS}{\nb}{\xone \xtwo}{}[(\ioc{\Ac}{n}{\k}{\d})^{\sve_1} 
\ioc{\Bc}{n}{\d}{\k}, \ioc{\Cc}{n}{\k}{\k}] (\ve_2) 
.  
\hspace{-0.5cm}  &
\end{flalign}	
\end{subequations}
These expansions are restricted to shells $\nb > n$, since
they are seeded by operators
defined on the $\K\K$ sector of shell $n$ (their 
$\X\Xb \neq \K\K$ matrix elements vanish for all
shells $\le n$). The expansions \eqref{eq:TnKKDc}  and
\eqref{eq:TnKDKc} yield high resolution of the 
smaller frequency in the regimes $|\ve_1| \!<\! |\ve_2|$
or $|\ve_2| \!<\! |\ve_1|$, respectively. 
The diagram below
depicts the iterative refinement \eqref{eq:TnKDKc} of the $\K\D\K$ sector 
for fixed $\underline{\ve}_1$, represented
by the slanted blue plane (viewed from somewhat below, again with the $E_{\underline{1}}$-$E_{\underline{2}}$ 
plane at the front):

\noindent \includegraphics[width=0.98\linewidth]{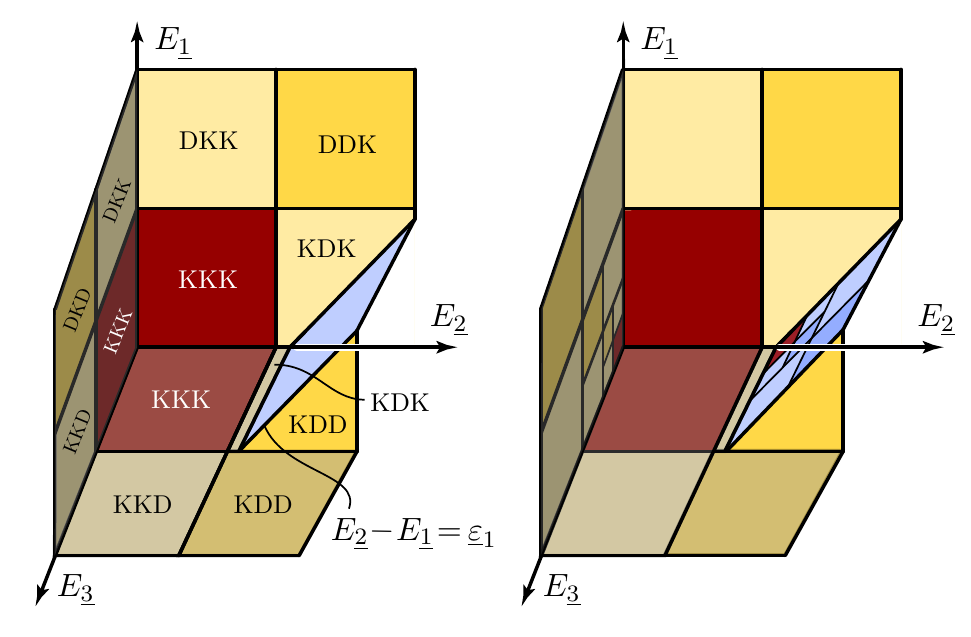}

It is instructive to visualize our treatment of $3$p PSFs by depicting the trace contraction patterns for \Eq{eq:SnProjectorProduct3}
and the summands of Eqs.~\eqref{eq:TnKKDc} and \eqref{eq:TnKDKc} 
diagrammatically:\\
\includegraphics[width=\linewidth]{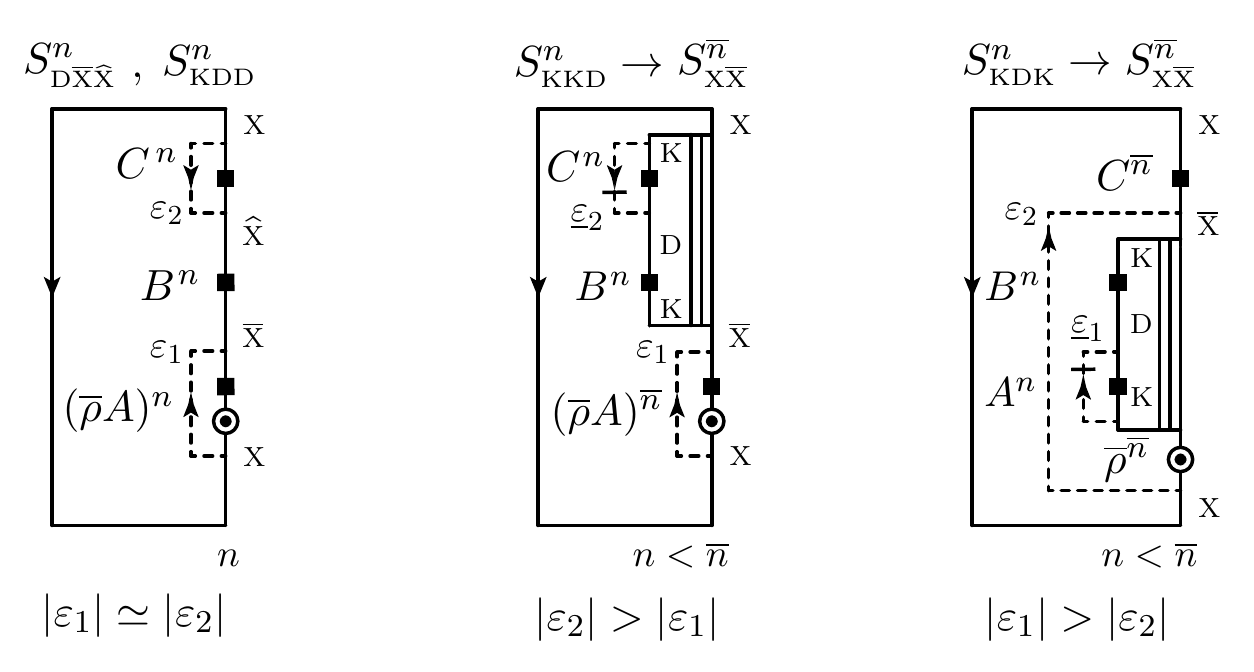} 
The middle and right-hand diagrams involve projections of $\K\K$ 
composite operators from shell $n$ onto later shells $\nb$,
e.g., $\bigl(\ioc{\varrho}{\nb}{\x}{\x}\iP{\nb}{\x} (\ioc{\Ac}{n}{\k}{\d})^{\sve_1} \ioc{\Bc}{n}{\d}{\k}
\iP{\nb}{\xb}\bigr){}^{\ve_2}$
for $\ioc{\eS}{n}{\k\d\k}{}$.
Such projections  can be computed recursively, 
as shown here: \\
\includegraphics[width=\linewidth]{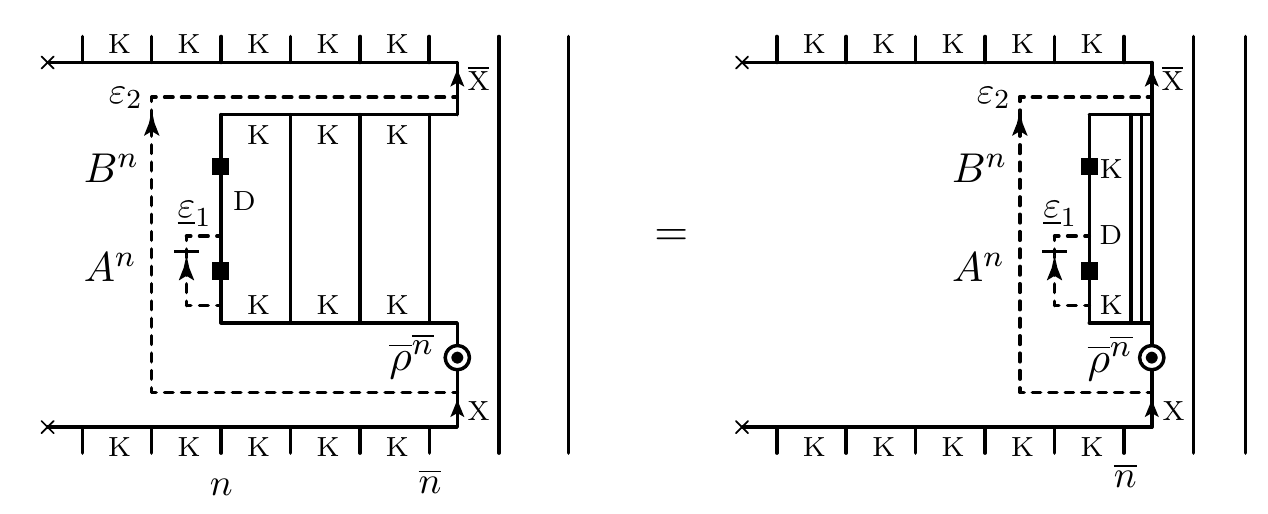}
It mimics the diagram after \Eq{eq:OperatorEarlyToLateProjection}
for finding $\Oc^n$ from $\Oc^{n_0}$, but is seeded by 
$(\ioc{\Ac}{n}{\k}{\d})^{\sve_1}$ and $\ioc{\Bc}{n}{\d}{\k}$, 
the former sliced. 
The triple line on the right edge of the effective matrix element signifies a contraction over the ladder from sites $n$ to $\nb$. 
The frequency dependence 
enters shell diagonally for both $\ve_1$  on site
$n$ and $\ve_2$ on site $\nb$, with resolution $\sim\! \Lambda^{-n/2}$ or $\Lambda^{-\nb/2}$, respectively. 
For $\nb$ much larger than $n$, such matrix elements
contribute only for frequencies satisfying $|\ve_1| \gg |\ve_2|$.
In physical terms, they give 
the amplitudes for the composite operator $\Ac \Bc$ 
to cause low-energy transitions with energy cost
$\ve_2$ via  virtual intermediate states with energy cost $\ve_1$. 
The $\ve_2$ dependence may be either binned
(as here) or sliced (as in Sec.~\ref{sec:l=4PartialSpectralFunctions}).

The expansion \eqref{eq:l=3PSFsumovern}, with summands $ \ioc{\eS}{n}{\x \xb}{\xh} [\Ac,\Bc,\Cc](\vec{\ve})$ given by \Eq{eq:SnProjectorProduct3} for $\D\Xb\Xh$ and $\K\D\D$, \Eq{eq:TnKKDc} for $\K\K\D$, and \Eq{eq:TnKDKc} for $\K\D\K$,
constitute our final formulas for 
$\cor{\eS}(\vec{\ve})$.  They express it through $\sum_n$ sums of both 
adequately refined three-operator traces and 
$2$p PSFs enabling further refinement.
The numerical cost for computing all adequately refined
terms scales as $\sim\! N$. 
The cost for computing the refinements of
all $2$p PSFs scales as $\sim\! N^2$,
requiring  a $\sum_{\nb > n}$ sum for every $n$, but this can be done in a nested fashion, reusing code written for $2$p PSFs. 
In other words, though \Eqs{subeq:TnKKDKDKc} and their
grapical depictions are instructive for understanding 
the structure of the refinement procedure, they need not
be coded explicitly; instead, $\ioc{\eS}{n}{\k\k\d}{}$ and
$\ioc{\eS}{n}{\k\d\k}{}$ can be computed using the $2$p PSF subroutine
implementing \Eq{eq:l=2ABFinalResult}, with some sliced matrix elements
as input.

\subsection{Partial spectral functions: \texorpdfstring{$\ell=4$}{l=4}}
\label{sec:l=4PartialSpectralFunctions}

Finally, we consider $4$p PSFs, $\eS(\vec{\ve})$ with  
$\vec{\ve} = (\ve_1, \ve_2, \ve_3)$.
Evoking \Eq{eq:4pt_tilde_spectral_rep_compact}, we will use the
five equivalent forms
\begin{flalign}
\label{eq:ABomegaCompactl=4}
& \cor{\eS}[\Ac,\Bc,\Cc,\Dc](\vec{\ve}) 
= \Tr [\bigl(\varrho \hspace{0.2mm} (\Ac^{\ve_1} \Bc)^{\ve_2} \Cc \bigr)^{\ve_3}
\Dc ]   \hspace{-1cm} &
\\ \nonumber
& = \Tr [(\varrho \hspace{0.2mm} 
\Ac^{\ve_1} \Bc)^{\ve_2} \Cc
\Dc^{-\ve_3} ] 
= \Tr [\bigl(\hspace{-0.2mm} 
(\varrho \Ac)^{\ve_1} \Bc\bigr)^{\ve_2} \Cc
\Dc^{-\ve_3} ] 
\\ \nonumber
&   = \Tr [(\varrho \Ac)^{\ve_1} \Bc (\Cc
\Dc^{-\ve_3})^{\! -\ve_2}]   
 = \Tr [(\varrho \Ac)^{\ve_1} \Bc^{\, \ve_2 - \ve_1} \, \Cc
\Dc^{-\ve_3}] 
.  
\hspace{-1cm} 
&
\end{flalign}
There is some freedom in choosing which form to use
in a given situation. Our choices below 
are convenient for expressing some contributions to $4$p PSFs through $2$p PSFs and $3$p PSFs involving composite operators containing sliced matrix elements, facilitating their numerical computation.

Expanding the third form 
via \Eqs{eq:OperatorQuadrupleProductRefinement} yields 
\begin{align}
\label{eq:sumnABomegaCompactl=4}
\cor{\eS}(\vec{\ve}) 
& = \sum_n \sum_{\x \xb \xh \xt}^{\neq \k\k\k\k} 
\ioc{\eS}{n}{\xone \xtwo \xthree \xfour}{} (\vec{\ve}) 
, 
\end{align}
with shell-diagonal traces $\eS^n$ defined as 
\begin{align}
\label{eq:ProjectorProduct4}
\ioc{\eS}{n}{\xone \xtwo \xthree \xfour}{} (\vec{\ve}) 
= \Tr \pig[
\pigl(\hspace{-0.4mm} 
\bigl(\hspace{-0.2mm} \ioc{(\varrho \Ac)}{n}{\xone}{\xtwo}\bigr)^{\ve_1} \, 
\!\ioc{\Bc}{n}{\xtwo}{\xthree}\pigr)^{\!\ve_2} 
\ioc{\Cc}{n}{\xthree}{\xfour}
(\ioc{\Dc}{n}{\xfour}{\xone})^{\! - \ve_3}
\pig]  
. 
\end{align}
Out of the 15 contributions from the sum over sectors,
nine, namely $\D\Xb\Xh\Xt$ and $\K\D\D\D$, do not involve products of $\K\K$ sectors of the three frequency-dependent operators $(\varrho \Ac)^{\ve_1}$, $\Bc^{\, \ve_2}$, $\Dc^{-\ve_3}$. 
These nine are adequately resolved and can be computed as shell-$n$ traces using \Eq{eq:LocalExpectationShelln} 
while binning all three frequencies.
By contrast, the other six contributions need  
$\K\K$ top-down refinements via PSFs of composite operators. 
The three terms involving two $\K$'s and two $\D$'s can be 
expressed through $2$p PSFs 
of a single operator and a composite triple, or of 
two composite doubles. For the $\K\K\D\D$ term, e.g.,
we express \Eq{eq:ProjectorProduct4}
through the fourth form in \Eq{eq:ABomegaCompactl=4}
and then slice the dependence on 
$\ve_2$ and $\ve_3$, entering via $\K\D$ or $\D\K$ sectors:
\begin{subequations}
\label{subeq:KKDD-KDKD-KDDK}
\begin{flalign}
\nonumber
\ioc{\eS}{n}{\k\k\d\d}{} (\vec{\ve}) 
& =  
\Tr \big[
\bigl(\hspace{-0.2mm}\ioc{(\varrho \Ac)}{n}{\k}{\k}\bigr)^{\ve_1} 
\ioc{\Bc}{n}{\k}{\d} \bigl(\hspace{-0.2mm}
 \ioc{\Cc}{n}{\d}{\d}
(\ioc{\Dc}{n}{\d}{\k})^{\! - \ve_3}
\bigr)^{-\ve_2} 
\big]  
\\ \label{eq:TnKKDD}
& \simeq \eS
[\ioc{\Ac}{n}{\k}{\k} , 
\ioc{\Bc}{n}{\k}{\d} \bigl(\hspace{-0.2mm}
 \ioc{\Cc}{n}{\d}{\d}
(\ioc{\Dc}{n}{\d}{\k})^{\! - \sve_3}
\bigr)^{-\sve_2} 
] (\ve_1) 
. 
\end{flalign}
Here, we view
$\ioc{\Bc}{n}{\k}{\d} \bigl(\hspace{-0.2mm}
\ioc{\Cc}{n}{\d}{\d}
(\ioc{\Dc}{n}{\d}{\k}){}^{- \sve_3}
\bigr){}^{-\sve_2} $ as the $\K\K$ sector  of 
a composite triple, sliced with respect to both $\sve_2$ and $\sve_3$, 
and the second line as
a $2$p PSF of two shell-$n$ operators [cf.~\Eq{eq:ABomegaCompactl=2}]. 
The $\K\D\K\D$ and $\K\D\D\K$ terms can be treated analogously, 
using the second and first forms of \Eq{eq:ABomegaCompactl=4}:
\begin{flalign}
\nonumber
\ioc{\eS}{n}{\k\d\k\d}{} (\vec{\ve}) 
& = 
\Tr \big[
\bigl(\ioc{\varrho}{n}{\k}{\k} (
\ioc{\Ac}{n}{\k}{\d})^{\ve_1} 
\ioc{\Bc}{n}{\d}{\k} \bigl)^{\ve_2} 
 \ioc{\Cc}{n}{\k}{\d}
(\ioc{\Dc}{n}{\d}{\k})^{ - \ve_3} 
\big]  \hspace{-1cm} & 
\\ \label{eq:TnKDKD}
& \simeq \eS[
(\ioc{\Ac}{n}{\k}{\d})^{\sve_1} 
\ioc{\Bc}{n}{\d}{\k} ,
\ioc{\Cc}{n}{\k}{\d}
(\ioc{\Dc}{n}{\d}{\k})^{ - \sve_3} 
] (\ve_2 ) 
, 
\hspace{-1cm} &
\\[0.5mm]
\nonumber\ioc{\eS}{n}{\k\d\d\k}{} (\vec{\ve}) 
& =  
\Tr \pig[
\pigl(\hspace{-0.4mm} \ioc{\varrho}{n}{\k}{\k} \bigl(\hspace{-0.2mm} (
\ioc{\Ac}{n}{\k}{\d} )^{\ve_1} 
\ioc{\Bc}{n}{\d}{\d} \bigr)^{\ve_2} 
\ioc{\Cc}{n}{\d}{\k} \pigl)^{\! \ve_3}
\ioc{\Dc}{n}{\k}{\k}
\pig]  \hspace{-1cm} & 
\\ \label{eq:TnKDDK}
& \simeq
\eS
\big[ \bigl(\hspace{-0.2mm}  (
\ioc{\Ac}{n}{\k}{\d} )^{\sve_1}
\ioc{\Bc}{n}{\d}{\d} \bigr)^{\sve_2} 
\ioc{\Cc}{n}{\d}{\k} , 
\ioc{\Dc}{n}{\k}{\k}
\big] (\ve_3) 
. 
\hspace{-1cm} &
\end{flalign} 
\end{subequations}

Each of the three $2$p PSFs defined in \Eqs{subeq:KKDD-KDKD-KDDK} depends on a single, nonsliced frequency argument, with the other two 
frequencies entering via slicing of composite operators. The
dependence on the nonsliced frequency can be further refined via a subsequent $\sum_{\nb > n}\ioc{\eS}{\nb}{\x}{\xb}$ expansion, as discussed in earlier sections. 
The representations used for \Eqs{eq:ProjectorProduct4} and
\eqref{subeq:KKDD-KDKD-KDDK} can be depicted as follows, \\
\includegraphics[width=\linewidth]{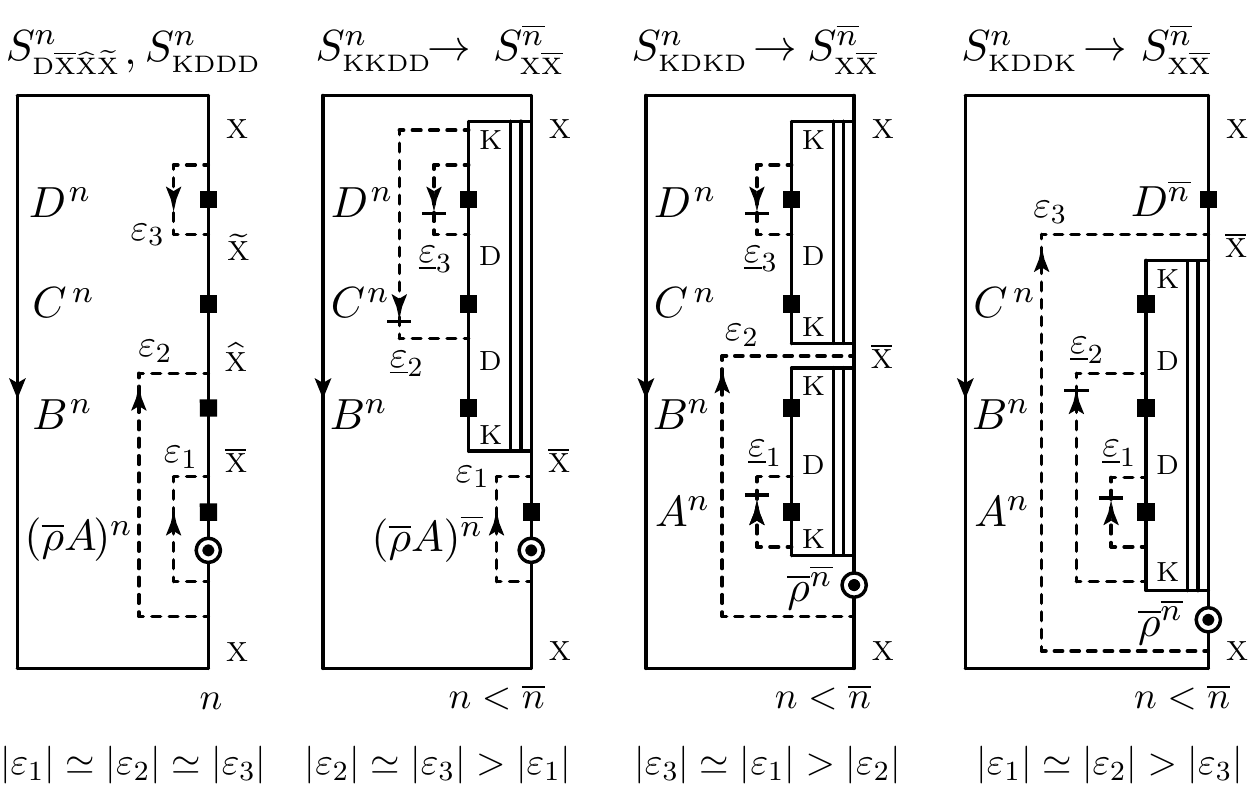} 
with composite triple matrix elements found iteratively: \\
\includegraphics[width=\linewidth]{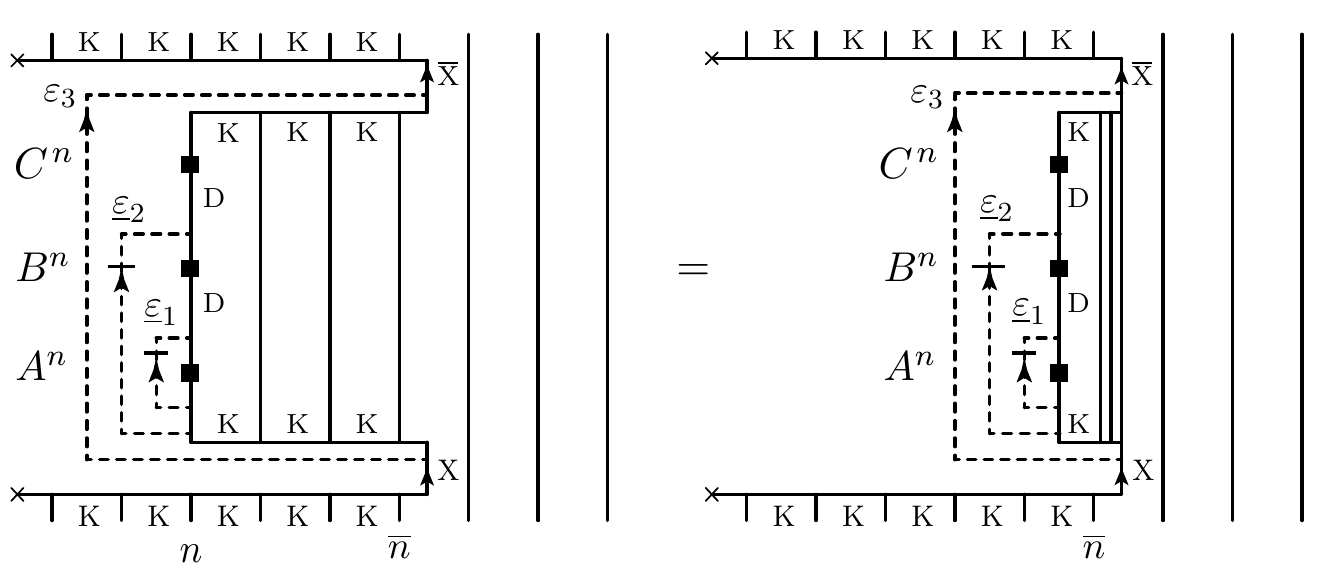}

The remaining three contributions to \Eq{eq:sumnABomegaCompactl=4}, involving three $\K$'s and one $\D$, can be expressed 
through  $3$p PSFs of two single operators and one composite double. 
Using the fourth, fifth, or first forms of \Eq{eq:ABomegaCompactl=4},
we obtain 
\begin{subequations}
\label{subeq:KKKD-KKDK-KDKK}
\begin{flalign}
\nonumber
\ioc{\eS}{n}{\k\k\k\d}{} (\vec{\ve})
& = 
\Tr \big[
\bigl(\hspace{-0.2mm}\ioc{(\varrho \Ac)}{n}{\k}{\k}\bigr)^{\ve_1} 
\ioc{\Bc}{n}{\k}{\k} 
 \bigl(\ioc{\Cc}{n}{\k}{\d}
(\ioc{\Dc}{n}{\d}{\k})^{ - \ve_3} \bigr)^{-\ve_2}
\big]  \hspace{-1cm} & 
	\\ \label{eq:TnKKKD}
& \simeq 
\eS [
  \ioc{\Ac}{n}{\k}{\k} , 
\ioc{\Bc}{n}{\k}{\k} , 
\ioc{\Cc}{n}{\k}{\d}
(\ioc{\Dc}{n}{\d}{\k})^{ - \sve_3} ](\ve_1, \ve_2)  , 
\hspace{-1cm} &
\\[1mm]
\nonumber
\ioc{\eS}{n}{\k\k\d\k}{} (\vec{\ve})
& = 
\Tr \big[
\bigl(\hspace{-0.2mm}\ioc{(\varrho \Ac)}{n}{\k}{\k}\bigr)^{\ve_1} 
(\ioc{\Bc}{n}{\k}{\d})^{\ve_2-\ve_1}  
\ioc{\Cc}{n}{\d}{\k}
(\ioc{\Dc}{n}{\k}{\k})^{ - \ve_3} 
\big]  \hspace{-1cm} & 
\\ \label{eq:TnKKDK}
	& \simeq \eS[
\ioc{\Ac}{n}{\k}{\k},  \!
(\ioc{\Bc}{n}{\k}{\d})^{\underline{\ve_2-\ve_1}}  
\ioc{\Cc}{n}{\d}{\k}, \!
\ioc{\Dc}{n}{\k}{\k}
] (\ve_1,\ve_3)  , \hspace{-1cm} & 
\\[1mm]
\nonumber
\ioc{\eS}{n}{\k\d\k\k}{} (\vec{\ve})
& = 
\Tr \pig[
\pigl(\hspace{-0.4mm} \ioc{\varrho}{n}{\k}{\k} \bigl(\hspace{-0.2mm} (
 \ioc{\Ac}{n}{\k}{\d} )^{\ve_1} 
\ioc{\Bc}{n}{\d}{\k} \bigr)^{\ve_2} 
 \ioc{\Cc}{n}{\k}{\k} \pigl)^{\! \ve_3}
\ioc{\Dc}{n}{\k}{\k}
\pig]  \hspace{-2cm} & 
\\ \label{eq:TnKDKK}
&  \simeq  \eS [
(\ioc{\Ac}{n}{\k}{\d})^{\sve_1} 
\ioc{\Bc}{n}{\d}{\k}
,\ioc{\Cc}{n}{\k}{\k}
,\! \ioc{\Dc}{n}{\k}{\k}
] (\ve_2, \ve_3)  , \hspace{-1cm} & 
\end{flalign} 
\end{subequations}
identifying $3$p PSF by 
evoking \Eq{eq:ABomegaCompactl=3b} for the first two  cases
and \Eq{eq:ABomegaCompactl=3a} for the last one.
For each $3$p PSF defined in \Eqs{subeq:KKKD-KKDK-KDKK}, the dependence on the two nonsliced frequencies
can be refined top down via an expansion $\sum_{\nb > n}\ioc{\eS}{\nb}{\xone \xtwo \xthree}{}$ as in \Eq{eq:l=3PSFsumovern}, but restricted to $\nb>n$. 
The resulting terms can be depicted as follows: \\
\includegraphics[width=\linewidth]{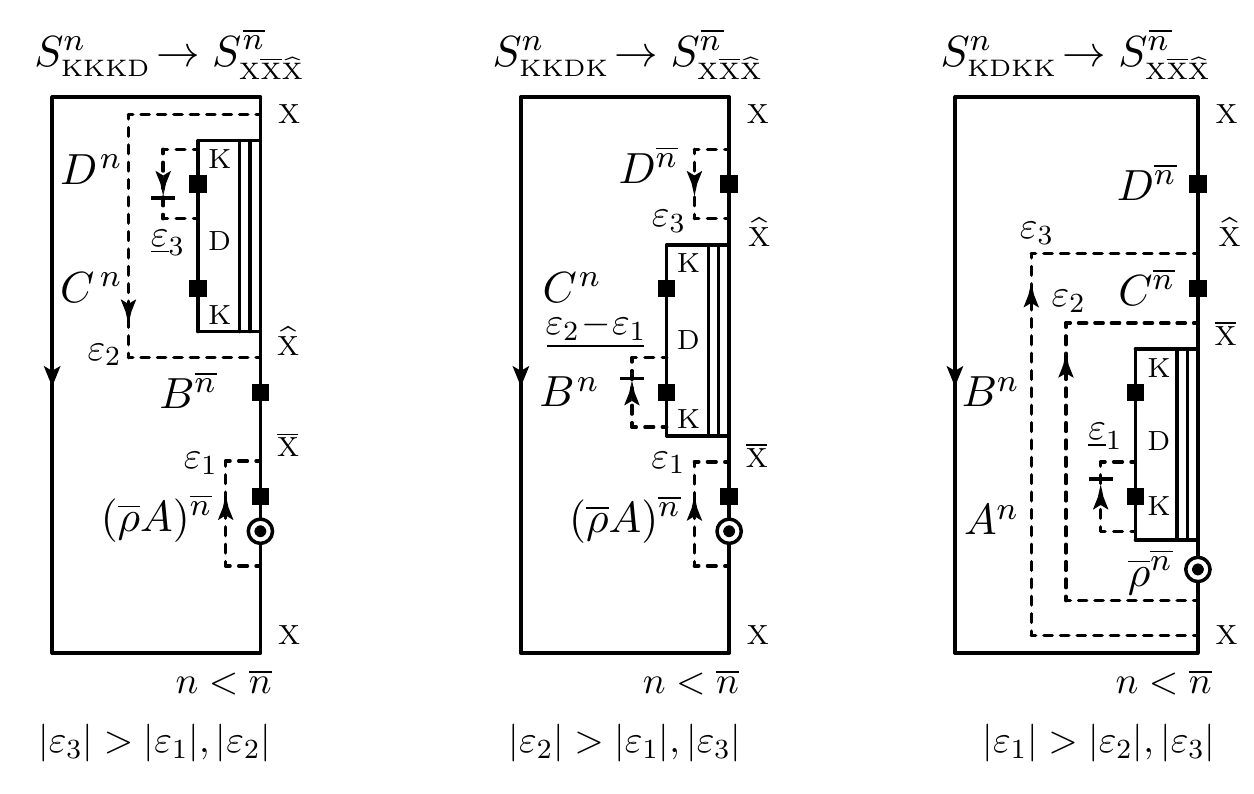}

Each such $\ioc{\eS}{\nb}{\xone \xtwo \xthree}{}$ is treated as
discussed in Sec.~\ref{sec:l=3PartialSpectralFunctions}: the  $\D\Xb\Xh$ and $\K\D\D$ contributions
via shell-$\nb$ traces, the $\K\K\D$ and $\K\D\K$ contributions
via identification of $2$p PSFs and their top-down expansions over shells 
$\nh \!>\! \nb$, thereby achieving
independent resolutions for the smaller two frequencies.
The $\nb \!>\! n$ expansion of $\ioc{\eS}{n}{\k\k\k\d}{}$, e.g., contributes
mainly for $|\ve_3| \!>\! |\ve_1|, |\ve_2|$. Each
$\ioc{\eS}{\tilde n}{\k\k\d}{}$ and $\ioc{\eS}{\tilde n}{\k\d\k}{}$ 
term in this expansion can be refined via a further top-down
$2$p PSF expansion 
$\sum_{\nh > \nb} \ioc{\eS}{\nh}{\x}{\xb}$, which mainly contributes for 
$|\ve_3| \!>\! |\ve_2| \!>\! |\ve_1|$ or
$|\ve_3| \!>\! |\ve_1| \!>\! |\ve_2|$, respectively: \\ 
\includegraphics[width=\linewidth]{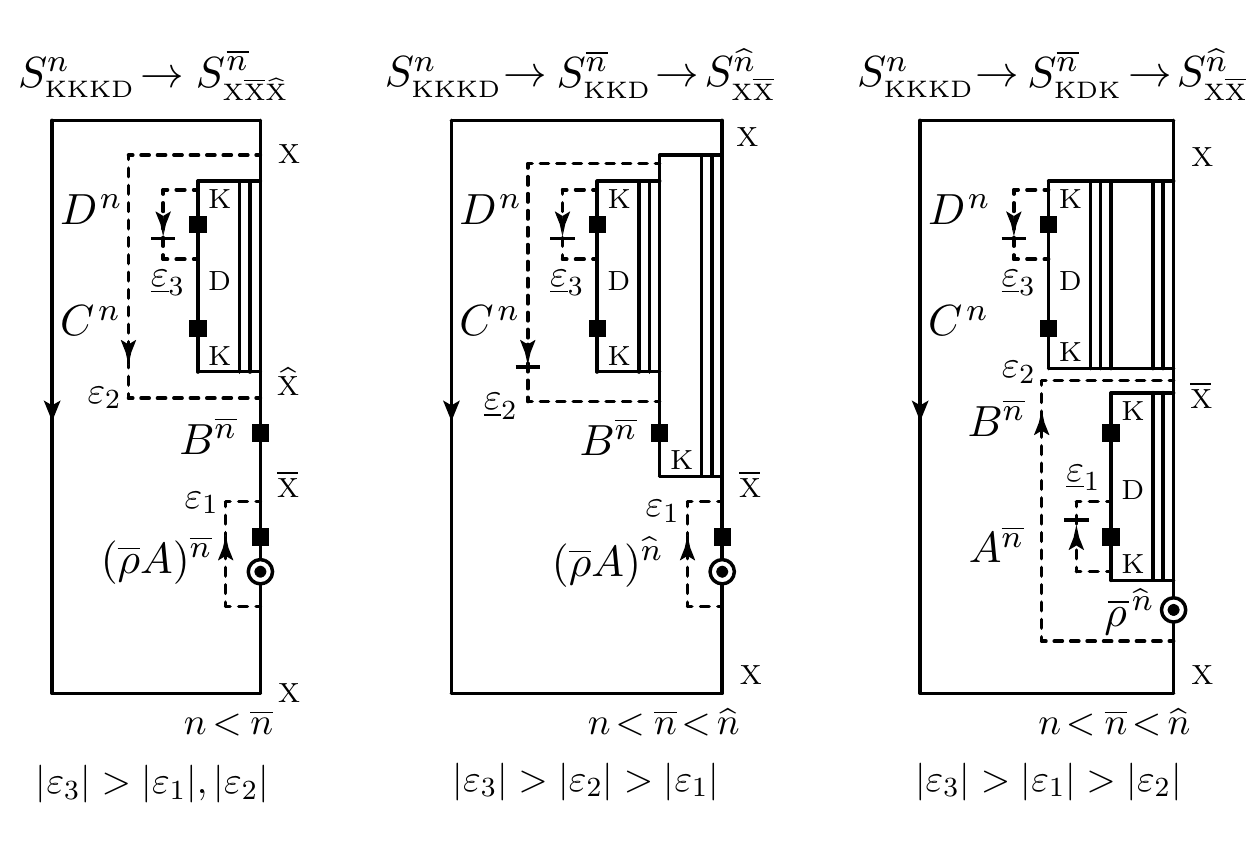} 
Similarly, the refinement of $\ioc{\eS}{n}{\k\k\d\k}{}$
yields proper resolution for
$|\ve_2| \!>\! |\ve_1 | , |\ve_3|$,
and of $\ioc{\eS}{n}{\k\d\k\k}{}$ for $|\ve_1| \!>\! |\ve_2| , |\ve_3|$.
The matrix elements of composite triples arising
in  such $\nh \!>\! \nb \!>\! n$ expansions are built from nested
combinations of matrix elements of composite doubles, such as \\
\includegraphics[width=\linewidth]{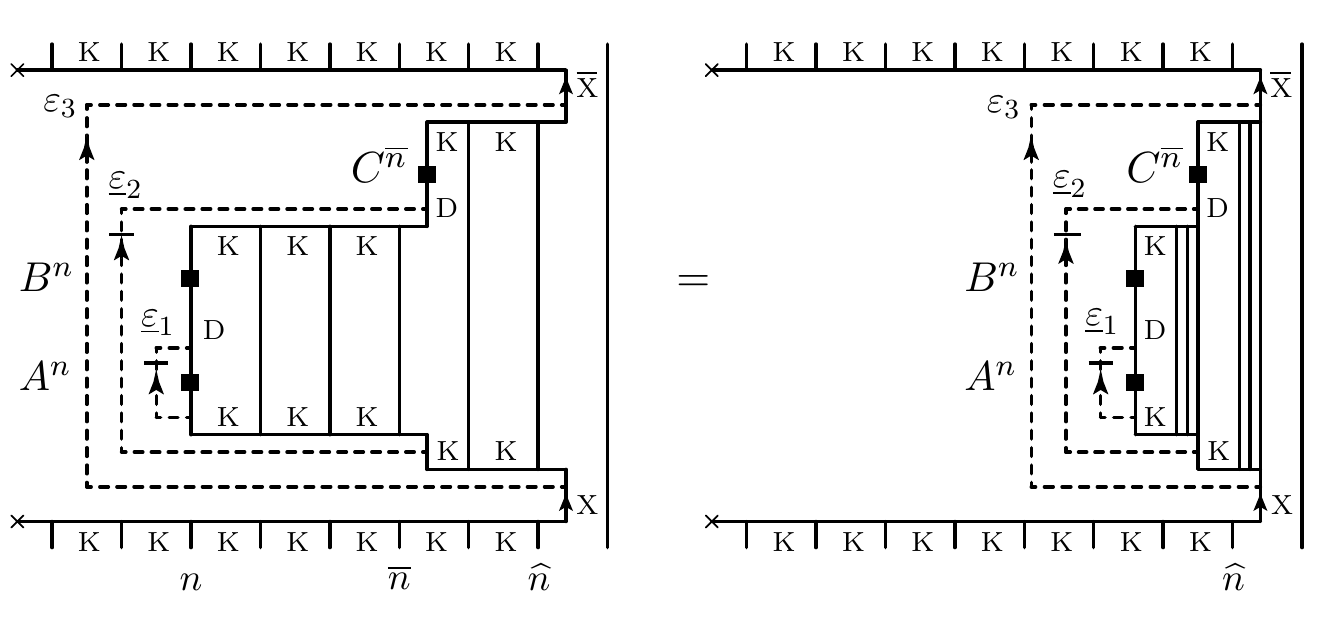}

\Equs{eq:sumnABomegaCompactl=4}--\eqref{subeq:KKKD-KKDK-KDKK}
are our final formulas for computing $4$p PSFs. The 
cost for computing all the adequately resolved 
$\D\Xb\Xh\Xt$ and $\K\D\D\D$ contributions scales
as $\sim\! N$, for all $\K\K\D\D$, $\K\D\K\D$, $\K\D\D\K$ as $\sim\! N^2$
(computation of a $2$p PSF for every $n$), 
and for all $\K\K\K\D$, $\K\K\D\K$ and $\K\D\K\K$ as
$\sim\! N^3$ (a $3$p PSF for every $n$).  Though costly, these computations can be performed
in a systematic, nested fashion, reusing the subroutines for $2$p PSFs when
computing $3$p PSFs, and the $2$p and $3$p PSF subroutines when computing $4$p PSFs. 
Of course, this presumes subroutines which 
work irrespective of how their input operators have been generated and whether they are sliced or not. 
There is then no need to write separate routines for all 
those diagrams above depicting $\sum_{\nb> n}$ or $\sum_{\nh > \nb}$ refinements;
instead, their contributions are generated automatically through nested function calls.

\section{From discrete partial spectral functions to connected correlators}
\label{sec:from-PSF-to-G}

The PSF computations described in the preceding sections lead to 
numerical results with the following structure. For a given permutation $p$, the PSF $\eS_p(\vec{\ve})$ defined in \Eq{eq:define-tilde-S-early} 
is represented as 
\begin{align}
\label{eq:define_Sgrid}
\eS_p(\vec{\ve}) = {\textstyle \sum_{\vec{\sve}}} \delta(\vec{\ve} - \vec{\sve})
\eS_p (\vec{\sve}) ,
\end{align}
where $\vec{\ve}$ and $\vec{\sve}$ are $(\ell\!-\! 1)$-tuples, 
the former containing the continuous variables $\ve_i =
\omega'_{\ovb{1} \cdots \ovb{i}}$ ($i \! <  \! \ell$), 
the latter  consisting of discrete frequencies
defined on the logarithmic grid used for binning and slicing, and
 $\delta(\vec{\ve} - \vec{\sve}) = \prod_{i=1}^{\ell-1} \delta(\ve_i - \sve_i)$. The coefficients $\eS_p (\vec{\sve})$ on the right, distinguished from the function $\eS_p (\vec{\ve})$ on the left by having arguments with underbars, are stored as an $(\ell\!-\!1)$-dimensional histogram. 

Such PSFs have to be convolved with a suitable kernel to compute
$\ell$p correlators. 
If the spectral representations of Sec.~\ref{subsec:SummarySpectralRepresentations} are expressed
through $\ve_i$ variables,
the convolution integrals $\int \md^{\ell-1} \omega'_p$ over
functions of $\vec{\omega}'_p$ (subject to $\omega'_{\ovb{1} \cdots \ovb{\ell}}=0$) turn into $\int \md^{\ell-1} \ve$ integrals
over functions of $\vec{\ve}$, yielding
$\sum_{\vec{\sve}}$ sums upon insertion of 
\Eq{eq:define_Sgrid} for $S_p(\vec{\ve})$. 
For \ZF/ correlators, e.g., we obtain [cf.~\Eqs{eq:G_w_ZF} and \eqref{eq:K_w_ZF}]
\begin{subequations}
\label{subeq:G-epsilon-S-ZF}
\begin{align}
\label{eq:G-epsilon-S-ZF}
G(\vec{\omega}) 
& = 
\sum_p \vec{\zeta}^p 
\nint \md^{\ell-1} \ve \, 
K(\vec{\omega}_p, \vec{\ve}) \eS_p (\vec{\ve}) 
\\ 
\label{eq:G-discretized-S-ZF}
& = 
\sum_p \vec{\zeta}^p  
\sum_{\vec{\sve}}
K(\vec{\omega}_p, \vec{\sve}) \eS_p (\vec{\sve})  ,
\\ 
\label{eq:G-discrete-Kernel-S-ZF}
K(\vec{\omega_p}, \vec{\sve}) & = 
\prod_{i=2}^{\ell} 
\big( -\omega_{\ovb{i} \cdots \ovb{\ell}} + 
\mi \gamma_{\ovb{i} \cdots \ovb{\ell}} + \sve_i \big)^{-1}  
. 
\end{align}
\end{subequations}
Similar expressions are obtained for \MF/ and \KF/ correlators. Evidently,
each grid point $\vec{\sve}$ defines the position of a pole
for the kernel. To avoid a proliferation of poles, we choose the same $\vec{\sve}$ grid for all permutations $p$. 

Given a set of PSF histograms, the numerical evaluation of the above expressions 
is conceptually simple but requires great care in practice. 
One reason is that the \ZF/ and \KF/ kernels have poles lying  very close to the real axis. The corresponding correlators, expressed as discrete $\sum_{\vec{\sve}}$ sums
over such kernels as in \Eq{eq:G-discretized-S-ZF}, therefore do not yield smooth functions of frequency. To recover \ZF/ or \KF/ correlators 
with a smooth frequency dependence, as desired for a system whose original bath (before discretization) is continuous, 
the $\delta$ peaks of the PSFs have to be broadened. 
However, the broadening scheme should not spoil delicate cancellations in the sum $\sum_p$ over permutations---a challenging requirement.

By contrast, \MF/ correlators need no broadening:
their frequency dependence is smooth, since it enters
as $\mi \vec{\omega}$, so that the kernel denominators 
are always sufficiently far from zero. 
Vanishing bosonic frequencies in denominators
are avoided, as their contributions are encoded
in anomalous terms [cf.\ \Eq{eq:K_Omega_IF}]. 
 
The second major challenge arises if one is interested
in computing the connected part, $G^\connected = G - G^\disconnected$, of a correlator. It describes mutual correlations between particles (in contrast to their independent propagation), and is  needed for extracting a corresponding vertex by amputating the external legs. Since the disconnected part $G^\disconnected$
has very large extremal values,
its subtraction has to be performed very accurately, 
and in such a manner that the discrete poles of $G$ and $G^\disconnected$ are aligned. 

The present section describes how we deal with these challenges,
with focus on computing $G^\connected$ for fermionic $4$p correlators. 
We employ a multistep procedure: 
(A) Disconnected contributions are subtracted
already at the level of PSFs. 
(B1) The PSFs are broadened in a Lorentzian manner and (B2) convolved with kernels to obtain $G^\connected$; 
in the \MF/, (B1) is omitted. 
(C) An equation-of-motion (EOM) scheme is used to improve the quality of $G^\connected$;
this eliminates small disconnected contributions which remain after step (A) due to numerical inaccuracies. 
(D) External legs are amputated to obtain the vertex $F$. 
For the \ZF/ and \KF/ cases, the external legs are computed using the same Lorentzian broadening scheme as in (B1); 
improved results for $G^\connected$ can be obtained 
by subsequently reattaching external legs, computed using a more refined log-Gaussian broadening scheme. 
In the ensuing sections, we motivate and describe these steps in detail. Numerical results for 
$G^\connected$ are presented in Sec.~\ref{sec:Results-Connected} below;
results for $F$ can be found in \paperI.

\subsection{Subtracting disconnected parts}
\label{sec:ConnectedPart}

Any multipoint correlator can be split into a connected and a disconnected part, 
$G = G^\connected + G^\disconnected$.
For example, for \ZF/ 4p correlators involving four fermionic operators, 
$\vec{\Oc} = (\Oc^1, \Oc^2, \Oc^3, \Oc^4)$, the disconnected part is ($\zeta \!=\! -1$)
\begin{flalign}
G^\disconnected [\vec{\Oc}] & (\vec{\omega}) 
= 
-2 \pi \mi \pig\{
G [\Oc^1, \Oc^2] (\omega_1) \, G[\Oc^3, \Oc^4] (\omega_3) \, \delta (\omega_{12}) 
\hspace{-0.5cm} &
\nonumber
\\
& \ + 
\zeta 
G [\Oc^1, \Oc^3] (\omega_1) \, G[\Oc^2, \Oc^4] (\omega_2) \, \delta (\omega_{13}) 
&
\nonumber
\\
& \ + 
G [\Oc^1, \Oc^4] (\omega_1) \, G[\Oc^2, \Oc^3] (\omega_2) \, \delta (\omega_{14}) 
\pig\}
,
&
\label{eq:Gdis}
\end{flalign}
with each summand a product of two $2$p correlators. The subtraction of $G^\disconnected$ from the full $G$ is numerically nontrivial for two reasons. First, the extremal values of $G^\disconnected$ are generally much larger in magnitude than $G^\connected$:
In the \ZF/ and \KF/, the 2p frequency conservation inherent 
in $G^\disconnected$ is implemented via Dirac $\delta$ functions 
[cf.\ \Eq{eq:Gdis}] whose peak heights, even if regularized, are very large. 
In the \MF/, the Dirac $\delta$'s are replaced by Kronecker $\delta$'s,
with a potentially large prefactor $1/T$.
Second, the $\delta$'s (Dirac or Kronecker) 
implementing 2p frequency conservation restrict the three-dimensional (3D) argument of $G(\vec{\omega})$ to a 2D plane. This 2D dependence is incompatible with the broadening scheme used in this work, which broadens $4$p PSFs along all three frequency directions (see \Sec{sec:broadening} below).
Indeed, the 2D dependence is a peculiarity of the disconnected part;
if a $4$p correlator does not involve operators 
describing decoupled degrees of freedom, 
its connected part should have a full 3D structure.

Given these challenges, it is advisable to subtract the disconnected
part \textit{prior} to broadening, already at the level of PSFs. 
To this end,
we exploit the fact \cite{Kugler2021} that the decomposition 
$G = G^\connected + G^\disconnected$ has a counterpart at the level of PSFs, 
$S_p= S_p^\connected + S_p^\disconnected$ for each permutation $p$. 
The disconnected PSFs $S^\disconnected[\vec{\Oc}_p] (\vec{\omega}_p)$ corresponding to $G^\disconnected [\vec{\Oc}]  (\vec{\omega}) $ are defined  in direct analogy  to \Eq{eq:Gdis} 
[see Eq.~\eqdisconnectedpaperI\ in \paperI].
Expressed in the notation of \Eq{eq:define-tilde-S-early},
with $\ve_i = \omega_{\ovb{1} \cdots \ovb{i}}$ and 
$\omega_{\ovb{1}} = \ve_1$, $\omega_{\ovb{2}} = \ve_2-\ve_1$, $\omega_{\ovb{3}} = \ve_3 - \ve_2$, $\omega_{\ovb{4}} 
= - \omega_{\ovb{1} \ovb{2}  \ovb{3}} = - \ve_3$, they read
\begin{flalign}
\label{eq:Sc_w-dis-epsilon}  
S^\disconnected_p (\vec{\ve})
& = 
S [\Oc^{\ovb{1}},\Oc^{\ovb{2}}](\ve_1) 
S [\Oc^{\ovb{3}},\Oc^{\ovb{4}}](\ve_3 - \ve_2)
\delta(\ve_2)
&
\\ 
& \ \ + 
\zeta
S [\Oc^{\ovb{1}},\Oc^{\ovb{3}}](\ve_1)
S [\Oc^{\ovb{2}},\Oc^{\ovb{4}}](\ve_3) 
\delta(\ve_1 - \ve_2 + \ve_3)
& 
\nonumber
\\ 
& \ \ + 
S [\Oc^{\ovb{1}},\Oc^{\ovb{4}}](\ve_1)
S [\Oc^{\ovb{2}},\Oc^{\ovb{3}}](\ve_2 - \ve_1)
\delta(\ve_1 - \ve_3) .
\hspace{-1cm}
&
\nonumber
\end{flalign}

Rather than computing the $2$p PSFs occurring herein anew, 
we extract them from the (full) $4$p PSFs via generalized sum rules, without extra NRG calculations.
This facilitates the cancellation between $S_p$ and $S^\disconnected_p$,
as then all ($4$p and $2$p) PSFs are obtained with 
the same accuracy. The sum rules are of the type
\begin{align}
\nint \! \md \varepsilon_3 
S [\Oc^1 \!\!\!\;,\!\!\; \Oc^2 \!\!\!\;,\!\!\; \Oc^3 \!\!\!\;,\!\!\; \Oc^4] 
(\varepsilon_1 \!\!\;,\!\!\; \varepsilon_2 \!\!\;,\!\!\; \varepsilon_3) 
& \!=\! 
S [\Oc^1 \!\!\!\;,\!\!\; \Oc^2 \!\!\!\;,\!\!\; \Oc^3 \Oc^4] 
(\varepsilon_1 \!\!\;,\!\!\; \varepsilon_2)
, 
\nonumber
\\
\nint \md \varepsilon_2 
S [\Oc^1 \!\!\!\;,\!\!\; \Oc^2 \!\!\!\;,\!\!\; \Oc^3] 
(\varepsilon_1 \!\!\;,\!\!\; \varepsilon_2) 
& \!=\! 
S [\Oc^1 \!\!\!\;,\!\!\; \Oc^2 \Oc^3] (\varepsilon_1)
. 
\label{eq:PSFsumrules}  
\end{align}
These rules, and the (anti)commutation relation 
$d_{\sigma}^\pdag d_{\sigma}^\dagger - \zeta  
d_{\sigma}^\dagger d_{\sigma}^\pdag = 1$,
imply that a 2p PSF 
can be obtained from a double integral over two $4$p PSFs, e.g.,
\begin{align}
\label{eq:4PSFto2PSF}
& 
S[\Oc^1 \!\!\!\;,\!\!\; \Oc^2] (\varepsilon_1) 
= 
\iint \md\varepsilon_2 \, \md\varepsilon_3 
\\
& \quad 
\pig\{
S [\Oc^1 \!\!\!\;,\!\!\; \Oc^2, d_{\sigma}^\pdag, d_{\sigma}^\dagger] (\vec{\varepsilon})  
- 
\zeta S [\Oc^1 \!\!\!\;,\!\!\; \Oc^2, 
d_{\sigma}^\dagger, d_{\sigma}^\pdag] (\vec{\varepsilon}) 
\pig\} 
. 
\nonumber
\end{align}

As consistency checks on our numerical data, we 
verified that the 2p PSFs  
$S [d_\sigma^\pdag,d^\dagger_\sigma](\sve_1)$ and $S [d^\dagger_\sigma, d_\sigma^\pdag](\sve_1)$ obtained from 4p PSFs
(i) satisfy the sum rule 
$\sum_{\sve_1} \! \bigl\{ S [d_\sigma^\pdag,d^\dagger_\sigma](\sve_1) \! + 
\! S [d_\sigma^\dagger,d_\sigma^\pdag](\sve_1) \! \bigr\} \! = \! 1$ 
with machine precision
and (ii) match those obtained
from direct 2p computations with an absolute accuracy of $10^{-3}$
(see Fig.~\ref{fig:Adisc2} in \App{app:1d-from-3d-histograms}).
Property (i) is expected, since our iterative scheme for computing $\ell$p PSFs conserves sum rules by construction. 
The quality of agreement for (ii) is very satisfying,
given the complexity of the computation. 
Nevertheless, small discrepancies remain, 
as discussed in \App{app:1d-from-3d-histograms}. 
Therefore, it is preferable to compute $S^\disconnected_p$ using 2p PSFs obtained not directly, but from precisely the 4p PSFs involved in $S_p$. 
Then, numerical inaccuracies inherent in $S_p$ are passed on to $S_p^\disconnected$,
facilitating the proper removal of disconnected 
contributions in $S_p - S^\disconnected_p$.

Inserting discrete 2p PSFs, 
$S[\Oc^1 \!\!\!\;,\!\!\; \Oc^2] (\ve_1) 
= 
\sum_{\sve_1} \delta(\ve_1 - \sve_1)   
S[\Oc^1 \!\!\!\;,\!\!\; \Oc^2] (\sve_1)$,
into \Eq{eq:Sc_w-dis-epsilon}, 
and using sums on $\sve_1$ and $\sve_3$ for the first and second 
factors in each line, we obtain
\begin{align}
\label{eq:Sp_disconnected_binned}
& S_p^\disconnected  (\vec{\varepsilon}) = 
\\
& 
\textstyle
\sum_{\sve_1, \sve_3} 
\pig\{ 
S[\Oc^{\ovb{1}}, \Oc^{\ovb{2}}] (\sve_1) S[\Oc^{\ovb{3}}, \Oc^{\ovb{4}}] (\sve_3) \, \delta\big(\vec{\varepsilon} - (\sve_1, 0, \sve_3)\big) 
\nonumber
\\
& + \zeta S[\Oc^{\ovb{1}}, \Oc^{\ovb{3}}] (\sve_1) S[\Oc^{\ovb{2}}, \Oc^{\ovb{4}}] (\sve_3) \, 
\delta\big(\vec{\varepsilon} - (\sve_1, \sve_1 + \sve_3, \sve_3)\big) 
\nonumber
\\
& + S[\Oc^{\ovb{1}}, \Oc^{\ovb{4}}] (\sve_1) S[\Oc^{\ovb{2}}, \Oc^{\ovb{3}}] (\sve_3) \, 
\delta\big(\vec{\varepsilon} - (\sve_1, \sve_1 + \sve_3, \sve_1)\big) 
\pig\}
.
\nonumber
\end{align}
We seek to express $S_p^\connected = S_p - S_p^\disconnected$ 
in the form \eqref{eq:define_Sgrid}, with histograms 
$S_p (\vec{\sve})$ and $S_p^\disconnected(\vec{\sve})$ defined on 
\textit{matching} 3D $\vec{\sve}$ grids, so that their difference directly
yields the histogram $S_p^\connected(\vec{\sve})$. 
(Different grids for $S_p$ and $S_p^\disconnected$ would yield
nonmatching pole positions for $G$ and $G^\disconnected$
and a less reliable subtraction.) 
In \Eq{eq:Sp_disconnected_binned}, the first term has support 
in the $\sve_2 = 0$ plane of the 3D grid. 
The second and third terms, however, 
have support at $\ve_2 = \sve_1 + \sve_3$, 
defining slanted planes incommensurate with the 3D grid. 
To associate these terms with points \textit{on} the 3D grid, 
we rebin $\sve_1 + \sve_3$ onto the $\sve_2$ grid, 
e.g., assigning a histogram element
associated with the transition energies $(\sve_1, \sve_1 + \sve_3, \sve_i)$ ($i \!=\! 3,1$) to the grid point $(\sve_1, \sve_2, \sve_i)$ closest to 
it on a logarithmic scale 
(i.e., with $\ln|\sve_2|$ closest to $\ln|\sve_1 + \sve_3|$). 

Since rebinning, needed to align the poles of
$G$ and $G^\disconnected$, slightly shifts some transition
energies without recomputing the corresponding matrix elements, 
it introduces a non-negligible numerical error. Furthermore, at finite temperature, the effective finite length of the Wilson chain~\cite{Weichselbaum2007,Weichselbaum2012b} 
introduces a low-energy cutoff,
which limits the spectral resolution below $T$. 
More specifically, the thermal density matrix $\ioc{\varrho}{n}{\d}{\d}$ is mostly populated at shells 
$n \sim n_T = -2 \log_\Lambda (T/D)$ \cite{Weichselbaum2007,Weichselbaum2012b}. 
Therefore, transitions originating from very low-lying energy levels with 
$E_\ub{1} \!\ll\! T$, 
resolved only within later shells with $n \!>\! n_T$, 
are not captured accurately.
Because of these numerical artifacts, arising from rebinning and finite temperature,
the disconnected part in $S$ is not canceled exactly.
For example, \Figs{fig:Adisc4}(b) and \ref{fig:Adisc4_same_spin}(b) below show results for 
$S^\connected$ for a noninteracting system; 
it should vanish by Wick's theorem but does not. 
We remedy this by removing the remaining disconnected contributions with a EOM-based strategy, described in  Sec.~\ref{sec:EOM} below. 
Yet, before that, a broadening procedure, described next,
is required for \ZF/ and \KF/ kernels.

\subsection{Broadening}
\label{sec:broadening}

For \ZF/ and \KF/ correlators, whose kernels have poles lying very close to the real axis, 
the PSFs must be broadened before convolving them with kernels, as mentioned above. 
To this end, each Dirac $\delta(\vec{\ve} - \vec{\sve})$ in $S_p(\vec{\ve})$  is replaced by a peak-shaped broadening kernel, $\delta_\broad(\vec{\ve}, \vec{\sve}) = \prod_{i=1}^{\ell-1} \delta_\broad(\ve_i, \sve_i) $. This amounts to replacing the convolution
kernels, e.g., $K(\vec{\omega}_p,\vec{\sve})$ in \Eq{eq:G-discretized-S-ZF},
by broadened versions, 
\begin{align}
\label{eq:broadened-kernels-general}
K_\broad(\vec{\omega},\vec{\sve}) = 
\nint \md^{\ell-1} \ve \, K(\vec{\omega}_p,\vec{\ve}) \,
\delta_\broad(\vec{\ve}, \vec{\sve}) 
. 
\end{align}
For each grid direction $i$, the 
broadening kernel $\delta_\broad(\ve_i, \sve_i)$
is characterized by the shape and width of its peak when 
viewed as a function of $\ve_i$ for given $\sve_i$. 
We first discuss the requirements for its width, say $\gamma_{\broad, i}$.
(The meaning of ``width'' of course depends on the peak shape, but the ensuing
arguments are generic.) Because of the logarithmic bath discretization, 
the dominant contributions to the coefficients $S_p (\vec{\sve})$ lie within \textit{clusters} on the $\vec{\sve}$ grid, 
with each cluster involving transitions within a specific combination of NRG shells.  When plotted using logarithmic grid axes, the clusters are spaced by 
$\sim\! \ln \Lambda$, along each grid axis (see, e.g., \Fig{fig:Adisc4} below),
reflecting the single-particle level spacing of the discretized bath.
The width of the broadening peak, plotted on the same axes, should be 
comparable to this in order to smooth out the clusters, i.e., $\ln\bigl((|\sve_i| + \gamma_{\broad,i})/|\sve_i|\bigr)
\lesssim \ln \Lambda$. This requires a peak width proportional to frequency, $\gamma_{\broad,i} = \bfactor |\sve_i|$, with 
$\bfactor \lesssim \Lambda - 1$. Smaller choices for $b$ are possible if one averages over several, say $n_z$, discretization grids (``$z$ averaging'' \cite{Bulla2008}).

We next discuss the choice of peak shape for $\delta_\broad (\ve, \sve_i)$. 
NRG calculations of $2$p functions conventionally use a log-Gaussian broadening
kernel~\cite{Bulla2008,Weichselbaum2007}, i.e.,
a Gaussian function of $\ln |\ve_i / \sve_i|$ (for details, see \App{app:logGaussian}). It has the useful property
of not overbroadening low-energy features resolvable only on a logarithmic scale.

For 4p correlators, we were not able to obtain
satisfactory results using log-Gaussian broadening. By contrast, 
a Lorentzian broadening kernel with width $\bfactorL|\sve_i|$, 
\begin{equation}
\delta_\Lorentz  (\ve_i, \sve_i) = \frac{\bfactorL | \sve_i | / \pi}{(\ve_i - \sve_i)^2 + (\bfactorL | \sve_i |)^2} ,
\label{eq:LorentzianKernel}
\end{equation}
yielded very good results for  benchmark checks against various analytic results for the 4p vertex \cite{Kugler2021}.  Via \Eq{eq:broadened-kernels-general}, it leads to broadened \ZF/ and \KF/ kernels of the form
\begin{subequations}
\label{eq:K_w_Lorentz}
\begin{align}
K_\Lorentz (\vec{\omega}_p, \vec{\sve}) &=
\prod_{i = 2}^{\ell} \widetilde{\xi}_{\ovb{i}\cdots\ovb{\ell}}^{-1}
,
\label{eq:K_w_ZF_Lorentz} \\
K^{\vec{k}_p}_\Lorentz (\vec{\omega}_p, \vec{\sve})  &=
\sum_{\lambda = 1}^{\ell}
(-1)^{k_{\ovb{1}\cdots \ovb{\lambda-1}}} 
\frac{ 1 + (-1)^{k_\ovb{\lambda} }}{2}
\nonumber \\
& \quad \times
\bigg[
\prod_{i = 1}^{\lambda-1} \widetilde{\xi}_{\ovb{1}\cdots\ovb{i}}^{-1}
\prod_{j = \lambda+1}^{\ell} \widetilde{\xi}_{\ovb{j}\cdots\ovb{\ell}}^{-1}
\bigg],
\label{eq:K_w_KF_Lorentz}
\end{align}
which match Eqs.~\eqref{eq:K_w_ZF} and \eqref{eq:K_w_KF}, respectively, except for the replacements,
\begin{equation}
\begin{aligned}
\xi_{\ovb{1}\cdots\ovb{i}} &\mapsto 
\widetilde{\xi}_{\ovb{1}\cdots\ovb{i}} = 
-\omega_{\ovb{1}\cdots\ovb{i}} + \sve_i + \mr{i} (\gamma_{\ovb{1}\cdots\ovb{i}} + b_\Lorentz | \sve_i | ) ,\\
\xi_{\ovb{j}\cdots\ovb{\ell}} &\mapsto
\widetilde{\xi}_{\ovb{j}\cdots\ovb{\ell}} = 
-\omega_{\ovb{j}\cdots\ovb{\ell}} - \sve_{j-1} + \mr{i} (\gamma_{\ovb{j}\cdots\ovb{\ell}} + b_\Lorentz | \sve_{j-1} | ) .
\end{aligned}
\label{eq:K_w_xi_Lorentz}
\end{equation}
\end{subequations}%
Note that $\widetilde{\xi}_{\ovb{1}\cdots\ovb{i}}$ and $\widetilde{\xi}_{\ovb{j}\cdots\ovb{\ell}}$ are the broadened version of ${\xi}_{\ovb{1}\cdots\ovb{i}}$ and ${\xi}_{\ovb{j}\cdots\ovb{\ell}}$, respectively, 
and not the sum of individual $\widetilde{\xi}_{m}$'s.
Thus, Lorentzian broadening modifies the kernels only by shifting their poles away from the real axis by $\pm \mi \bfactorL |\sve_i|$. 
Importantly, however, it does not change their analytic structure. As a result, the intricate cancellation patterns between different PSFs in permutation sums $\sum_p$ largely remain intact during broadening. 
We believe this to be the main reason why, for 4p correlators, Lorentzian broadening outperforms log-Gaussian broadening.

The disadvantage of Lorentzian broadening, of course, is that Lorentzian
peaks have long tails which very strongly overbroaden low-energy features---indeed, this is the reason why Lorentzian broadening is not used for 2p correlators. For 4p correlators, the effects of overbroadening appear to
be more tolerable. Moreover,  when computing the 4p vertex, some of their
consequences can be canceled out by also using Lorentzian broadening for 
the external-leg 2p functions being divided out.
To calculate the latter, we use the kernel~\eqref{eq:K_w_Lorentz} for $\ell \!=\! 2$, but there is one subtlety:
While we use the same value of $\gamma_i \!=\! \gamma_0$ for all $1 \!\leq\! \ell \!\leq\! 4$ when computing
4p correlators, the values of $\gamma_{i = 1,2}$ for each 2p correlator depend on which leg it is attached to.
This is explained in more detail in Sec.~\ref{sec:amputation}.

The values used for the broadening parameters $\bfactorL$ and $\gamma_0$ depend on the discretization parameter $\Lambda$, and on $n_z$ if $z$ averaging is used. Artifacts appearing
for frequencies lower than temperature can be smeared out by 
choosing $\gamma_0$ to be $\sim T$. 
In practice, we choose $\bfactorL$ and $\gamma_0$
just large enough to yield smooth results for broadened correlators.
For the ZF and KF results in this paper, we used $\Lambda \!=\!  4$, 
$n_z \!=\! 4$, $\gamma_0 \!=\! 3T$ and $\bfactorL \!=\! 0.6$.
(These values were also used for  the results presented 
in \paperI, with two exceptions: we used 
$\bfactorL \!=\! 1.4$, $\gamma_0 \!=\! T/2$ for the x-ray--edge singularity, 
and $\bfactorL \!=\! 0$, $\gamma_0 \!=\!  T$ for the Hubbard atom.
Though we treated the former in the ZF, we chose a small but finite $T\!=\!10^{-5}D$ to define the density matrix and $\gamma_0$.)

\subsection{Equation of motion}
\label{sec:EOM}

When computing the connected contributions of PSFs, 
as described in Sec.~\ref{sec:ConnectedPart},
some disconnected parts remain, due to numerical inaccuracies.
To remove these, we employ an EOM strategy 
\cite{Hafermann2012} which 
expresses $G^\connected$ through two 4p correlators whose
disconnected parts mutually cancel.
(This step is not needed if the computation of PSFs is exact,
as was the case for benchmark results on the Hubbard atom in \paperI.) 
We first explain the EOM strategy in the \MF/,
then discuss its transcription to the \ZF/ and \KF/.

In NRG computations of 2p correlators for quantum impurity models, 
it is common practice to improve the results by computing the self-energy via an EOM scheme \cite{Bulla1998}. 
In the \MF/, QMC computations of
local 4p correlators use a similar strategy, 
employing EOMs to obtain improved results for the connected
part, $G^\connected$ \cite{Hafermann2012,Kaufmann2019}. 
Transcribed to our situation, 
Eq.~(26) of Ref.~\onlinecite{Hafermann2012} reads 
\begin{align}
\label{eq:MF_4p-EOM}
G^{\connected} [d_{\sigma}^\pdag, d_{\sigma}^\dagger, d_{\sigma'}^\pdag, d_{\sigma'}^\dagger] (\mi \vec{\omega}) 
& = 
\\ 
G[d_{\sigma}^\pdag, d_{\sigma}^\dagger] (\mi \omega_1) \,
&
\Gtilde \big[ \dtilde_{\sigma}^\pdag, d_{\sigma}^\dagger, d_{\sigma'}^\pdag, d_{\sigma'}^\dagger
\big] 
(\mi \vec{\omega}) 
\nonumber
\\
- \,
\Gtilde \big[ d_{\sigma}^\pdag, \dtildedag_\sigma \big] 
(\mi \omega_1) \,
&
G [d_{\sigma}^\pdag, d_{\sigma}^\dagger, d_{\sigma'}^\pdag, d_{\sigma'}^\dagger] 
(\mi \vec{\omega}) 
,  \nonumber
\end{align}
where $\dtilde_\sigma^\pdag = [ d_{\sigma}^\pdag, \Hc_\interaction]$ and $\dtildedag_\sigma$ are composite operators, and we put tildes on 
correlators containing one of them. 
Suppressing arguments to reveal the structure, this reads
\begin{align}
\label{eq:MF_4p-EOM-compact}
G^\connected_\fourpoint  = 
G_\twopoint \Gtilde_\fourpoint - 
\Gtilde_\twopoint G_\fourpoint
. 
\end{align}
Equation~\eqref{eq:MF_4p-EOM-compact} correctly yields
$G^\connected_\fourpoint \!=\! 0$ if $\Hc_\interaction \!=\!0 $, since then $\Gtilde_\fourpoint \!=\! \Gtilde_\twopoint \!=\! 0$ by construction.
In the QMC study of Ref.~\onlinecite{Hafermann2012}, 
computing $G^\connected_\fourpoint$ via \Eq{eq:MF_4p-EOM-compact}, 
rather than $G_\fourpoint - G^\disconnected_\fourpoint$, indeed led to markedly improved results. The reason is that the statistical errors for $G_\fourpoint$ and $G^\disconnected_\fourpoint$ are very different, whereas those for
$G_\fourpoint$ and $\Gtilde_\fourpoint$ are similar. 
Our NRG computations of $G_\fourpoint$ and $G^\disconnected_\fourpoint$
are also not perfectly matched---before subtracting the latter, rebinning
of discrete spectral data is needed, see Sec.~\ref{sec:ConnectedPart}.
Thus, computing $G^\connected_\fourpoint$ via \Eq{eq:MF_4p-EOM-compact} is advisable in our case, too. 

Benchmark checks against analytical results (e.g., for weak
interactions) show that a direct application of \Eq{eq:MF_4p-EOM-compact},
with full 4p correlators as input on the right, does not yield optimal results
for $G^\connected$. Instead, we use \textit{connected} 4p correlators as input, available via the subtraction scheme of 
\Sec{sec:ConnectedPart}. 
Expressing both 4p correlators on the right of \Eq{eq:MF_4p-EOM-compact} 
as sums of connected and disconnected parts,
we obtain
\begin{align}
G^\connected_\fourpoint  
& =  G_\twopoint \Gtilde^\connected_\fourpoint - 
\Gtilde_\twopoint G_\fourpoint^\connected
,
\label{eq:MF_4p-EOM-connected}
\end{align}
as the disconnected parts mutually cancel,
$G_\twopoint \Gtilde^\disconnected_\fourpoint - 
\Gtilde_\twopoint G_\fourpoint^\disconnected  = 0$.
This cancellation follows by using \Eq{eq:Gdis}
and
$\Gtilde \big[ \dtilde_{\sigma}, d_{\sigma}^\dagger \big] =
\Gtilde \big[ d_\sigma, \dtildedag_{\sigma} \big]$.
The latter equality stems from two different EOM derivations for 
$G_\twopoint$,
differentiating it with respect to either the first or the second time argument~\cite{Hafermann2012} and using 
$\partial_{t_1} G_\twopoint  = - \partial_{t_2} G_\twopoint $.

We evaluate \Eq{eq:MF_4p-EOM-connected} for $G^\connected_\fourpoint$ by inserting, on its right, results for $G^\connected_\fourpoint$ and
$\Gtilde^\connected_\fourpoint$ obtained by subtracting
disconnected parts as described in Sec.~\ref{sec:ConnectedPart}. 
Benchmark checks show that the output $G^\connected_\fourpoint$ is improved
relative to the input $G^\connected_\fourpoint$. The reason is likely that remnant disconnected contributions, still contained in the input correlators
due to numerical inaccuracies, tend to cancel in \Eq{eq:MF_4p-EOM-connected}. 

We now turn to real-frequency EOMs for \ZF/ and \KF/ correlators.
Apart from the fact that, here, input correlators must first be broadened, 
one proceeds analogously. 
The \ZF/, just as the \MF/, involves time-ordered correlators. 
Hence, \ZF/ EOMs can be obtained from \MF/ ones by simply replacing  
real by imaginary times. 
In the frequency domain, this yields 
the EOM \eqref{eq:MF_4p-EOM} with
$\mi \vec{\omega}$ replaced by $\vec{\omega}$.

In the \KF/ in the Keldysh basis, a derivation analogous to that of Ref.~\onlinecite{Hafermann2012}, but tracking Keldysh indices, yields 
the same operator and frequency structure as \Eq{eq:MF_4p-EOM-compact}:
\begin{align}
\label{eq:KF_4p-EOM}
 G^{\connected, \vec{k}}_\fourpoint  & = 
 (G_\twopoint 
 \sigma_x 
 \Gtilde_\fourpoint  - \Gtilde_\twopoint 
 \sigma_x  
 G_\fourpoint)^{\vec{k}} \, .
 \end{align}
 The only complication is that the $\omega_1$
 leg of each $4$p correlator is contracted with the Pauli matrix $\sigma_x = \bigl({\scalefont{0.5}{\begin{matrix} 0 \; \; 1 \\  1 \; \; 0 \end{matrix}}} \bigr)$ via 
$(G_\twopoint  \sigma_x
 \Gtilde_\fourpoint)^{k_1 k_2 k_3 k_4}  =   \sum_{k'_1 k''_1} G^{k_1 k'_1}_\twopoint \sigma_x^{k'_1 k''_1}  \Gtilde^{k''_1 k_2 k_3 k_4}_\fourpoint $.
As a consistency check, we note that the \ZF/ version of \Eq{eq:MF_4p-EOM-compact}
 can be obtained
from \Eq{eq:KF_4p-EOM} by transforming the latter to the contour basis 
and considering only components restricted to the forward branch of the Keldysh contour.

The mutual cancellation of disconnected parts
discussed above 
applies to the real-frequency EOMs, too. 
For the \ZF/, \Eq{eq:MF_4p-EOM-connected} applies
unchanged. For the \KF/, it becomes
\begin{align}
\label{eq:KF_4p-EOM-connected}
G^{\connected, \vec{k}}_\fourpoint  
=  
(G_\twopoint 
\sigma_x  
\Gtilde^\connected_\fourpoint - 
\Gtilde_\twopoint \sigma_x  
G_\fourpoint^\connected)^{\vec{k}}
,
\end{align}
while the cancellation of disconnected parts can be verified using
$\Gtilde^\vec{k} \big[ \dtilde_{\sigma}, d_{\sigma}^\dagger \big]
= 
\Gtilde^\vec{k} \big[ d_\sigma, \dtildedag_{\sigma} \big]$,
abbreviated $\Gtilde^\vec{k}_\twopoint$,
and $(G_\twopoint \sigma_x \Gtilde_\twopoint)^\vec{k}
=
(\Gtilde_\twopoint \sigma_x G_\twopoint)^\vec{k}$.
The latter relation follows by evoking the fluctuation-dissipation theorem,
$G^{22}(\omega) = \tanh(\beta\omega/2) (G^{21} - G^{12})(\omega)$, 
for both $G_\twopoint$ and $\Gtilde_\twopoint$.

The EOM scheme described above treats the frequency arguments
of 4p correlators asymmetrically, 
in that $\omega_1$ receives special treatment. 
For \KF/ correlators $G^{\vec{k}}$, this leads to the fact
that the $\vec{k} \!=\! 2111$ component
is obtained with better quality than others. 
This problem can be remedied using a symmetric EOM scheme
presented recently in the \MF/ in Ref.~\onlinecite{Kaufmann2019}. 
We leave its transcription to the \KF/ and use for NRG computations 
to the future.

\subsection{Amputating and reattaching external legs}
\label{sec:amputation}

The strategies described in the preceding sections yield 
a connected 4p correlator, $G^\connected$, free from 
disconnected parts. To obtain the corresponding 4p vertex $F$, 
the external legs are amputated by dividing out $2$p correlators for 
all frequencies $\omega_i$. The \MF/ vertex, e.g., is 
\begin{flalign}
F (\mi \vec{\omega}) & = 
G^\connected (\mi \vec{\omega}) / 
\big( 
G (\mi \omega_1) G (-\mi \omega_2) G (\mi \omega_3) G (-\mi \omega_4)   
\big)
,
\hspace{-0.5cm}
& 
\label{eq:amputation}
\end{flalign}
with a positive (negative) sign in the propagator of an outgoing (incoming) leg,
associated with an annihilation (creation) operator in the definition of the 4p correlator.

The  \ZF/ and \KF/ vertices are obtained using similar expressions
(see \paperI\ for details), but require special care. First, 
the choice of imaginary shifts for the arguments 
of the external legs $G(\omega_i)$ have to be chosen 
consistent with those of $G^\connected (\vec{\omega})$ (see Appendix D of \paperI).
We choose $\gamma_1 \!=\! \gamma_0$ and $\gamma_2 \!=\! 3\gamma_0$ ($\gamma_1 \!=\! 3\gamma_0$ and $\gamma_2 \!=\! \gamma_0$) for the outgoing (incoming) legs,
where $\gamma_0$ is the value of all $\gamma_i$'s in the kernel~\eqref{eq:K_w_Lorentz} used for computing the 4p correlator.
Second, since real-frequency versions of $G^\connected$ are computed using Lorentzian-broadened kernels, we compute the 2p correlators in the denominator using the same Lorentzian broadening scheme~\eqref{eq:K_w_xi_Lorentz}.
Hence, the broadening width parameter $b_\Lorentz$ is chosen identical to the one of the 4p correlator.
This ameliorates the undesired overbroadening effects of Lorentzian broadening, in that they tend to
cancel out in the ratio \eqref{eq:amputation}.
This strategy is essential  for obtaining the correct large-frequency behavior for $F$, since the large-frequency
behavior for $G^\connected$ is dominated by the 2p correlators. 

Having computed $F$, a yet-again improved version of $G^\connected$ can be obtained by multiplying $F$ with external-leg 2p correlators,  
now computed using the customary log-Gaussian broadening scheme (cf.~\App{app:logGaussian}). This ensures that those features dominated by the external legs are not overbroadened.
In other words, to obtain a \ZF/ or \KF/ $G^\connected$ most accurately, we first compute the $F$ vertex using Lorentzian-broadened ingredients, and then reattach the external legs through optimally broadened $2$p correlators.
This strategy for optimizing the treatment of external legs is particularly useful from the perspective that experimental probes typically measure response functions corresponding to 
correlators \textit{with} external legs.

\section{Results: Connected Correlators}
\label{sec:Results-Connected}

To establish the power of our multipoint NRG scheme, 
we have performed detailed benchmark computations
for the paradigmatic AIM.
We presented \MF/ and \KF/ results for its $4$p vertex in \paperI. 
To complement these, we here analyze both the underlying PSFs 
and the connected $4$p correlators $G^\connected$ obtained
from them by convolution with suitable \MF/, \ZF/, or \KF/ kernels.

\subsection{Model}
\label{sec:AIM}

The Hamiltonian of the AIM in standard notation is
\begin{align}
\Hc_\siam
& = 
\sum_{\epsilon \sigma}  \epsilon \, c_{\epsilon \sigma}^\dag c^\pdag_{\epsilon \sigma} 
+ 
 \sum_\sigma \epsilon_d \, d^\dagger_\sigma d^\pdag_\sigma 
+
U d_{\uparrow}^\dag d^\pdag_{\uparrow} d_{\downarrow}^\dag d^\pdag_{\downarrow} 
\nonumber \\
& 
\quad + 
\sum_{\epsilon \sigma}  (V_\epsilon d^\dagger_\sigma c^\pdag_{\epsilon \sigma} + \textrm{H.c.} ) \, .
\label{eq:H_AIM}
\end{align}
It describes a band of spinful electrons, hybridizing with 
a local level with energy $\epsilon_d$ and Coulomb repulsion $U$. 
We take $\epsilon_d \!=\! -U/2$.
The coupling between impurity and bath is fully characterized by the hybridization function 
$\Delta (\nu) \!=\! \sum_{\epsilon} \pi |V_\epsilon|^2 \delta(\nu - \epsilon)$,
which we choose box-shaped, $\Delta (\nu) = \Delta \, \theta(D \!-\! |\nu|)$, 
with half-bandwidth $D$. Our goal is to compute  the connected part of the local $4$p correlator $G [d^\pdag_\sigma  , d_\sigma^\dagger , d^\pdag_{\sigma'} , d_{\sigma'}^\dagger] (\vec{\omega})$.

For the NRG calculation,
the bath is represented on a discrete, logarithmic grid, with grid intervals bounded by the points $\pm D \Lambda^{-k-z}$. 
Here, $k \geq 0$ is an integer and $z \!\in\! (0,1]$ a shift parameter.
A discrete energy representing each grid interval is chosen
using the prescription proposed in Ref.~\onlinecite{Zitko2009}.
We use $\Lambda \!=\! 4$, and average
results from $n_z$ different discretization grids, with $z = 1/n_z, 2/n_z, \ndots, 1$.
We choose $n_z \!=\! 2$ for MF correlators, which 
are less sensitive to discretization artifacts (for the same reason 
as why MF PSFs need not be broadened; see \Sec{sec:broadening}),
and $n_z \!=\! 4$ for the more challenging real-frequency correlators.
The Wilson chain is diagonalized iteratively by keeping $300$ multiplets 
respecting $\mr{U}(1)$ charge and $\mr{SU}(2)$ spin symmetries. 
We generate and manipulate non-Abelian symmetric tensors using 
the QSpace library developed by Weichselbaum \cite{Weichselbaum2012a,Weichselbaum2020}.
By exploiting non-Abelian symmetries, the different spin components of the $4$p correlator are all obtained simultaneously.

\begin{figure}[t]
\includegraphics[width=\linewidth]{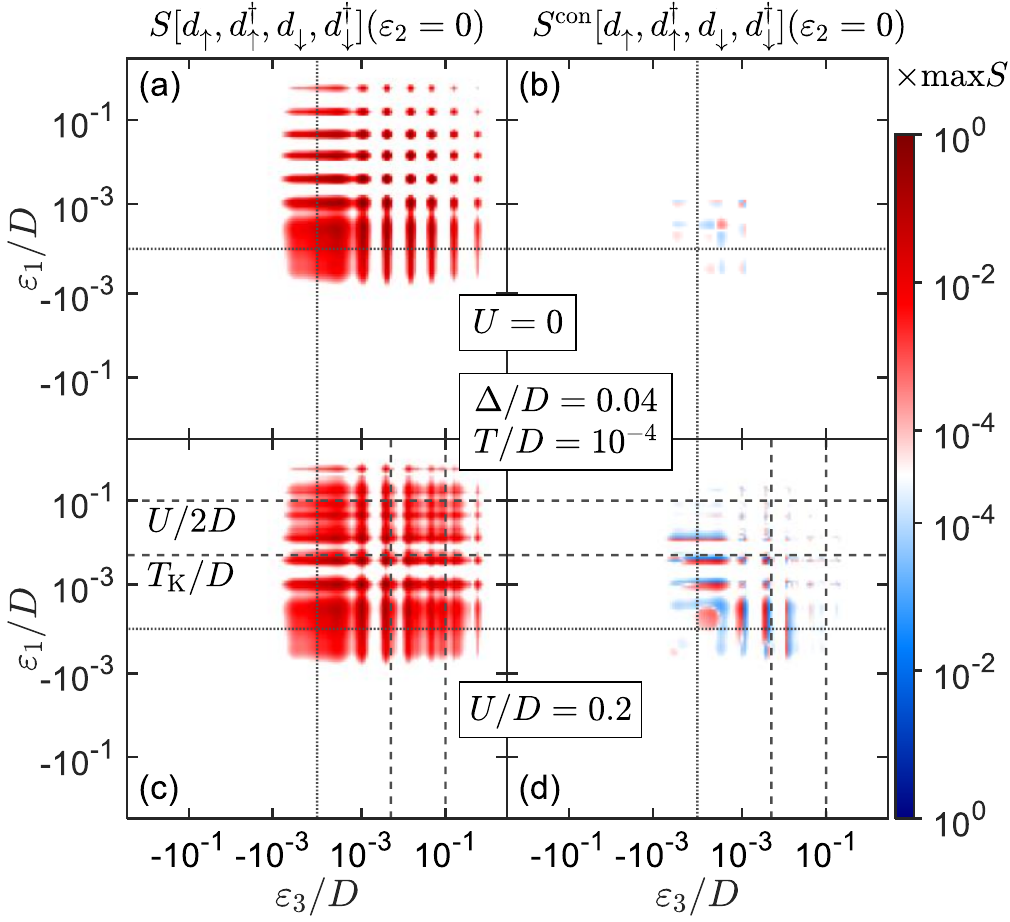}
\caption{%
An exemplary opposite-spin $4$p PSF of the AIM,
$S[d_\uparrow^\pdag, d_\uparrow^\dag, d_\downarrow^\pdag, d_\downarrow^\dag] (\vec{\ve})$
(out of the 24 PSFs $S_p$),
as a function of $\varepsilon_1$ and $\varepsilon_3$ at $\varepsilon_2 \!=\! 0$.
The top and bottom rows show noninteracting ($U \!=\! 0$)
and strongly interacting $(U/\Delta \!=\! 5)$ cases, the left and right 
columns the full PSF $S$ and its connected part $S^\connected$, 
respectively. For $U \!=\! 0$, $S^\mathrm{con}$ should vanish by Wick's theorem; 
the fact that
small but nonzero values remain in (b)
is due to numerical artifacts discussed in Sec.~\ref{sec:ConnectedPart}.
For $U \!\neq\! 0$, the distribution of spectral weight reflects the energy scales $U/2$ and 
$\Tk \!\simeq\! D/200$,
and $S^\mathrm{con}$ 
reveals mutual two-particle correlations.
These PSFs were computed using $\Lambda \!=\! 4$ and $z \!=\! 1$.
For visualization, the discrete data was minimally broadened, using the log-Gaussian--Fermi kernel \eqref{eq:logGaussFermibroadening} 
\cite{Lee2016} with narrow width parameters.
The intensity values for each row are normalized by the maximum magnitude of 
$S$ on the left.}
\label{fig:Adisc4} 
\end{figure}

\begin{figure}[t]
\includegraphics[width=\linewidth]{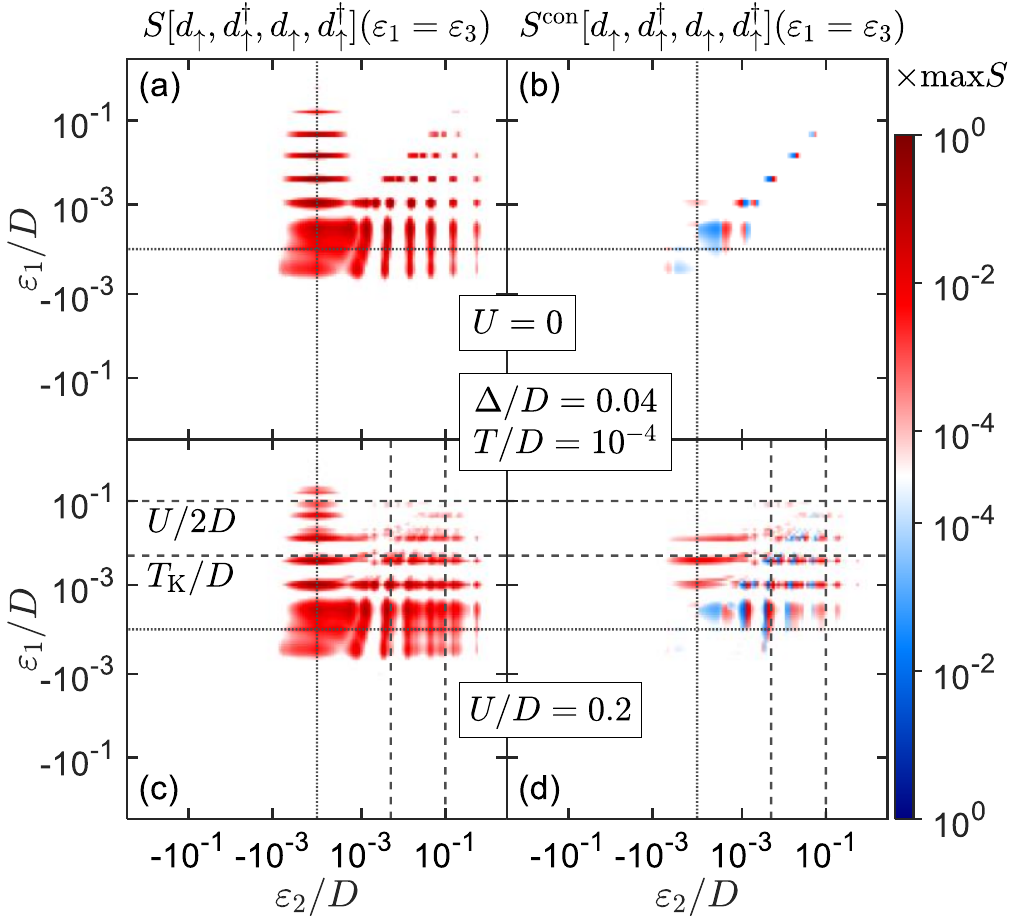}
\caption{%
Analogous to \Fig{fig:Adisc4}, but for a same-spin 4p PSF
$S[d_\uparrow^\pdag, d_\uparrow^\dag, d_\uparrow^\pdag, d_\uparrow^\dag] (\vec{\ve})$, 
shown as a function of $\ve_1$ and $\ve_2$ at $\ve_3 \!=\! \ve_1$.
In (b), the narrow adjacent red and blue regions along the diagonal $\ve_1 \!\simeq\! \ve_2 \!>\! 0$ reflect contributions from $S$ and $S^\disconnected$ which fail to cancel fully, due to rebinning involved
for $S^\disconnected$.}
\label{fig:Adisc4_same_spin} 
\end{figure}

\subsection{Discrete partial spectral functions}

The first output of our multipoint NRG calculations are discrete $4$p PSFs.
Figures \ref{fig:Adisc4}(a) and \ref{fig:Adisc4}(c) [\Figs{fig:Adisc4}(b) and \ref{fig:Adisc4}(d)] display the full [connected] part
of the PSF $S[d^\pdag_\uparrow, d_\uparrow^\dag, d^\pdag_\downarrow, d_\downarrow^\dag] (\vec{\ve})$, computed for 
$U = 0$ in \Figs{fig:Adisc4}(a) and \ref{fig:Adisc4}(b) [$U \neq 0$ in \Figs{fig:Adisc4}(c) and \ref{fig:Adisc4}(d)]
and shown as a function of $\ve_1$ and $\ve_3$ at $\ve_2 \!=\! 0$.
Generally, we observe that the spectral contributions are spread over a range of energies,
from $-T$ to the largest energy scale in the system (here $D$).
As $4$p PSFs describe \textit{excitation} spectra---%
their arguments are transition energies, 
$\vec{\ve} = (E_{\ub{2} \ub{1}}, E_{\ub{3} \ub{1}}, E_{\ub{4} \ub{1}})$---%
the spectral weight predominantly lies at positive energies.
The spectral weight at negative energies, $\ve \ll -T$,
is exponentially suppressed by the Boltzmann factor.

For noninteracting systems, Wick's theorem implies that 
$4$p PSFs 
consist of only disconnected contributions,
so that exact numerics would yield $S \!=\! S^\disconnected$,
with
$S^\disconnected [d^\pdag_\uparrow, d_\uparrow^\dagger, d^\pdag_\downarrow, d_\downarrow^\dagger] (\vec{\ve}) = S[d^\pdag_\uparrow, d_\uparrow^\dagger] (\ve_1) \, S[d^\pdag_\downarrow, d_\downarrow^\dagger] (\ve_3) \, \delta(\ve_2)$, cf.\ \Eq{eq:Sc_w-dis-epsilon}.
This factorization into two 2p PSFs
is visible in \Fig{fig:Adisc4}(a), as the spectral weight is
distributed on an equidistant, square grid for $\ln \ve_1$ and $\ln \ve_3$, with grid spacing $\ln \Lambda$.

Though $S^\connected \!=\! S \!-\!  S^\disconnected$ 
should vanish for $U \!=\! 0$,
we numerically obtain a small but nonzero result, 
see \Fig{fig:Adisc4}(b). One reason
stems from the fact that 
discrete spectral data at low energies, $| \sve_i | \!\lesssim\! T$, 
are inaccurate. We therefore broaden them, 
using a broadening width $\sim\! T$ \cite{Lee2016},
to obtain $S^\connected \!\simeq\! 0$ at energies $| \ve_i | \!\lesssim\! T$.
The finite but minuscule value of $S^\connected$ for $|\ve_i|$ slightly larger than $T$ 
[note the logarithmic color scale in \Fig{fig:Adisc4}(b)] 
comes from the tail of the inaccurate low-energy region, 
which is not fully smeared out through broadening.

Additional numerical artifacts arise from  the rebinning of the sum of frequencies $\sve_1 + \sve_3$ into $\sve_2$ bins during the computation of $S^\disconnected$ (cf.~\Sec{sec:ConnectedPart}). This is exemplified in 
\Fig{fig:Adisc4_same_spin}, showing $S[d_\uparrow^\pdag, d_\uparrow^\dag, d_\uparrow^\pdag, d_\uparrow^\dag] (\vec{\ve})$ as a function of 
$\ve_1$ and $\ve_2$ at $\ve_3=\ve_1$, with a layout analogous to 
Fig.~\ref{fig:Adisc4}. The rebinning artifacts, seen 
in \Fig{fig:Adisc4_same_spin}(b) along the diagonal, are small as well.
For example, the maximum magnitude of the data of \Fig{fig:Adisc4_same_spin}(b) is about $2.5\%$ of that of \Fig{fig:Adisc4_same_spin}(a).
Meanwhile, Fig.~\ref{fig:Adisc4} is free from rebinning artifacts, because  $\sve_2=0$ there.
Generally, rebinning artifacts are
smeared out by broadening (cf.~\Sec{sec:broadening}).
They can be further reduced via the EOM strategy described in Sec.~\ref{sec:EOM}. 
(For $U=0$, the property $G^\connected=0$ is then obtained exactly.)

\begin{figure}
\includegraphics[width=\linewidth]{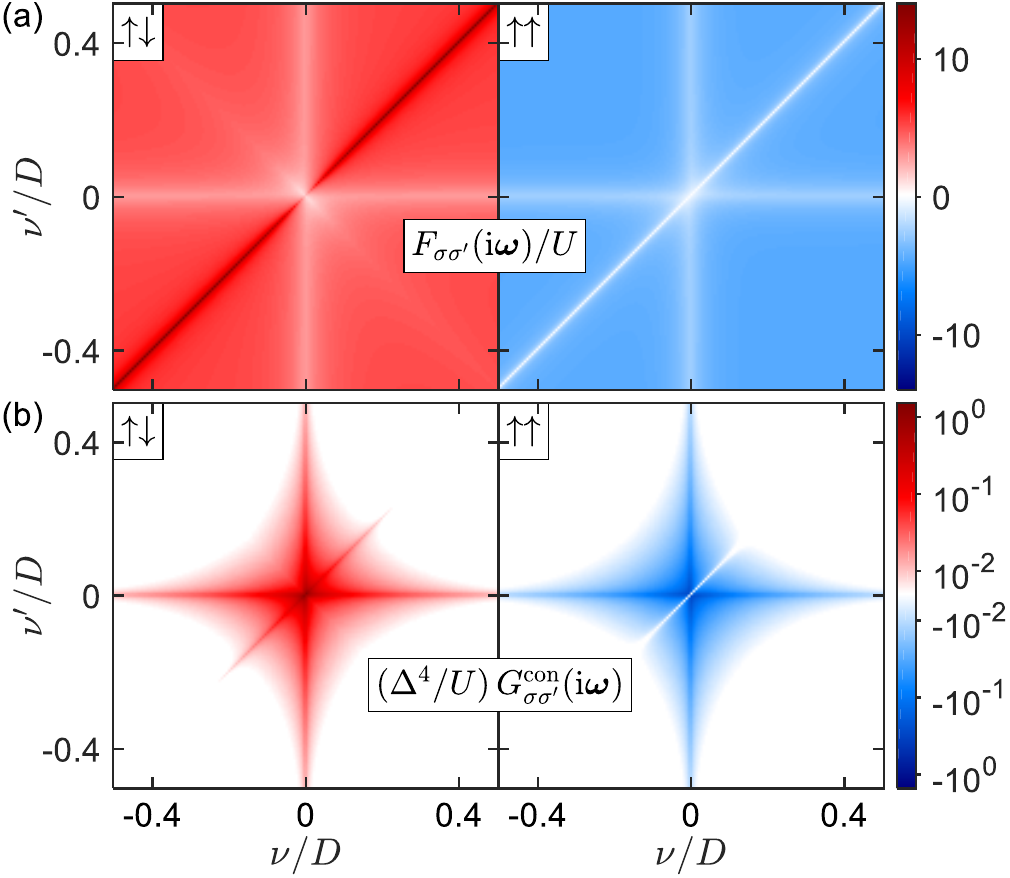}
\caption{%
(a) \MF/ $4$p vertex $F_{\sigma \sigma'} (\mi\vec{\omega})/U$, and (b)
MF connected $4$p correlator $(\Delta^4 /U) \, G_{\sigma \sigma'}^\connected (\mi \vec{\omega})$, for the strongly interacting AIM ($U/\Delta\!=\!5$, $\Delta / D \!=\! 1/25$, $T /D\!=\! 10^{-4}$),
as a function of $\nu$ and $\nu'$ at $\omega \!=\! 0$.
The left (right) panels show $\sigma \!\neq\! \sigma'$ ($\sigma \!=\! \sigma'$).}
\label{fig:IF-connected-GF-AIM} 
\end{figure}

\begin{figure*}
\includegraphics[width=\linewidth]{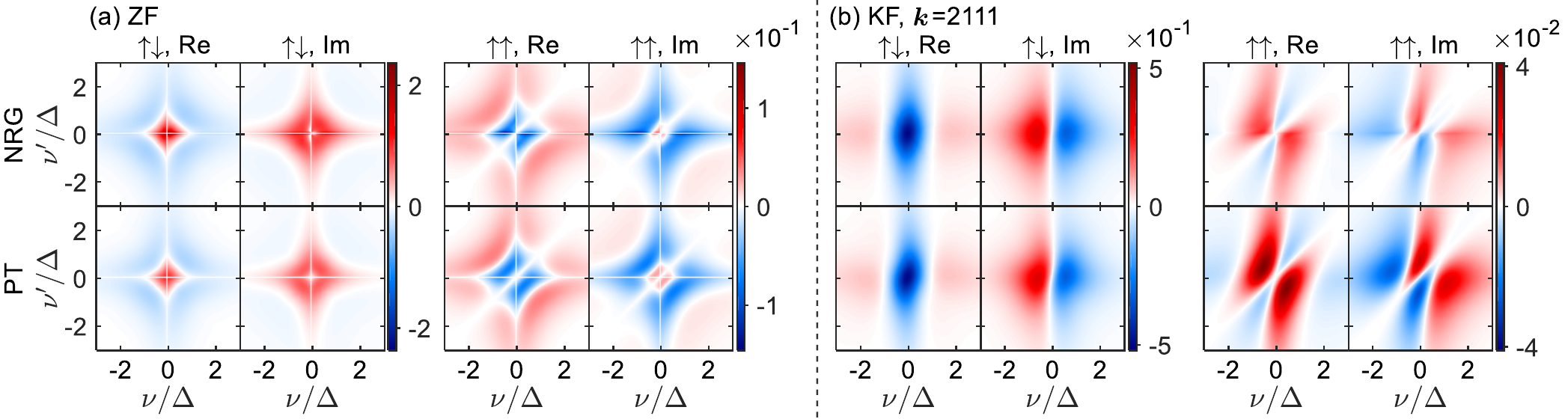}
\caption{%
Weak-coupling benchmark comparison of real-frequency correlators. 
We show the connected
$4$p correlator $(\Delta^4 /U) \, G_{\sigma \sigma'}^{\connected}$ at $\omega \!=\! 0$, obtained by NRG (top row) or second-order perturbation theory (PT)
(bottom row), for the weakly interacting AIM ($U/\Delta\!=\!1/2$, 
$\Delta/D\!=\!1/10$, $T\!=\!10^{-3} D$).
For both (a) the ZF and (b) the KF ($\vec{k} \!=\! 2111$) correlators, 
the NRG and PT results show very good qualitative agreement.
Perfectly quantitative agreement cannot be expected, due
to the broadening needed to obtain smooth results.
The retarded KF component $\vec{k} \!=\! 2111$ looks intriguingly different from the ZF correlator.
}
\label{fig:ZF-KF-connected-GF-SOPT}
\end{figure*}

In the presence of interactions, the $4$p PSFs exhibit a rich structure; 
see \Figs{fig:Adisc4}(c), \ref{fig:Adisc4}(d), \ref{fig:Adisc4_same_spin}(c), and \ref{fig:Adisc4_same_spin}(d).
The spectral distribution patterns in Figs.~\ref{fig:Adisc4}(c) and \ref{fig:Adisc4_same_spin}(c) change noticeably across characteristic energy scales induced by interactions, such as the Hubbard-band position $U/2$ 
and the Kondo temperature $\Tk$.
The connected part $S^\connected$, describing mutual two-particle correlations, now has considerable weight over a wide frequency range,
see Figs.~\ref{fig:Adisc4}(d) and \ref{fig:Adisc4_same_spin}(d).
It is most pronounced at energies below $U/2$ and around $\Tk$
but very small at large energies $\ve_i  \!>\! U$;
hence, two-particle correlations are negligible there.
The full and connected PSFs for the other permutations exhibit similar behavior; see \Fig{fig:Adisc4_all_permutation} in 
\App{app:Example_4p_PSFs}.

\subsection{Connected 4p correlators for AIM}

We next present results for the connected correlator $G^\connected$,
computed from PSFs using the strategies described in Sec.~\ref{sec:from-PSF-to-G}, including the amputation and reattachment of 
external legs (Sec.~\ref{sec:amputation}). 
We start with MF results, as these can also be obtained 
by means of ED or QMC,
and consider the AIM at strong interaction 
$U/\Delta \!=\! 5$, with $U/D \!=\! 1/5$. 
In \paperI, we benchmarked the NRG $4$p vertex,
$F_{\sigma\sigma'} (\mi \vec{\omega})$, against QMC 
at an intermediate temperature $T \!=\! 10^{-2} D$,
finding relative deviations on the level of $1\% $.
(For that benchmark, both QMC and NRG results were generated
without using the EOM strategy of Sec.~\ref{sec:EOM}.)
We  also presented results for the 4p vertex at much lower temperatures,
out of reach for QMC algorithms.
For convenience, Fig.~\ref{fig:IF-connected-GF-AIM}(a) reproduces
the results from Fig.~8 of \paperI\ for 
$F_{\sigma\sigma'} (\mi \vec{\omega})$, now obtained by using the EOM strategy.
Figure~\ref{fig:IF-connected-GF-AIM}(b) shows
the corresponding connected correlator $G^\connected_{\sigma\sigma'}(\mi\vec{\omega})$ at $T \!=\! 10^{-4} D$
as a function of $\nu$ and $\nu'$ at $\omega \!=\! 0$,
in the particle-hole representation \cite{Kugler2021,Rohringer2018}, 
\begin{equation}
\vec{\omega} = (\nu, -\nu-\omega, \nu'+\omega, -\nu'),
\label{eq:PHconvention}
\end{equation}%
obtained by reattaching external legs to the vertex.
These external legs ensure that $G^\connected$ decays in all directions
(note the logarithmic color scale), contrary to the 
finite background value of $F$.
There is a strong signal at the origin and along the axes
$\nu^{(\prime)} \!=\! 0$.
Just as for the vertex \cite{Kugler2021}, the equal-spin component vanishes
for $\nu \!=\! \nu'$ and $\omega \!=\! 0$ by symmetry.
These features are mostly known already;
\Fig{fig:IF-connected-GF-AIM} demonstrates that NRG 
provides a viable tool to compute MF $4$p correlators at very low temperatures.

For the real-frequency frameworks, ZF and KF, we cannot resort as easily to previous results.
Hence, we next benchmark our NRG ZF and KF data against perturbation theory (PT),
for the AIM at weak interaction, $U/\Delta \!=\! 1/2$.
In \paperI, we discussed second-order PT for the vertex in the MF and KF
(one proceeds analogously in the ZF);
results for $G^\connected$ follow after reattaching external legs.
Figure~\ref{fig:ZF-KF-connected-GF-SOPT}(a) shows $G^\connected_{\sigma\sigma'}(\vec{\omega})$
for the ZF,
again at $T \!=\! 10^{-3} D$ and $\omega \!=\! 0$.
There is very good agreement between NRG and PT.
The color plots, now involving real and imaginary parts,
show features somewhat similar to the MF results discussed above:
a decay in all directions
and a plus-shaped structure around a prominent signal at the origin.
The two spin components mostly have opposite signs;
however, contrary to the MF, the ZF data also involve sign changes within a spin component.
Further, for these weak-interaction results, 
the $\uparrow\downarrow$ component is larger in magnitude 
and extends to larger $\nu^{(\prime)}$ values
than the $\uparrow\uparrow$ component, since
the former has a first-order contribution
while the latter starts at order $U^2$.

Turning to KF results, 
our scheme of extracting $G^\connected$ out of the full PSFs $S$ works best for the retarded Keldysh components,
where the Keldysh indices $\vec{k}$ of a correlator contain only one entry equal to $2$.
The reason is that we use a simple but \textit{asymmetric} EOM (see \Sec{sec:EOM}).
Figure~\ref{fig:ZF-KF-connected-GF-SOPT}(b) shows the retarded 
correlator $G^{\vec{k};\mathrm{con}}_{\sigma\sigma'}(\vec{\omega})$
with $\vec{k} \!=\! 2111$ for the weakly interacting AIM
(identical parameters as above).
Again, there is good agreement between NRG and PT.
Similar to the ZF results, we observe that the $\uparrow\downarrow$ component
exceeds the $\uparrow\uparrow$ component in magnitude.
However, the structure of these $\nu$-$\nu'$ color plots at $\omega \!=\! 0$
is markedly different between the ZF and KF:
whereas the ZF $\uparrow\downarrow$ component is mirror and point symmetric,
the retarded counterpart is only point symmetric.
The mirror symmetry of the ZF $\uparrow\uparrow$ component is merely broken by the vanishing diagonal;
the retarded counterpart, sharing the vanishing diagonal, is generally only point symmetric.
The reduced symmetry of the KF correlator is due to the fact the Keldysh indices
$\vec{k} \!=\! 2111$ single out the first index, related to the frequency $\nu$.
The $\nu$-$\nu'$ structure at $\omega \!=\! 0$ of other retarded components not shown,
i.e., $\vec{k} \!\in\! \{1211, 1121, 1112\}$,
follows from the $\vec{k} \!=\! 2111$ component by $90^\circ$ rotations in either 
the real part, the imaginary part, or both.

\begin{figure}
\includegraphics[width=\linewidth]{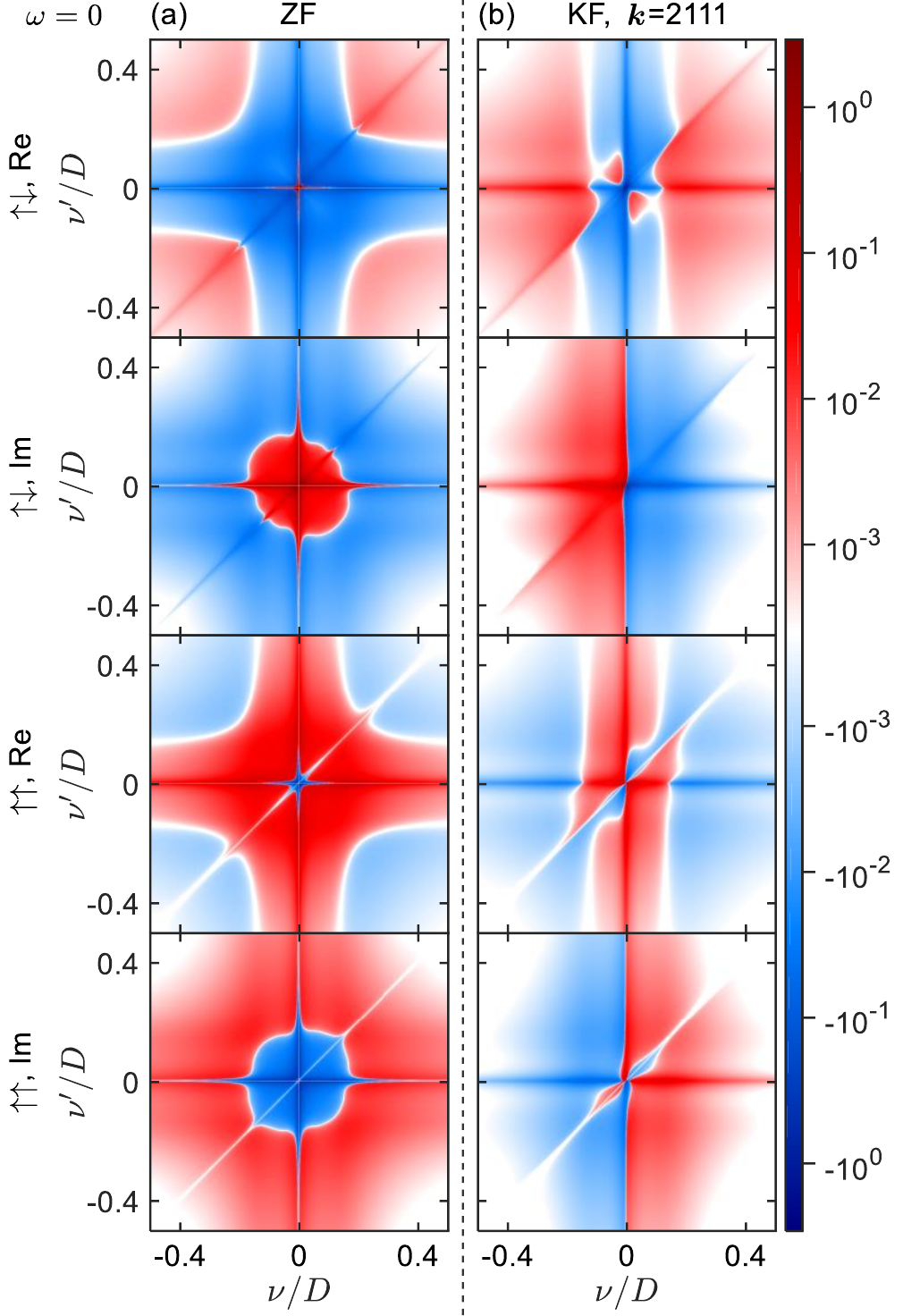}
\caption{%
Real-frequency correlators
at strong coupling. (a) ZF and (b) KF ($\vec{k} \!=\! 2111$) 
connected $4$p correlator,
$(\Delta^4 /U) \, G_{\sigma \sigma'}^{\connected}$,
of the strongly interacting AIM ($U/\Delta \!=\! 5$, $\Delta / D \!=\! 1/25$, $T \!=\! 10^{-4} D$, $\Tk / D \!\simeq\! 1/200$).
In both panels, the real-frequency structure is much richer than that 
obtained at weak interaction (Fig.~\ref{fig:ZF-KF-connected-GF-SOPT}), and also than that of the  imaginary-frequency \MF/ results (\Fig{fig:IF-connected-GF-AIM}) obtained for the same system parameters.
}
\label{fig:ZF-KF-connected-GF-AIM} 
\end{figure}

Finally, we turn to real-frequency connected correlators for the AIM at strong interaction,
setting $U / \Delta \!=\! 5$ as for the MF results.
Figures~\ref{fig:ZF-KF-connected-GF-AIM}(a) and \ref{fig:ZF-KF-connected-GF-AIM}(b) 
show the ZF and retarded KF ($\vec{k} \!=\! 2111$)
connected correlators $G^\connected_{\sigma\sigma'}$, respectively.
We see that the symmetry considerations from above still hold, and 
the characteristic structure with distinct values at the origin, along the $\nu^{(\prime)}$ axes,
and at the diagonal (with zero values for $\sigma \!=\! \sigma'$) persists.
Apart from these similarities, the real-frequency
structure seen in \Fig{fig:ZF-KF-connected-GF-AIM} 
is much richer than that of the weak-interaction results of \Fig{fig:ZF-KF-connected-GF-SOPT} or the imaginary-frequency MF results of \Fig{fig:IF-connected-GF-AIM}. The latter
were obtained not only with identical system parameters, 
but from the \textit{same} discrete data for the connected PSFs
as the ZF and KF results of \Fig{fig:ZF-KF-connected-GF-AIM}.
This striking increase in complexity  of real-frequency versus imaginary-frequency data illustrates the benefits of having direct access to real-frequency data---%
attempting to recover it from imaginary-frequency data via analytic continuation would be extremely challenging.

Altogether, the collection of plots in this section
illustrates the power of the 
spectral representation of multipoint correlators
and our NRG scheme to evaluate them:
The dynamic information of the system is 
fully and compactly encoded in the PSFs (exemplified in Figs.~\ref{fig:Adisc4}, \ref{fig:Adisc4_same_spin}, and \ref{fig:Adisc4_all_permutation}).
The diverse correlators of the 
\MF/ [\Fig{fig:IF-connected-GF-AIM}],
\ZF/ [\Fig{fig:ZF-KF-connected-GF-AIM}(a)], 
and \KF/ [\Fig{fig:ZF-KF-connected-GF-AIM}(b)]
all follow upon convolution of the same PSFs with appropriate kernels.

\section{Results:  RIXS}
\label{sec:RIXS}

\begin{figure}
\includegraphics[width=\linewidth]{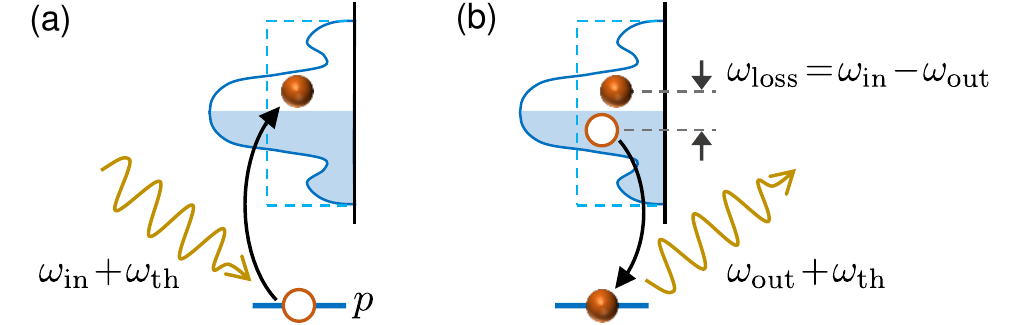}
\caption{%
Schematic depiction of XAS and RIXS.
(a) A photon of energy $\omegain + \omega_\threshold$ is absorbed (with absorption threshold $\omega_\threshold$), transferring an electron from the core level ($p$) to a valence orbital
at the same location.
For the Mahan impurity model (MIM), the local density of states of the valence orbital is flat (dashed box); for the augmented Anderson impurity model (AAIM), it has structure (solid line), including a Kondo resonance.
(b) An electron at the valence orbital relaxes to the core level to fill the hole, emitting a photon of energy $\omegaout \!+\! \omega_\threshold $ and resulting in an excitation of energy $\omegaloss \!=\! \omegain \!-\! 
\omegaout$ in the solid. XAS measures the probability for (a) as a function of $\omegain$; RIXS measures the joint probability for (a) and (b) as a function of
$\omegain$ and $\omegaloss$.}
\label{fig:RIXS_setup} 
\end{figure}

We conclude our presentation of numerical results with
an application of great physical interest: resonant inelastic x-ray scattering
(RIXS). This type of measurement involves photon-in--photon-out spectroscopy.
An incident x-ray photon excites an electron from a core level into a valence orbital [Fig.~\ref{fig:RIXS_setup}(a)]. Then,  a valence electron relaxes to fill the core hole, emitting a photon, 
while the difference between the energies of the incident and emitted photons,  $\omegaloss = \omegain - \omegaout$, is transferred into the solid [Fig.~\ref{fig:RIXS_setup}(b)]. Measuring  $\omegaloss$ thus yields spectroscopic information about the excitations of the solid.

Below, we compute RIXS spectra by convolving a single 4p PSF for a 
specified set of four operators with a suitably chosen convolution kernel (different from those used hitherto). By using NRG, we achieve 
a fine resolution of power-law singularities and characteristic low-energy scales not accessible by ED methods.

\subsection{Models}
\label{sec:XAS-AAIM}

We have performed proof-of-principle computations for two minimal models.
The first, to be called Mahan impurity model (MIM), was used by Mahan for describing x-ray absorption spectroscopy (XAS) in metals, manifesting the celebrated x-ray--edge singularity~\cite{Mahan1967,Roulet1969,Nozieres1969a,Nozieres1969b} in the absorption spectrum. Its Hamiltonian is 
\begin{equation}
\label{eq:XES-model}
\mc{H}_\mim = 
\sum_\epsilon \epsilon \, c_{\epsilon}^\dagger c^\pdag_{\epsilon} 
+ | \epsilon_p |  \, p p^\dagger
- \Up \, c^\dagger c \, p p^\dagger ,
\end{equation}
where $c \!=\! \sum_{\epsilon} c_\epsilon$. 
The  first term describes a flat, half-filled band of spinless electrons with $\epsilon \!\in\! [-D,D]$ and Fermi energy at $\epsilon \! =\! 0$, the second  a localized core level with energy $\epsilon_p \!\ll\! -D$, usually occupied. An x-ray absorption process, described by the transition operator $\Tc^\dagger \!=\! c^\dagger p$, transfers an electron from the core level into the conduction band. The resulting core hole, with hole number operator $pp^\dagger$, induces an attractive, local scattering potential $ -\Up < 0$ for the band (with $\Up \!\ll\! |\epsilon_p|$), described by the third term. Its sudden switch-on changes the wave functions of
all conduction electrons, so that their initial and final ground states 
become orthogonal. This is Anderson's orthogonality catastrophe \cite{Anderson1967,Anderson1967a,Weichselbaum2011}.
It is also responsible \cite{Hopfield1969,Muender2012} for the singular
behavior in the XAS spectrum \cite{Mahan1967,Roulet1969,Nozieres1969a,Nozieres1969b}, accessible by NRG \cite{Oliveira1981,Helmes2005,Muender2012},
and also to the RIXS spectrum \cite{Ting1972,Nozieres1974b}. 
To our best knowledge, the singularity-related features of the latter have not yet been studied numerically.

The second model describes x-ray absorption for an augmented Anderson impurity model (AAIM). 
 We take the AIM Hamiltonian \eqref{eq:H_AIM}, augmented by core level terms,
\begin{equation}
\mc{H}_{\aaim} = \mc{H}_\siam +|\epsilon_p| \, p p^\dagger - \Up \sum_\sigma d_\sigma^\dagger d^\pdag_\sigma p p^\dagger .
\end{equation}
We ignore the spin of the core electron, since the interaction between the core and impurity levels is spin independent and the core-hole number cannot be larger than one. The transition operator describing x-ray absorption now reads 
$\Tc_{\sigma}^\dagger = d_{\sigma}^\dagger p$. The XAS and RIXS spectra 
will have a richer structure than those for the first model,
reflecting properties of the local density of states of the AIM, such as the presence of a Kondo resonance and Hubbard side bands. 

We note in passing that a model of this type has been used
to describe optical absorption for quantum dots with Kondo correlations
\cite{Helmes2005,Tureci2011}, predicting 
x-ray--edge-type lineshapes that have been verified experimentally \cite{Latta2011}. Moreover,
our methodology is applicable also if the local density of states is computed self-consistently via DMFT, as done for RIXS computations for material systems 
\cite{Hariki2018,Wang2019:RIXS}. We leave such applications, and generalizations to nonlocal RIXS processes 
with momentum-dependent spectra, for future work. Here, our goal is to use the above two minimal models to highlight features in RIXS spectra that emerge when truly low-energy scales, beyond the reach of ED~\cite{Tsutsui2003,Ishii2005,Chen2010,Stavitski2010,Kourtis2012,Haverkort2012,Wohlfeld2013,
Haverkort2014,Green2016,Jia2016,Hariki2018,Wang2019:RIXS} or Bethe--Salpeter   treatments \cite{Vinson2011,Gilmore2015,Gilmore2021}, can be resolved. 

\subsection{Partial spectral functions for XAS and RIXS}
\label{sec:how-to-compute-RIXS}

We now turn to the computation of XAS and RIXS spectra, starting with 
the former. XAS measures the probability for the absorption
of an x-ray photon [Fig.~\ref{fig:RIXS_setup}(a)], the first part of a RIXS process, without tracking what happens subsequently. This probability
is given by~\cite{Ament2011,Hariki2020} 
\begin{align}
I^\xas_{\sigma} (\omegain) &=  
\frac{-1}{\pi} \mr{Im} \sum_{\ub{1}\ub{2}} \frac{\rho_{\ub{1}} (\Tc^\pdag_{\sigma})_{\ub{1}\ub{2}} (\Tc_{\sigma}^\dagger)_{\ub{2}\ub{1}}}{\omegain + \omega_\threshold - E_{\ub{2}\ub{1}} + \mi \Gammap} \, ,
\label{eq:XAS-textbook}
\end{align}
where $\omegain + \omega_\threshold$ is the incident photon energy, $\omega_\threshold$ the absorption threshold, and $\Gammap$
the inverse lifetime of the core hole.  Using notation encompassing both models, we write the transition operator as 
$\Tc_{\sigma}^\dagger = x^\dagger_\sigma p$, 
with $x^\dagger = c^\dagger$ and the spin index omitted for the MIM, or
$x^\dagger_\sigma = d^\dagger_\sigma$ for  the AAIM. 
The transition operator links two Hilbert subspaces, to be 
called ``no hole'' and ``one hole.'' These are not coupled by  
the Hamiltonian, $[p^\dagger p, \Hc] = 0$, and can be diagonalized separately.
The sums $\sum_{\ub{2}}$ or $\sum_{\ub{1}}$
run over all energy eigenstates of the one-hole or no-hole subspaces, respectively. 
The difference between their
ground-state energies  defines $\omega_\threshold$.

Equation~\eqref{eq:XAS-textbook} has the form of a convolution of a  2p PSF, $S^\xas_\sigma (\ve) = S [\Tc^\pdag_{\sigma}, \Tc_{\sigma}^\dagger] (\ve)$,
with a Lorentzian peak: 
\begin{align}
I^\xas_{\sigma} (\omegain)  
&=  \int \md \ve \, 
\frac{ (\Gammap/\pi)
S_\sigma^\xas(\ve)}{(\omegain + \omega_\threshold - \ve)^2 + \Gammap^2} .
\label{eq:XAS}
\end{align}
The x-ray absorption spectrum thus measures the 2p PSF
$S^\xas_\sigma (\omegain + \omega_\threshold)$, broadened by $\Gammap$.

Next, consider RIXS spectra. They measure the probability for the absorption and subsequent reemission of x-ray photons with energies
$\omegain + \omega_\threshold$ and
$\omegaout + \omega_\threshold$, respectively, with 
$\omegaloss \!=\! \omegain - \omegaout$ (Fig.~\ref{fig:RIXS_setup}). 
This probability is given by the  Kramers--Heisenberg formula~\cite{Kotani2001,Ament2011,Ament2009},
\begin{flalign}
\label{eq:RIXS_conventional}
& I^\rixs_{\sigma \sigma'} (\omegain, \omegaloss) & \\
\nonumber
& \; = 
\sum_{\ub{1}\ub{3}} \rho_{\ub{1}} 
\, \pigg|
\sum_{\ub{2}} 
\frac{  (\Tc^\pdag_{\sigma})_{\ub{1}\ub{2}} (\Tc_{\sigma'}^\dagger)_{\ub{2}\ub{3}} }{\omegain + \omega_\threshold - E_{\ub{2}\ub{1}} - \mi \Gammap} 
{\pigg|}^2
\delta(\omegaloss - E_{\ub{3}\ub{1}} )  &
\\ \nonumber
& \; = \sum_{\ub{1}\ub{2}\ub{3}\ub{4}}  
\frac{\rho_{\ub{1}} (\Tc^\pdag_{\sigma})_{\ub{1}\ub{2}} (\Tc_{\sigma'}^\dagger)_{\ub{2}\ub{3}} \delta(\omegain \!-\! \omegaout \!-\! E_{\ub{3}\ub{1}} )
(\Tc^\pdag_{\sigma'})_{\ub{3}\ub{4}} (\Tc_{\sigma}^\dagger)_{\ub{4}\ub{1}} }{(\omegaout \!+\! \omega_\threshold \!-\! E_{\ub{2}\ub{3}} \!-\! \mi \Gammap)
(\omegain \!+\! \omega_\threshold \!-\! E_{\ub{4}\ub{1}} \!+\! \mi \Gammap)}  . &
\end{flalign}
In the second line, the expression within the square is the
amplitude for a second-order process, involving  
up- and downward transitions from the no-hole to one-hole subspaces and back.
The third line explicates the dependence on the
frequencies $\omegain$ and $\omegaout$.
Transitions with $\omegain = E_{\ub{4}\ub{1}} -
\omega_\threshold$ and $\omegaout = E_{\ub{2}\ub{3}} -
\omega_\threshold$ conserve energy and are ``real.''  If both
equalities hold (at $T=0$ this requires $\omegain > 0$),
the net result is a resonant fluorescence-type process with a 
diverging amplitude due to vanishing energy denominators, requiring regularization  by $\Gammap$. If one or both equalities are violated, the net result is an off-resonance Raman-type process involving virtual intermediate states. Nevertheless, its amplitude can be sizable provided that  
$\omegaloss =  E_{\ub{3}\ub{1}}$ (at $T=0$ this requires $\omegaloss > 0$), corresponding to a real initial-to-final transition between two no-hole states.
Hence, $I^\rixs$ has support for both positive and negative $\omegain$, but only for $\omegaloss \gtrsim - T$, i.e., the dependence on $\omegaloss$ has threshold character.
Sharp features in the $\omegain$ dependence of $I^\rixs$ are smeared
out by $\Gammap$. By contrast, the $\omegaloss$ dependence enters separately from $\Gammap$ and $\omegain$ in \Eq{eq:RIXS_conventional};
hence, it is \textit{not} smeared out and can contain sharp, distinctive features even if $\omegaloss \ll \Gammap, |\omegain|$.

We can express \Eq{eq:RIXS_conventional} as a convolution involving a $4$p PSF, $ S_{\sigma \sigma'}^\rixs (\vec{\ve}) = S [\Tc^\pdag_{\sigma}, \!\!\Tc_{\sigma'}^\dagger, \!\!\Tc^\pdag_{\sigma'}, \!\! \Tc_{\sigma}^\dagger] (\vec{\ve}) $ and suitable kernel:
\begin{flalign}
\label{eq:RIXS}
& I^\rixs_{\sigma \sigma'} (\omegain, \omegaloss) & 
\\ \nonumber 
& \quad  = \int \! \md^3 \ve \, 
\frac{S_{\sigma \sigma'}^\rixs (\vec{\ve}) \, 
 \delta(\omegaloss \!-\! \ve_2)}{(\omegain \!+\! \omega_\threshold \!-\! \ve_1 \!-\! \mi \Gammap) (\omegain \!+\! \omega_\threshold \!-\! \ve_3 \!+\!
  \mi \Gammap)} .
\end{flalign}
The PSF arguments $\ve_1$ and $\ve_3$ are connected
to $\omegain + \omega_\threshold$ by Lorentzian kernels,
$\ve_2$ to $\omegaloss$ by a Dirac $\delta$ function.
The dependence of $I^\rixs$ on $\omegain$ and $\omegaloss$ thus 
probes that of $S^\rixs$ on $\ve_1,$ $\ve_3$, and $\ve_2$, respectively, the former two broadened by $\Gammap$, the latter not. 
[Note that, if $\Gamma_p \to 0^+$, $I^\rixs$ can also be viewed as a component of the KF correlator in the contour basis,
$(1/4\pi) G^{\vec{c}} [\Tc^\pdag_{\sigma}, \!\!\Tc_{\sigma'}^\dagger, \!\! 
\Tc^\pdag_{\sigma'}, \!\! \Tc_{\sigma}^\dagger](\vec{\omega})$,
with $\vec{c}=+$%
$+$%
$-$%
$-$ and frequency arguments, in the particle-hole convention \eqref{eq:PHconvention}, given by  $\nu \! = \! \nu' = \omegain \! + \! \omega_\threshold$, $\omega \! = \! - \omegaloss$.]

For sufficiently large $|\omegain|$, the dependence of 
$I^\rixs$ on $\omegaloss$ mimics that of a dynamical susceptibility.
This can be seen as follows. 
When $|\omegain|$ is much larger than $\Gammap$ and also lies far outside the support of the PSF along the $\ve_1 - \omega_\threshold$ and $\ve_3 - \omega_\threshold$ axes, the denominator of the RIXS kernel in \Eq{eq:RIXS} can be approximated by $|\omegain|^2$, ignoring its $\ve_1$ and $\ve_3$ dependence.
According to the PSF sum rules \eqref{eq:PSFsumrules}, $I^\rixs$ then reduces to a 2p PSF, 
\begin{subequations}
\label{subeq:RIXS_largewin}
\begin{flalign}
I^\rixs_{\sigma \sigma'} (\omegain, \omegaloss) & \simeq 
|\omegain|^{-2}
S[ \Tc^\pdag_{\sigma} \Tc_{\sigma'}^\dagger , \Tc^\pdag_{\sigma'} \Tc_{\sigma}^\dagger ] (\omegaloss)  \hspace{-1cm} & 
\label{eq:RIXS_largewin}
\\ 
&   \simeq |\omegain|^{-2}  S[x^\pdag_{\sigma} x^\dagger_{\sigma'}, 
x^\pdag_{\sigma'} x^\dagger_{\sigma}] (\omegaloss)  .
\hspace{-1cm} & 
\label{eq:RIXS_largewin2}
\end{flalign}
\end{subequations}
The second line, following via 
$\Tc^\pdag_{\sigma} \Tc_{\sigma'}^\dagger = x^\pdag_{\sigma} x^\dagger_{\sigma'} p^\dagger p$, involves a $2$p PSF evaluated purely within the no-hole subspace.
It represents the imaginary part of an equilibrium 2p correlator, the ZF particle-hole susceptibility $\chi_{\sigma \sigma'}(\omegaloss) \!=\!  G[x_{\sigma}^\pdag x^\dagger_{\sigma'}, x_{\sigma'}^\pdag x^\dagger_{\sigma}] (\omegaloss)$,
considered for frequencies $\omegaloss \gtrsim T$.
[The latter restriction is needed because $\chi(\omega)$ involves PSFs for two permutations, whereas \Eq{eq:RIXS_largewin2} involves only one.]
Similarly, when $\Gammap$, instead of $|\omegain|$, is the largest scale in the kernel denominator, we have 
\begin{equation}
I^\rixs_{\sigma \sigma'} (\omegain, \omegaloss) \simeq \Gammap^{-2} \, S[x^\pdag_{\sigma} x^\dagger_{\sigma'}, 
x^\pdag_{\sigma'} x^\dagger_{\sigma}] (\omegaloss) .
\label{eq:RIXS_largeGammap}
\end{equation}

\begin{figure}
\includegraphics[width=\linewidth]{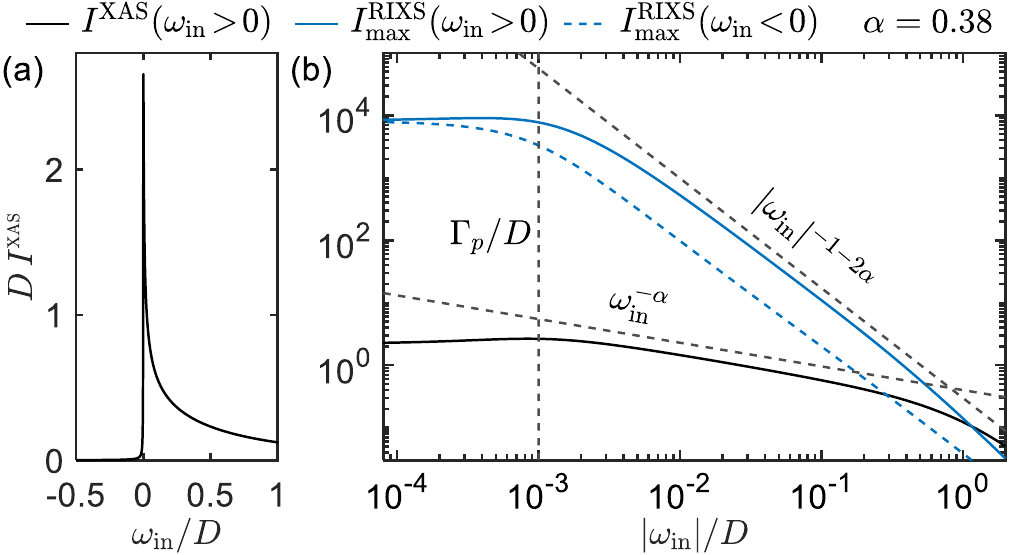}\vspace{-2mm}
\caption{
XAS spectrum $I^\xas(\omegain)$ (black lines)
for the MIM ($U_p/D \!=\! 0.5$,  $T/D \!=\! 10^{-5}$, $\Gammap/D \!=\! 10^{-3}$), using (a) linear  and (b) logarithmic 
scales.  The power law $I^\xas \sim \omegain^{-\alpha}$ observed for 
$\Gammap \lesssim \omegain \lesssim D$ quantitatively matches the prediction of \Eq{eq:XES_analytic}. For comparison, (b) also shows the RIXS maximum 
$I_{\max}^\rixs (\omegain)\!=\! \max_{\omegaloss} 
[I^\rixs(\omegain,\omegaloss)]$ (blue lines, solid and dashed),
diverging much more strongly as $\sim |\omegain|^{-1-2\alpha}$.
Here, $\epsilon_p$ merely shifts $\omega_\threshold \!\simeq\! -0.33 D +|\epsilon_p| \gg D$, without changing $I^\xas$ and $I^\rixs$ as a function of $\omegain$ or $\omegaloss$.}
\label{fig:XAS_XES}
\end{figure}

\begin{figure*}
\includegraphics[width=\linewidth]{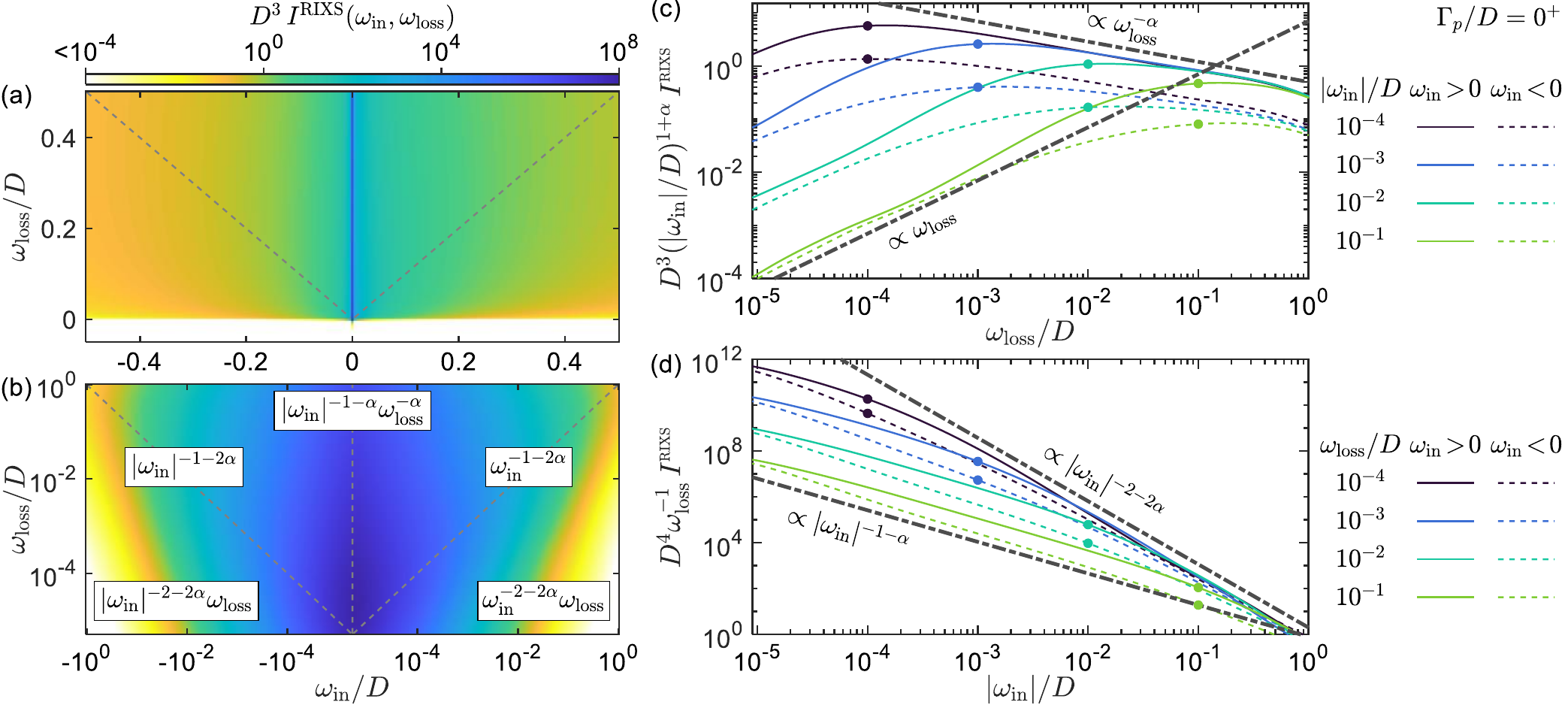}
\vspace{-4mm}
\caption{
RIXS spectrum $I^\rixs(\omegain,\omegaloss)$ 
for the MIM ($U_p/D \!=\! 0.5$, $T/D \!=\! 10^{-5}$, $\Gammap/D \!=\! 0^+$). 
 (a),(b) Color-scale plots using linear or logarithmic frequency scales. (c) Line cuts of $I^\rixs$ as a function of $\omegaloss$, for 
several fixed values of $\omegain$ (indicated by dots placed at 
$\omegaloss = |\omegain|$). (d) Analogous to 
(c), with the roles of $\omegaloss$ and $|\omegain|$ interchanged.
The power laws revealed in (c) and (d) are summarized in 
\Eq{eq:RIXS_XES_powerlaws},  and in (b) 
using legends placed at positions indicating
the corresponding values of $|\omegain|$ and $\omegaloss$.}
\label{fig:RIXS_XEM_unnormalized-colorscale}
\end{figure*}

We next discuss some computational details. As mentioned above,
the core-hole number is conserved; hence, the no-hole and one-hole subspaces can be diagonalized separately \cite{Weichselbaum2011,Muender2012,Weichselbaum2012b}. This yields two sets of AS bases, capable of accurately resolving low frequencies $|\omegain|, |\omegaloss| \ll D$.
Their direct sum yields an eigenbasis for the whole Hilbert space, 
suitable for computing PSFs as explained in Sec.~\ref{sec:PartialSpectralFunctions}. The density matrix is populated only
by no-hole states. We write the eigenenergies 
of the one-hole subspace as $E_\ub{i} + \omega_\threshold$. 
This splits off $\omega_\threshold$ from the arguments $\ve$
of  $S^\xas (\ve)$ and $\ve_1$ and $\ve_3$ of $S^\rixs
(\vec{\ve})$, canceling the $\omega_\threshold$ in the kernels of Eqs.~\eqref{eq:XAS} and \eqref{eq:RIXS}.
The discrete PSFs $S^\xas(\sve)$ and $S^{\rixs}(\vec{\sve})$
are saved as histograms. To focus on the inelastic part of $I^\rixs (\omegaloss \gneq 0)$, we exclude the spectral weights of the histogram $S^\rixs (\vec{\sve})$ at $\sve_2 = 0$. We use $\Lambda = 4$ and $n_z = 4$.
Broadened PSFs $S^\xas(\ve)$ and $S^{\rixs}(\vec{\ve})$
are obtained using centered log-Gaussian kernels, as
described at the end of \App{app:logGaussian}.
We have verified that our NRG code and a recent ED code~\cite{Wang2019:RIXS} give consistent result for Wilson chains short enough that exact diagonalization is possible.

\subsection{MIM: XAS and RIXS results}
\label{sec:XAS-RIXS-results-XES}

Figure~\ref{fig:XAS_XES} shows a typical XAS spectrum for the MIM.
$I^\xas(\omegain)$ features a threshold at $\omegain=0$, and, just above it,
the x-ray--edge singularity [Fig.~\ref{fig:XAS_XES}(a)]. 
Sharp features are smeared out by nonzero $\Gammap$ or $T$.
As these have analogous broadening effects, we take $\Gammap \! > \! T$ throughout,
except for the scaling analysis in the very next paragraph and \Fig{fig:RIXS_XEM_unnormalized-colorscale}.
In the following, we focus on the behavior 
outside the broadening region, $\omegain \! \gg \!  \Gammap$. 
A log-log plot for positive frequencies [Fig.~\ref{fig:XAS_XES}(b), black line] reveals power-law behavior  in the range 
$\Gammap \lesssim \omegain \lesssim 0.5 D$. 
Its form is consistent with the analytical prediction \cite{Mahan1967,Roulet1969,Nozieres1969a,Nozieres1969b}
\begin{align}
I^\xas(\omegain) \sim \omegain^{-\alpha}
, \quad
\alpha = 2 \frac{\delta}{\pi} - \frac{\delta^2}{\pi^2} ,
\quad  \tan \delta = \frac{\pi U}{2D} , 
\label{eq:XES_analytic}
\end{align}
and previous NRG studies \cite{Helmes2005,Muender2012}. Here, $\delta$ 
is the phase shift  for conduction electrons near the Fermi level induced by the core-hole scattering potential. It can also
be computed using $\delta = \pi \Deltah$, where $\Deltah$ is the charge drawn in toward the scattering site by the core hole \cite{Weichselbaum2011,Muender2012}. The power-law behavior 
can be traced back to Anderson orthogonality~\cite{Anderson1967,Anderson1967a}, see Ref.~\onlinecite{Muender2012} and \App{app:AndersonOrthogonality} for a heuristic discussion.
Such power-law behavior is inaccessible to ED methods, since the description of Anderson orthogonality in essence requires an infinite bath.

Figures~\ref{fig:RIXS_XEM_unnormalized-colorscale}(a) and \ref{fig:RIXS_XEM_unnormalized-colorscale}(b)
show color-scale plots of the RIXS spectrum, using 
(a) linear and (b)  logarithmic frequency axes. $I^\rixs(\omegain,\omegaloss)$ has support 
only at positive $\omegaloss$, and at both positive and negative $\omegain$, with somewhat more weight at positive than negative $\omegain$. It shows a power-law dependence
on both $\omegain$ and $\omegaloss$, culminating in a divergence
around $\omegain = 0$ 
(cut off by $\Gammap$; to better reveal it, we here chose $\Gammap=0^+$). The power-law exponents depend
on the relative values of $|\omegain|$ and $\omegaloss$, as indicated by
legends in \Fig{fig:RIXS_XEM_unnormalized-colorscale}(b).
They were identified by plotting $I^\rixs$ as a function of $\omegaloss$ or $\omegain$, respectively, for several fixed values of the other variable [Figs.~\ref{fig:RIXS_XEM_unnormalized-colorscale}(c) and \ref{fig:RIXS_XEM_unnormalized-colorscale}(d)]. We find (i)
$I^\rixs \sim |\omegain|^{-2 - 2 \alpha} \omegaloss$ for $\omegaloss 
\ll |\omegain|$; (ii) $I^\rixs \sim |\omegain|^{-1 - 2 \alpha}$ for 
$\omegaloss \simeq |\omegain|$; and 
(iii) $I^\rixs \sim |\omegain|^{-1-  \alpha} \omegaloss^{- \alpha}$ 
for $\omegaloss \gg |\omegain|$, with the same exponent
$\alpha$ as for XAS spectra. In summary, 
\begin{flalign}
\label{eq:RIXS_XES_powerlaws}
I^\rixs 
& 
\sim   \frac{f(\omegaloss/\omegain)}{|\omegain|^{1+2\alpha}}, 
\quad f(x)  \sim 
\begin{cases}
|x| &  (x \ll 1) \, ,   \\
1 & (x \simeq 1) \, ,  \\
|x|^{-\alpha} &  (x \gg 1 ) \, .
\end{cases} \hspace{-1cm} & 
\end{flalign}
The function $f$ is not fully symmetric, $f(x) \neq f(-x)$, having
the same asymptotic behavior but different prefactors for
arguments of opposite sign. Figure~\ref{fig:RIXS_XEM_unnormalized-colorscale}(c) also shows that $ I^\rixs(\omegain,\omegaloss)$, viewed as a function of $\omegaloss$ for fixed $\omegain$, has a maximum around $\omegaloss \simeq |\omegain|$. Its value at this maximum, 
$I^\rixs_\mr{max}(\omegain)$, scales as  $\sim |\omegain|^{-1-2\alpha}$
[Fig.~\ref{fig:XAS_XES}(b), blue lines], with a smaller prefactor (by about an order of magnitude) for negative than positive $\omegain$.

The XAS and RIXS power laws for the MIM are governed by the \textit{same} 
exponent $\alpha$ because both stem from Anderson orthogonality. This was 
recognized already in the early 1970s \cite{Ting1972}, leading to analytic predictions for RIXS exponents [cf.\ Eqs.~(37) and (41)--(43) of Ref.~\onlinecite{Nozieres1974b}]. These predictions match our numerics for case (i) above (numerically most challenging since $\omegaloss \ll |\omegain|$), but not for cases (ii) and (iii).  The discrepancies in the latter two cases, 
where $\omegaloss$ is not $\ll |\omegain|$, 
suggest that, in Ref.~\onlinecite{Nozieres1974b}, the treatment of excitations in the no-hole subspace [labeled $\ub{3}$ in \Eq{eq:RIXS_conventional}] was 
not sufficiently accurate. Clarifying this would require revisiting the analysis of Ref.~\onlinecite{Nozieres1974b}, which is, however, beyond the scope of this work. 

\begin{figure}
\includegraphics[width=\linewidth]{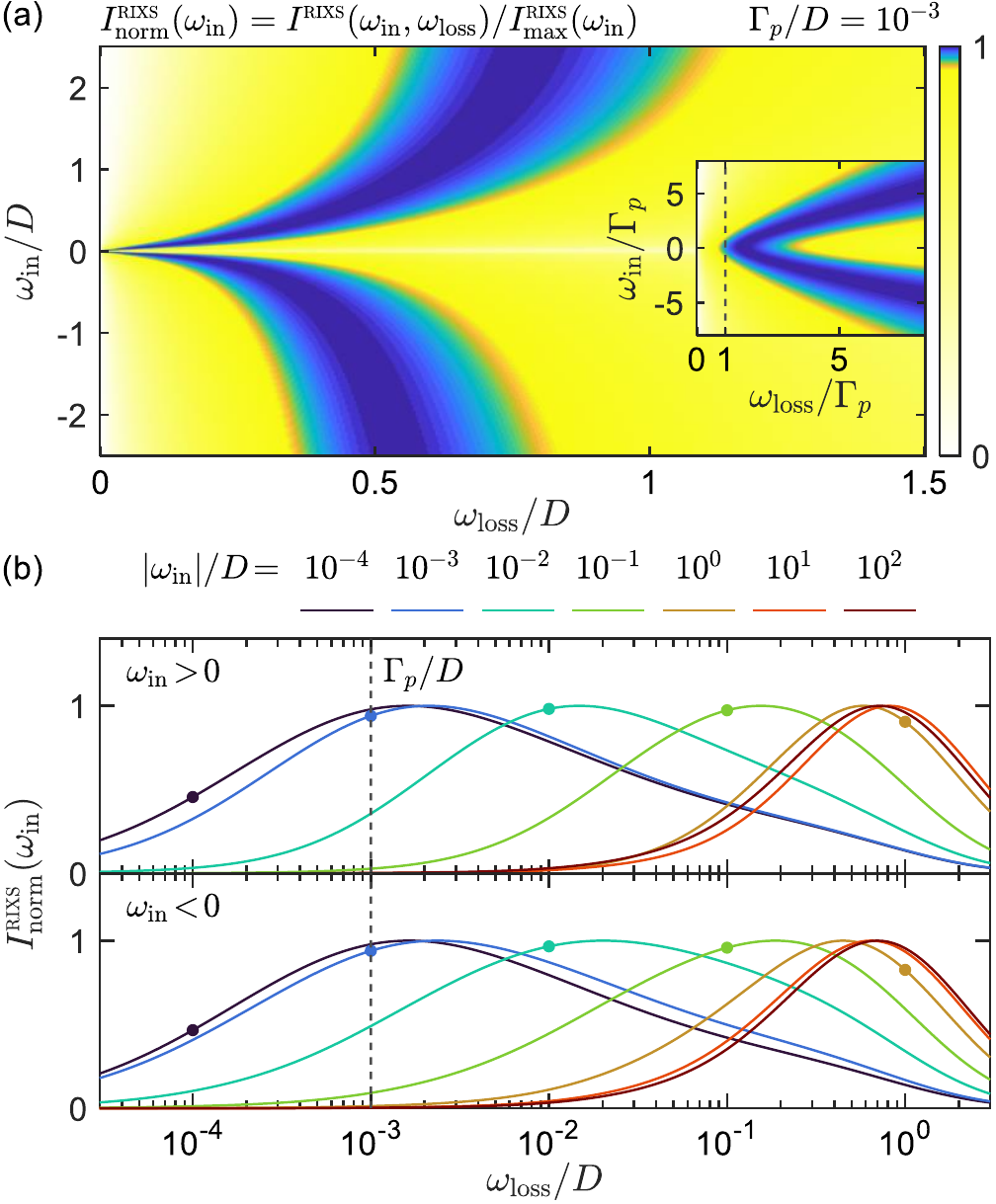}\vspace{-2mm}
\caption{%
Normalized RIXS spectrum $I^\rixs_\normalized(\omegain,\omegaloss) $ 
of the MIM (same parameters as for \Fig{fig:XAS_XES}). 
(a) A color-scale plot using linear frequency axes, and  (b) line plots as functions of $\omegaloss$ for selected values of $\omegain > 0$ (upper panel) and $\omegain < 0$ (lower panel), indicated by dots placed at $\omegaloss =|\omegain| $.
The vertical dashed lines mark the value of $\Gammap$.
Inset of (a) enlarges the low-energy region of (a).
In (b), the peak positions  along the $\omegaloss$ axis 
depend on $|\omegain|$, occurring at
(i) $\simeq \Gammap$ for $|\omegain| < \Gammap$;
(ii) $\simeq |\omegain|$ for $\Gammap < |\omegain| < D$;
(iii) $\simeq D$ for $0.5 D \lesssim \omegain$.}
\label{fig:RIXS_XEM} 
\end{figure}

Because of the strong dependence of the RIXS spectrum on $\omegain$,
it is convenient, when analyzing its dependence on $\omegaloss$, to first normalize the spectrum by its maximal value, 
$I^\rixs_\normalized(\omegain,\omegaloss) =
I^\rixs(\omegain,\omegaloss)/I_{\max}^\rixs(\omegain) $. 
A color-scale plot of $I^\rixs_\normalized$ 
is shown in \Fig{fig:RIXS_XEM}(a). It highlights the presence of a peak as a function of $\omegaloss$ for each value of $\omegain$. We already encountered this peak in \Fig{fig:RIXS_XEM_unnormalized-colorscale}(c). Its evolution with $\omegain$ is analyzed in \Fig{fig:RIXS_XEM}(b), showing very similar,
though slightly asymmetric behavior for positive and negative $\omegain$. 
For $|\omegain|$ in the range between $\Gammap\lesssim \omegain \lesssim 
0.5 D$, $I^\rixs_\normalized$ has a peak at $\omegaloss \simeq |\omegain|$.
For positive or negative $\omegain$, this implies that emission is strongest for
$\omegaout \simeq 0$ or $\omegaout \simeq 2 \omegain$, respectively.
In the former case, all the above-threshold incident energy
is absorbed by the Fermi sea, implying fluorescence-type behavior. The
latter case does not seem to have a simple interpretation, 
but for brevity we will call it ``fluorescence-like,'' too. For both cases,
the observed behavior implies strongly energy-dependent
matrix elements $\Tc_{\ub{i}\ub{j}}$ in \Eq{eq:RIXS_conventional},
reflecting Anderson orthogonality. 

For incident energies smaller than the hole decay rate, $|\omegain| < \Gammap$, the peak gets pinned to $\omegaloss \simeq \Gamma_p$ [\Fig{fig:RIXS_XEM}(b)], 
since the dependence on $\omegain$ becomes smeared.
For very large incident energies, $\omegain \gtrsim 0.5 D$,
the peak positions become pinned at $\omegaloss \simeq D$.
To understand the latter, we evoke single-particle arguments, valid at large
energies: The probability distributions for the excitation of a single particle
or hole are flat (following the local density of states), with support in the range $[0,D]$.
The distribution of a one-particle--one-hole excitation energy has range $[0,2D]$,
with a peak around $D$. 

This case study of the RIXS spectrum of the MIM shows (i) that our scheme is capable of resolving two-variable power-law behavior even if the two variables differ by orders of magnitude and (ii) that the normalized RIXS spectrum is dominated by a peak at $\omegaloss \simeq |\omegain|$ (provided that
$|\omegain| > \Gammap$), reflecting strongly energy-dependent transition matrix elements.

\subsection{AAIM: XAS and RIXS results}
\label{sec:XAS-RIXS-results-AAIM}

\begin{figure}
\includegraphics[width=\linewidth]{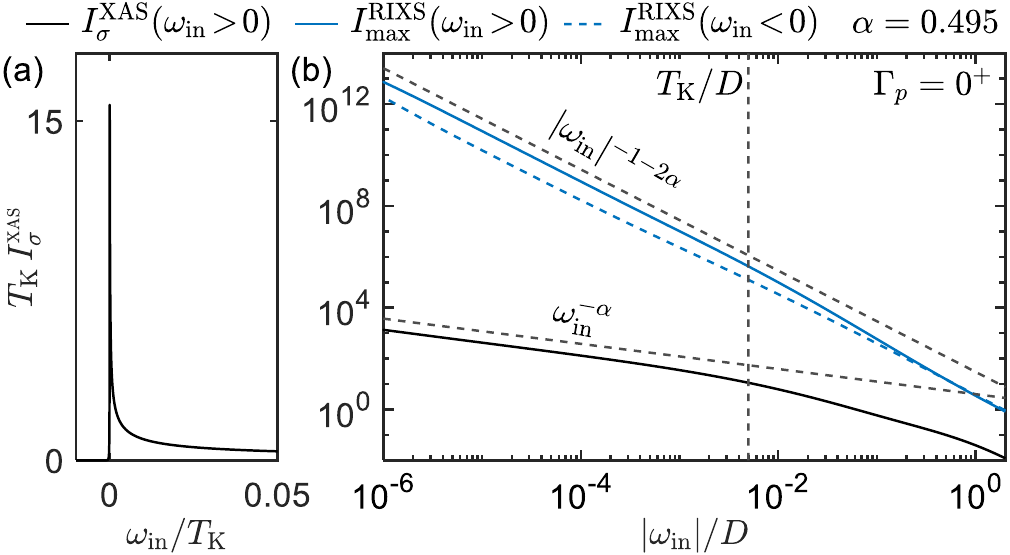}
\caption{
XAS spectrum $I_\sigma^\xas(\omegain)$ (black lines)
for the AAIM ($U/\Delta \!=\! 1/5$, $\Delta/D \!=\! 1/25$, $U_p/U \!=\! 3/2$, $\Tk/D \!\simeq\! 1/200$, $T /D\!=\! 10^{-7}$,
$\omega_\threshold \!\simeq\! -0.47D \!+\! |\epsilon_p| \!\gg\! D$).
The layout is the same as for Fig.~\ref{fig:XAS_XES}.
The power law $I^\xas \sim \omegain^{-\alpha}$ observed for 
$\omegain \lesssim \Tk$ quantitatively matches the prediction of \Eq{eq:XAS-AAIM-alpha}.}
\label{fig:XAS_AAIM}
\end{figure}

\begin{figure*}
\includegraphics[width=\linewidth]{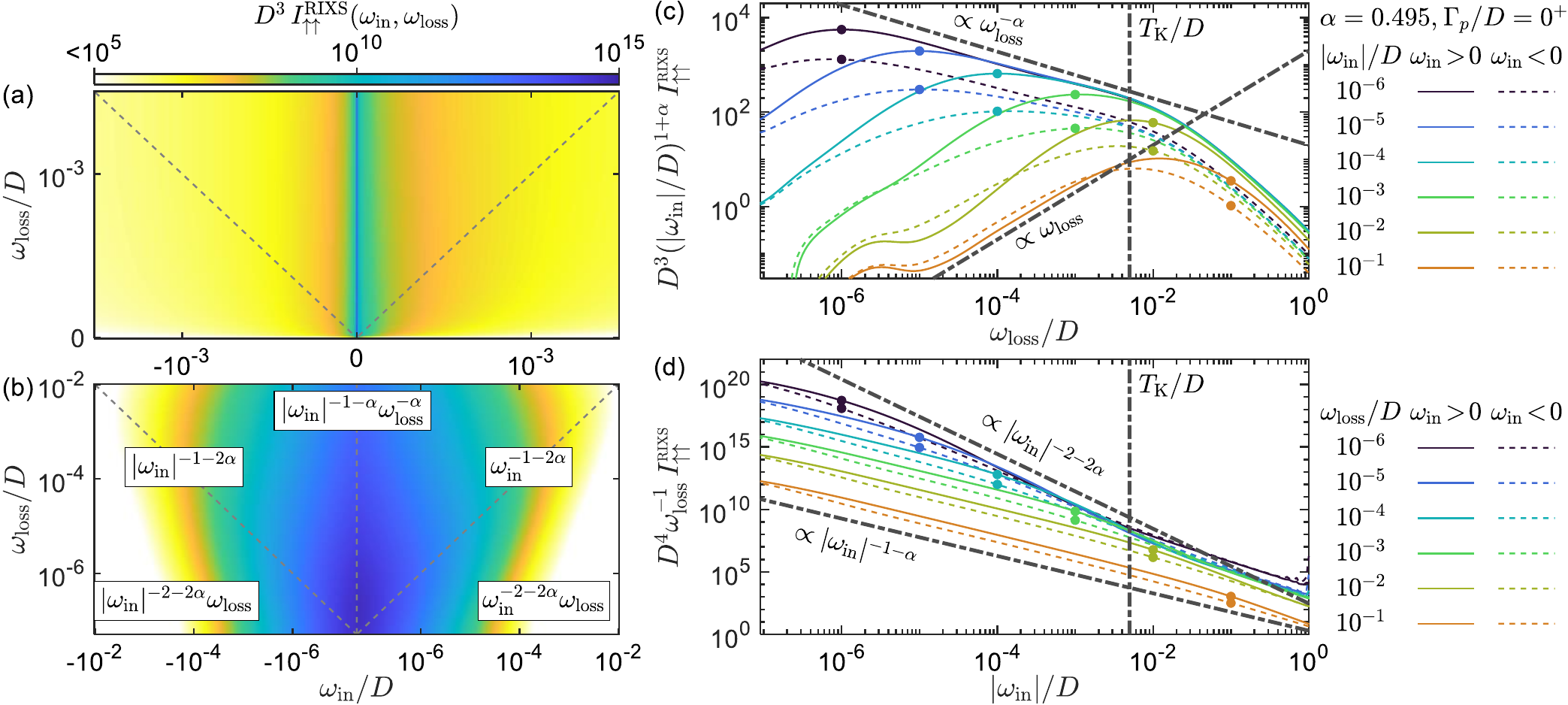}
\vspace{-4mm}
\caption{
RIXS spectrum $I_{\uparrow \uparrow}^\rixs(\omegain,\omegaloss)$ 
for the AAIM (same parameters as for Fig.~\ref{fig:XAS_AAIM}). 
The layout is the same as for \Fig{fig:RIXS_XEM_unnormalized-colorscale}.
The spectrum $I_{\uparrow \downarrow}^\rixs$ (not shown) is qualitatively similar, exhibiting the same power laws.}
\label{fig:RIXS_AAIM_unnormalized-colorscale}
\end{figure*}

Finally, we discuss our XAS and RIXS results for the AAIM.
The unnormalized XAS and RIXS spectra are shown in Figs.~\ref{fig:XAS_AAIM} and
\ref{fig:RIXS_AAIM_unnormalized-colorscale}, respectively. 
For $|\omegain| \ll \Tk$ and $\omegaloss \ll \Tk$, they---remarkably---exhibit power-law behavior of the \textit{same} functional form as for the MIM.

For the XAS spectrum, this is expected: Its power-law form can be understood
\cite{Hopfield1969,Muender2012} by assuming Fermi-liquid excitations in the low-energy regimes of the no-hole or one-hole subspaces, and this assumption is valid for both the MIM and the AAIM. (Low-energy means $\ve \le D$ for the MIM, and $\ve \le \Tk$ for the AAIM, where $\Tk$ is the Kondo temperature in the no-hole subspace.) The only difference is that, for the AAIM, the power-law exponent governing the XAS spectrum depends on spin, 
\begin{align}
\label{eq:XAS-AAIM-alpha}
I_\sigma^\xas(\omegain) \sim \omegain^{-\alpha_\sigma} \, ,  \quad 
\alpha_\sigma = 1 \! - \! (1\! - \! \Delta n_{d \sigma})^2 \! - \! 
\Delta n_{d, -\sigma}^2 \, , 
\end{align}
where $\Delta n_{d\sigma} = n_{d\sigma}^{\rm one} -  
n_{d\sigma}^{\rm no}$ is the 
difference in $n_{d\sigma} = \langle d_\sigma^\dagger d_\sigma^\pdag \rangle$, the 
$d\sigma$-level occupancy,
between the one-hole and no-hole ground states. 
Equation~\eqref{eq:XAS-AAIM-alpha} was derived heuristically in Refs.~\onlinecite{Tureci2011,Latta2011}. These describe theoretical predictions and experimental observations for
probing the Kondo effect in quantum dots by optical absorption experiments [cf.~Eq.~(8) of Ref.~\onlinecite{Tureci2011}].
For the parameters chosen here, 
the no- or one-hole subspaces have $n_{d\sigma}^{\rm no}=1/2$
or $n_{d\sigma}^{\rm one}= 0.949$, respectively,
the former enabling, the latter prohibiting Kondo correlations. The 
absorption-induced no-hole to one-hole transition thus amounts to
switching off Kondo correlations. They decay with time in power-law fashion, due to Anderson orthogonality \cite{Tureci2011},  leading to power-law behavior for the XAS spectrum. Equation~\eqref{eq:XAS-AAIM-alpha} yields spin-independent values of $\Delta n_{d\sigma} = 0.449$ 
and $\alpha_\sigma = 0.495$ (henceforth written as $\alpha$). In Refs.~\onlinecite{Tureci2011,Latta2011}, an exponent $\alpha$
close to $0.5$ was viewed as a fingerprint of Kondo correlations in absorption spectra. 

\begin{figure*}
\includegraphics[width=\linewidth]{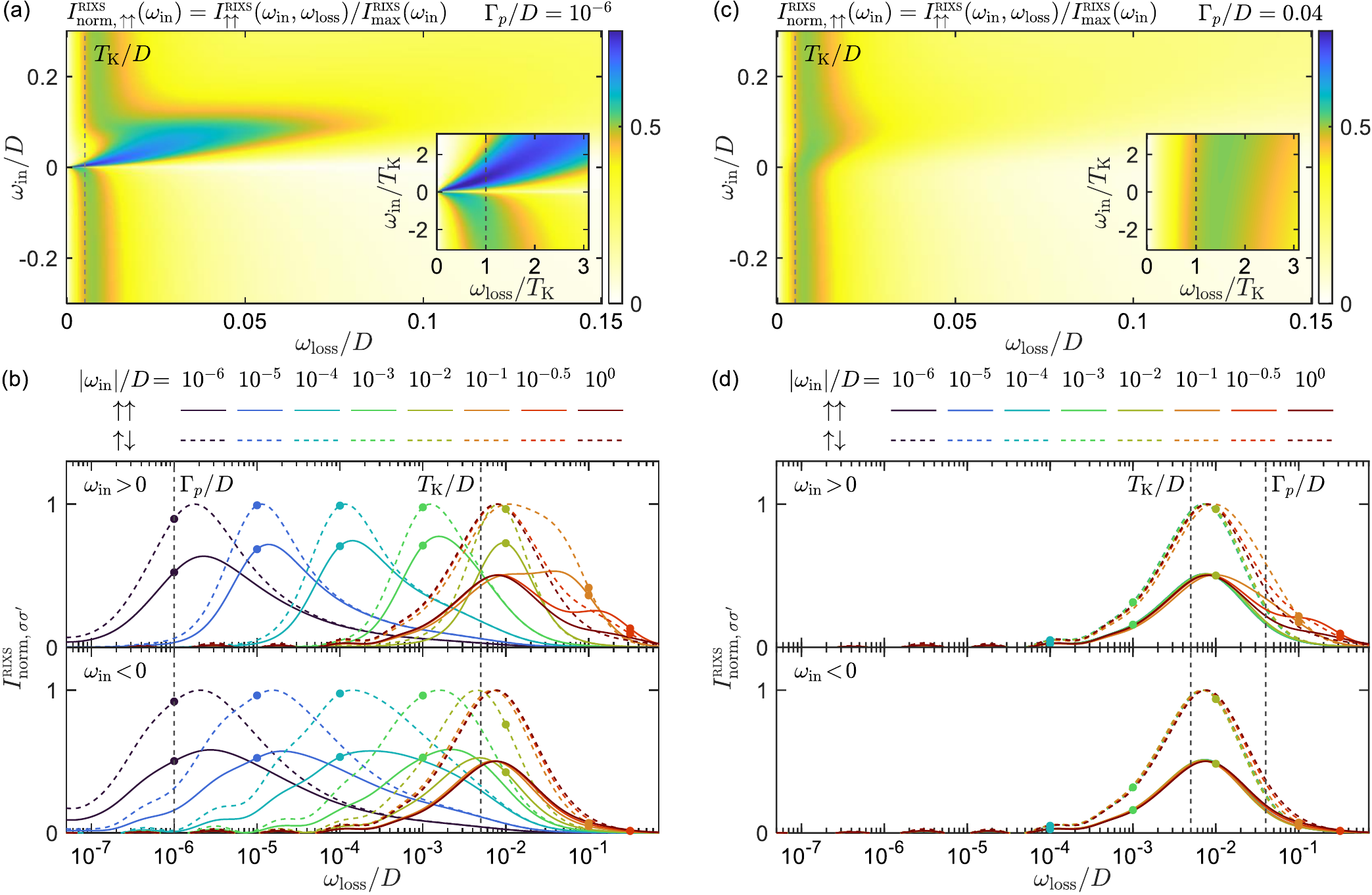}
\caption{%
Normalized RIXS spectrum $I_{\normalized, \sigma \sigma'}^\rixs$
of the AAIM, with the same parameters as for \Fig{fig:XAS_AAIM}, except
that $\Gammap$ is larger (see panel headers) and the broadening width 
smaller (see \App{app:logGaussian}).
We normalize $I_{\sigma \sigma'}^\rixs(\omegain,\omegaloss)$ by 
$I_{\max}^\rixs (\omegain)\!=\! \max_{\omegaloss} 
[I_{\uparrow\downarrow}^\rixs(\omegain,\omegaloss)]$, 
since $I_{\uparrow\downarrow}^\rixs$ has larger maxima than $I_{\uparrow\uparrow}^\rixs$ [compare dashed to solid lines in (b)].
We chose $\Gammap \ll \Tk$ in (a) and (b) and $\Gammap \gg \Tk$ in (c) and (d). For both cases, the same layout
is used as for  \Fig{fig:RIXS_XEM}: (a),(c) Color-scale plots using linear frequency scales, and (b),(d) line cuts using a logarithmic frequency scale
to show the $\omegaloss$ dependence for several fixed values of $\omegain$,
indicated by dots placed at $\omegaloss= |\omegain|$.
A fluorescence-type peak or shoulder is seen at $\omegaloss \simeq |\omegain|$
if $\Gammap \lesssim |\omegain| \lesssim \Tk$, and at $\omegaloss \simeq \omegain$ if
$\max\{\Tk,\Gammap\} \lesssim \omegain \lesssim D$.
Moreover, a Raman-type peak occurs at $\omegaloss \simeq \Tk$ for $\max\{|\omegain|,\Gammap\} \gtrsim \Tk$.
}
\label{fig:RIXS_AIM} 
\end{figure*}

While the XAS spectrum in Fig.~\ref{fig:XAS_AAIM} matches the findings of Refs.~\onlinecite{Tureci2011,Latta2011}, the RIXS spectrum of the AAIM, shown in Fig.~\ref{fig:RIXS_AAIM_unnormalized-colorscale}, is new. As mentioned above,
it shows power laws with the same functional form as for the MIM, \Fig{fig:RIXS_XEM_unnormalized-colorscale}. Presumably, this commonality again reflects the Fermi-liquid nature of the low-energy excitations of the MIM and AAIM. Understanding this in detail is left as a challenge for future work. From a purely computational perspective, Fig.~\ref{fig:RIXS_AAIM_unnormalized-colorscale} illustrates that our approach, 
applied to a model with strong interactions and subtle correlations, is able to probe energy regimes differing by many orders of magnitude, uncovering 
power-law behavior with a quantitatively consistent value for the exponent $\alpha$. 

Though the unnormalized RIXS spectra of the MIM and AAIM look similar,
their normalized forms exhibit differences. 
For the AAIM, we define $I_{\normalized, \sigma \sigma'}^\rixs = 
I_{\sigma \sigma'}^\rixs/I_{\max}^\rixs$, with 
$I_{\max}^\rixs (\omegain)\!= 
\! \max_{\omegaloss, \sigma \sigma'}
[I_{\sigma \sigma'}^\rixs (\omegain,\omegaloss)]$.
As an example, Fig.~\ref{fig:RIXS_AIM}  shows $I^\rixs_{\uparrow \uparrow}$.
Compared to Fig.~\ref{fig:RIXS_XEM}  for the MIM, it has a richer structure, reflecting the presence of Kondo correlations. Consider first the case that the inverse core-hole lifetime  is much smaller than the Kondo temperature,  $\Gammap \ll \Tk$ [\Figs{fig:RIXS_AIM}(a) and \ref{fig:RIXS_AIM}(b)].
At low incident energies, $\Gammap < |\omegain| < \Tk$, the spectrum shows fluorescencelike behavior, having a single peak at $\omegaloss \simeq |\omegain|$. If $|\omegain|$ is further increased above $\Tk$, a Raman-type peak remains pinned at $\Tk$, more pronounced for $I_{\normalized, \uparrow \downarrow}^\rixs$ than $I_{\normalized, \uparrow \uparrow}^\rixs$.
Moreover, for positive 
$\omegain$ (but not for negative $\omegain$) a second, shoulder-type structure  emerges,  moving upward with $\omegain$, in fluorescencelike fashion. (See the curves for $\omegain /D = 10^{-1}, 10^{-0.5}$.) The height of this shoulder decreases as $\omegain$ gets larger, 
and for $\omegain \simeq D$, the shoulder essentially disappears. 

Next, consider the case  $\Gammap \gg \Tk $  [\Figs{fig:RIXS_AIM}(c) and \ref{fig:RIXS_AIM}(d)].
Then, a fluorescence-type peak is found only at
positive incident energies, $\omegaloss \simeq \omegain$,
and, since it requires  $\Gammap \lesssim \omegain \lesssim D$, 
only for $\omegain$ well above $\Tk$; in fact, such a structure is
visible only as a weak shoulder on the right of \Fig{fig:RIXS_AIM}(d).
Importantly, however, the Raman-type peak at $\Tk$ persists---its position and shape are the same as for $\Gammap \ll \Tk$ [\Fig{fig:RIXS_AIM}(b)]. The reason is that $\omegaloss$ enters the Heisenberg--Kramers formula separately from $\Gammap$, as mentioned earlier.

The pinning of the peak at $\Tk$ for large $|\omegain|$ is consistent with the relation \eqref{eq:RIXS_largewin2} between $I^\rixs$ and dynamical susceptibilities. Expressed through $d$ operators, it reads 
\begin{subequations}
\label{subeq:RIXS_largewin_ud}
\begin{align}
|\omegain|^2 I_{\uparrow\uparrow}^\rixs (\omegain,\omegaloss) &\simeq S[d^\pdag_\uparrow d_\uparrow^\dagger, d^\pdag_\uparrow d_\uparrow^\dagger] (\omegaloss), 
\label{eq:RIXS_largewin_uu}\\
|\omegain|^2 I_{\uparrow\downarrow}^\rixs (\omegain,\omegaloss) &\simeq S[d^\pdag_\uparrow d_\downarrow^\dagger, d^\pdag_\downarrow d_\uparrow^\dagger] (\omegaloss).
\label{eq:RIXS_largewin_ud}
\end{align}
\end{subequations}
Using $d^\dagger_\uparrow d^\pdag_\uparrow  =
\hat{n}_d / 2 + \hat S_z$,
$d^\dagger_\uparrow d^\pdag_\downarrow =\hat S_d^+$, and 
$d^\dagger_\downarrow d^\pdag_\uparrow = \hat S_d^-$,
we may express \Eqs{subeq:RIXS_largewin_ud} through susceptibilities
involving the local charge, magnetization, 
and spin-flip operators.
(Here, $\hat{n}_d = \sum_\sigma d_\sigma^\dagger d_\sigma^\pdag$ is the particle number operator.)
Hence, for large $|\omegain|$, the different-spin 
spectrum  $I_{\uparrow\downarrow}^\rixs$ is proportional to the 
dynamical \textit{spin} susceptibility, probing local spin fluctuations,
which has a peak at $\Tk$~\cite{Hanl2014}. By contrast, the same-spin spectrum $I_{\uparrow\uparrow}^\rixs$ receives contributions from both the spin and charge susceptibilities, 
and hence has peaks at both $\Tk$ and $U/2$.
The peak height of the latter is very much smaller than that of the former,
since the spin and charge susceptibilities have peak heights 
inversely proportional to the peak positions. 

A more detailed discussion of the RIXS spectra of the AAIM will be published elsewhere. For present purposes, the main messages of the proof-of-principle data presented here are the following: (i) at large $|\omegain|$ or large $\Gammap$, RIXS spectra probe dynamical
susceptibilities; (ii) for models involving Kondo correlations, they exhibit a distinct 
Raman-type peak at $\omegaloss \simeq \Tk$, well-resolved 
irrespective of the relative size of $\Gammap$ and $\Tk$; 
and (iii) our multipoint NRG scheme is very well suited for uncovering such peaks, even for widely separate scales, $\omegaloss \ll |\omegain|$.

\section{Summary and Outlook}
\label{sec:SummaryOutlook}

\subsection{Summary}
\label{sec:Summary}

In this work, we showed how NRG can be used to compute local $3$p and $4$p correlators of quantum impurity models.
Building on the spectral representation introduced in the accompanying paper, \paperI,
we used NRG to compute the fundamental PSFs, 
and then convolved these with suitable kernels to obtain 
both imaginary- and real-frequency correlators.

To compute the PSFs, we first developed a refined
tensor-network diagrammatic notation and explained
how to expand general operators along the NRG Wilson chain
in the AS basis.
Then, we introduced a ``slicing'' technique that enables 
recursively expressing higher- through lower-point PSFs.
The resulting PSFs can be computed at any temperature
and contain spectral information over a wide range of energies,
from the bare energy scales (e.g., Coulomb repulsion, bandwidth) 
down to excitations even below emergent energy scales (such as the Kondo temperature).

Once the PSFs have been obtained, imaginary- and real-frequency correlators can be computed from the same PSFs by applying appropriate convolution kernels.
This is advantageous compared to other methods that mainly treat imaginary-frequency correlators, since the numerical analytic continuation of multipoint objects from imaginary to real frequencies is extremely challenging.
Like the PSFs, the correlators can be computed at arbitrary temperature and frequencies. Our framework also encompasses a suite of strategies, including real-frequency EOMs, to accurately obtain \textit{connected} correlators and vertex functions. Altogether, results of the former, shown here, and of the latter, shown in \paperI, pass numerous qualitative and quantitative benchmark tests.

In addition to standard fermionic $4$p functions, 
we also applied our method to calculate RIXS spectra.
To this end, 
we rephrased the Kramers--Heisenberg formula, 
a theoretical description of RIXS spectra, 
as a special convolution of a PSF involving transition operators.
By studying two minimal models, 
we could identify distinctive contributions to the RIXS spectra:
(i) a fluorescence-type peak at $\omegaloss \simeq |\omegain|$, seen if 
$\omegain > \Gammap$, or if $-\omegain$ is
larger than $\Gammap$ and smaller than the upper bound to (quasi)particle excitation energies (half-bandwidth $D$ for the bare particles in the MIM; Kondo temperature $\Tk$ for the quasiparticles in the AAIM);
and 
(ii) a Raman-type part, peaked at $\omegaloss \simeq \Tk$ for the AAIM, 
seen if $|\omegain | \gtrsim \Tk$ or if $\Gammap \gtrsim \Tk$.
This Raman-like part is not smeared out by large $\Gammap$, demonstrating
the suitability of RIXS for probing (potentially small) many-body correlation scales.
Our approach, which can treat strong correlations and low-energy excitations accurately, may provide a benchmark for more approximate, but computationally cheaper, methods.

\subsection{Outlook}
\label{sec:Outlook}

Our scheme is both general and powerful, 
lending itself to various interesting applications.
The local  $4$p vertex, which we analyze in detail in \paperI\ and here use to improve the accuracy of the connected $4$p correlators, 
is at the heart of Feynman-diagrammatic analyses
of strongly correlated systems.
It is an input for diagrammatic extensions of DMFT
aiming to incorporate nonlocal correlations~\cite{Rohringer2018}.
Real-frequency applications along these lines are discussed in Sec.~V B of \paperI.
Furthermore, the vertex is an input to the computation of nonlocal response functions,
including the magnetic structure factor~\cite{Park2011}
and transport properties~\cite{Georges1996}.
For the latter, it was recently found that vertex corrections 
can be sizable even if the self-energy is practically local, i.e., consistent with
the DMFT approximation~\cite{Vucicevic2019}.
Since a major complication for computing conductivities is the numerical analytic continuation, 
our real-frequency method promises to be a breakthrough tool in this regard.

Another interesting application is to scrutinize how RIXS spectra reflect emergent energy scales in correlated lattice systems.
Such scales indicate key physical mechanisms: For instance,
the Mott metal-to-insulator transition in the Hubbard model is accompanied with the decrease of the spin screening scale~\cite{Raas2009,Raas2009:prb,Lee2017},
and  Hund metals are characterized by spin-orbital separation~\cite{Stadler2015,Stadler2019,Kugler2019,Kugler2020,Walter2020,Wang2020}, i.e., the fact that spin and orbital screening scales are separated by orders of magnitude.
Such small, many-body energy scales are hard to capture with established RIXS methods such as ED, while our NRG scheme is ideally suited for them.

All applications will benefit from further methodological development.
The asymmetric EOM used in this work leads to 
significant improvement for only some components of Keldysh correlators.
For applications building on the vertex, improvement for all Keldysh components, through a KF analogue of the symmetric EOM scheme of Ref.~\onlinecite{Kaufmann2019}, will be necessary. Moreover, refined multipoint broadening schemes, rather than the Lorentzian broadening prone to overbroadening used here, should be devised to better resolve all multiparticle spectral features.

\section*{Acknowledgments}
We deeply thank Andreas Weichselbaum for numerous inspiring discussions on all aspects of NRG over many years. The prescient remarks on higher-order correlation functions in Sec.~III E of his 2012 NRG-MPS review \cite{Weichselbaum2012b} contain the seeds for the method presented in this paper. For our numerical computations, we exploited symmetries using the QSpace library developed by him \cite{Weichselbaum2012a,Weichselbaum2020}. We thank Yilin Wang for providing us with RIXS results, obtained by ED \cite{Wang2019:RIXS}, for benchmarking our NRG RIXS results.
We were supported by the Deutsche Forschungsgemeinschaft (DFG, German Research Foundation) under Germany's Excellence Strategy EXC-2111 (Project No.\ 390814868) and through Project No.\ 409562408. 
S.-S.B.L.\ acknowledges the DFG grant LE3883/2-1 (Project No.\ 403832751). 
F.B.K.\ acknowledges support by the Alexander von Humboldt Foundation through the Feodor Lynen Fellowship.

\appendix

\section{Converting shell-off-diagonal operator-product expansions into shell-diagonal form}
\label{sec:OPEAppendix}

In this Appendix, we discuss an alternative method of deriving
the shell-diagonal operator-product expansions \eqref{subeq:OperatorRefinements}. It starts from expanding each operator all the way up to site $N$, using \Eq{eq:Operator-expansion-n0-to-N}. The result then contains  products of 
operator projections 
of the form $\ioc{\Ac}{n}{\x}{\xb} \ioc{\Bc}{\nb}{\xb'}{\xh}$.
By the projector product identity
\eqref{eq:projector-properties}, these yield three types of terms,
\begin{alignat}{7}
\nonumber
\ioc{\Ac}{n}{\x}{\xb} \ioc{\Bc}{\nb}{\xb'}{\xh} 
& = 
\idsuper{n <}{\nb} \delta_{\xb \k}  
\ioc{\Ac}{n}{\x}{\k} \ioc{\Bc}{\nb}{\xb'}{\xh} 
+
\idsuper{n}{\nb}  \delta_{\xb \xb'}  
\ioc{\Ac}{n}{\x}{\xb} \ioc{\Bc}{n}{\xb}{\xh} 
\\ & 
\qquad +
\idsuper{n > }{\nb}  \delta_{\k \xb'}  
\ioc{\Ac}{n}{\x}{\xb} \ioc{\Bc}{\nb}{\k}{\xh} 
.
\label{eq:nonzeroproducts}
\end{alignat}
This equation implies two rules: 
\begin{itemize}[topsep=2pt,itemsep=0pt,parsep=0pt,partopsep=0pt,labelindent=-4mm,itemindent=!,leftmargin=!]
\item
for shell-off-diagonal products ($n \!\neq\! \nb$), the ``inner sector''  of the earlier shell must be $\K$;
\item
for shell-diagonal products ($n \!=\! \nb$), the inner sectors must match. 
\end{itemize}

The diagrammatic illustration of these statements is analogous to that shown 
below \Eq{eq:projector-properties}, now with suitably inserted matrix elements. For example,  the case $n \!=\! \nb$ is depicted
in the diagram after \Eq{subeq:OperatorRefinements}, and $n \!< \! \nb$
as follows:
\newline
\includegraphics[width=\linewidth]{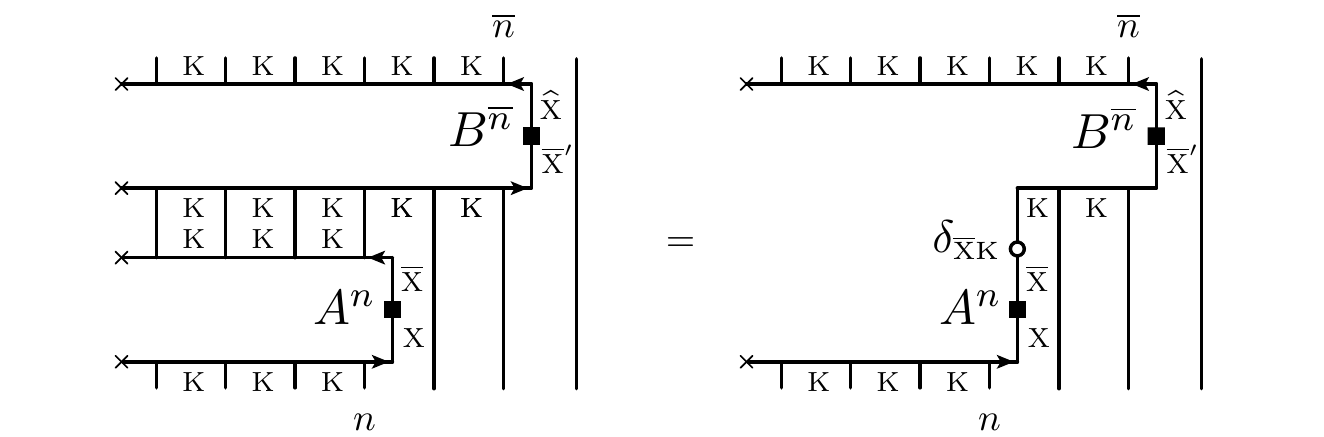}
Using \Eq{eq:nonzeroproducts}, all operator products can be reduced to nested sums over shells,  
$\sum_{n}\sum_{\nb \ge n} \sum_{\nh \ge \nb} \ndots $, with each summand involving only the building blocks of 
\Eq{eq:nonzeroproducts}, but containing
numerous shell-off-diagonal products. For the computation of thermal averages
using \Eq{eq:LocalExpectationShelln} though, we prefer a single
sum $\sum_n$, involving no shell-off-diagonal contributions.
This can be achieved by systematically using \Eq{eq:Operator-expansion-n-to-nhat}, 
$\sum_{\nh>\nb}\sum_{\x\xb}^{\neq \k\k} \ioc{\Oc}{\nh}{\x}{\xb} =
\ioc{\Oc}{\nb}{\k}{\k}$,  etc., to  collect all shell-off-diagonal contributions into $\K\K$ sectors of earlier shells.
The result is a sum $\sum_n$ containing, 
for given $n$, a sum over all possible products of  shell-$n$ projections, excluding the all-$\K$ case.  The latter is represented, in refined fashion, via later shells with larger $n$. This scheme
readily reproduces \Eqs{subeq:OperatorRefinements}. 
For example, expanding $\Ac \Bc$  using  
\Eq{eq:Operator-expansion-n0-to-N}, simplifying via 
\Eq{eq:nonzeroproducts}, and collecting shell-off-diagonal
terms using \Eq{eq:Operator-expansion-n-to-nhat},  we obtain 
\begin{flalign}
\label{eq:deriveABexpansion}
& \Ac \Bc = 
\sum_{n} \Bigl(\ioc{\Ac}{n}{\d}{\k} 
+ {\textstyle \sum_{\x}} \ioc{\Ac}{n}{\x}{\d} \Bigr)
\sum_{\nb} \Bigl(\ioc{\Bc}{\nb}{\k}{\d} 
+ {\textstyle \sum_{\xh}} \ioc{\Bc}{\nb}{\d}{\xh} \Bigr) \hspace{-1cm} & 
\\ \nonumber
& = \! \sum_{\nb} \Bigl[\ioc{\Ac}{\nb}{\d}{\k} \ioc{B}{\nb}{\k}{\d}
\!+\! 
{\textstyle \sum_{\x \xh}} \ioc{\Ac}{\nb}{\x}{\d} \ioc{B}{\nb}{\d}{\xh} 
\!+\! 
 \! \ioc{\Ac}{\nb}{\k}{\k} \ioc{\Bc}{\nb}{\k}{\d}\Bigr] 
\!+\! \sum_{n} \ioc{\Ac}{n}{\d}{\k} \ioc{\Bc}{n}{\k}{\k} ,   & 
\end{flalign}
which equals \Eq{eq:OperatorDoubleProductRefinement}. In 
the second line, the first two terms contain 
the nonzero shell-diagonal $n \!=\! \nb$ contributions; the third
term collects the shell-off-diagonal contributions
with $n \!>\! \nb$ into $\ioc{\Ac}{\nb}{\k}{\k}$, and the fourth 
those with 
$\nb \!>\! n$ into $\ioc{\Bc}{n}{\k}{\k}$. \Equs{eq:OperatorTripleProductRefinement} and
\eqref{eq:OperatorQuadrupleProductRefinement} can be found similarly.

\section{Derived 2p PSFs from 4p PSFs}
\label{app:1d-from-3d-histograms}
\begin{figure}[t]
\includegraphics[width=\linewidth]{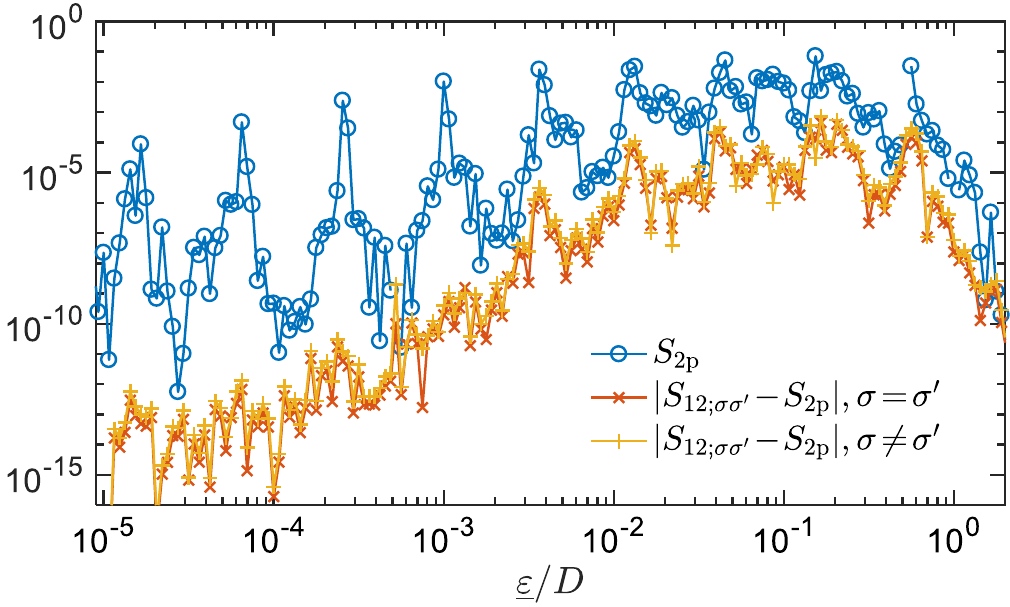}
\caption{%
Exemplary discrete $2$p PSFs of the AIM, with the same parameters as used in Figs.~\ref{fig:Adisc4}(a), \ref{fig:Adisc4}(b), \ref{fig:IF-connected-GF-AIM}, and \ref{fig:ZF-KF-connected-GF-AIM}.
Blue circles denote the $2$p PSF $S_{2\mr{p}} (\sve) \!=\! S [ d_{\sigma}, d_{\sigma}^\dagger ] (\sve)$ obtained directly by the $2$p shell-diagonal expansion.
The weights $S_{2\mr{p}} (\sve)$ are clustered.
Clusters are sharply concentrated (note the logarithmic scale of the ordinate), and the logarithms of their center positions are separated by $\sim \log \Lambda$.
Red $\times$'s and yellow $+$'s show the difference of the $2$p PSFs, derived from $4$p PSFs via sum rules, from $S_{2\mr{p}} (\sve)$.
The PSFs are obtained by using discretization parameters $\Lambda \!=\! 4$ and $z \!=\! 1$.
}
\label{fig:Adisc2} 
\end{figure}

To construct the disconnected PSFs of Sec.~\ref{sec:ConnectedPart}, we derive $2$p PSFs, which are the building blocks of the disconnected PSFs, from the full $4$p PSFs via sum rules.
For example, consider a system that preserves the $\mr{SU}(2)$ spin symmetry.
Then, a $2$p PSF $S[ d_{\sigma}^\pdag, d_{\sigma}^\dagger ] (\sve)$ 
can be derived from $4$p PSFs in four different ways, 
involving three different permutations of $\vec{\Oc} = (d_{\sigma}^\pdag, d_{\sigma}^\dagger, d_{\sigma'}^\pdag, d_{\sigma'}^\dagger)$:
One may sum either over  $(\sve_2, \sve_3)$ or over $(\sve_1, \sve_2)$, i.e., 
\begin{align*}
S_{12;\sigma\sigma'} (\sve = \sve_1) &= 
\textstyle
\sum_{\sve_2, \sve_3} S_{1234} (\vec{\sve}) + S_{1243} (\vec{\sve}), 
\\
S_{34;\sigma\sigma'} (\sve = \sve_3) &= 
\textstyle
\sum_{\sve_1, \sve_2} S_{1234} (\vec{\sve}) + S_{2134} (\vec{\sve}), 
\end{align*}
and do so for either the same-spin ($\sigma \!=\! \sigma'$) or different-spin ($\sigma \!\neq\! \sigma'$) configuration.
[Because of the $\mr{SU}(2)$ spin symmetry, the configurations are labeled only by the relation between $\sigma$ and $\sigma'$.]
If the PSFs were computed exactly, then all four $2$p PSFs would be identical.

However, with the $4$p PSFs obtained by NRG, there exist small but finite differences among $S_{12;\sigma\sigma'}$ and $S_{34;\sigma\sigma'}$.
In \Fig{fig:Adisc2}, we compare them with a $2$p PSF $S_{2\mr{p}}$ obtained by using our $2$p method [cf.~\Eq{eq:l=2ABFinalResult}], which is equivalent to fdm-NRG.
While $S_{34;\sigma\sigma'}$ is identical to $S_{2\mr{p}}$ up to double precision $\simeq 10^{-16}$ regardless of $\sigma\sigma'$ (hence not shown), $S_{12;\sigma\sigma'}$ differs from $S_{2\mr{p}}$, and also from each other depending on $\sigma\sigma'$.
The differences are smaller than the local maxima for individual clusters of spectral weights (which appear as peaks in the plot) by at least two orders of magnitude.

There are two reasons for these differences. First, the reduced density matrices in the kept sectors, $\ioc{\varrhored}{n}{\k}{\k}$, are not diagonal, since they are obtained by tracing out entangled states over a subsystem.
Second, the way of measuring the first argument $\sve_1$ of $4$p PSFs depends on the sectors.
For example, while the first operator $\Ac$ is sliced with respect to $\sve_1$ for computing $\ioc{\eS}{n}{\k\d\k\k}{} (\vec{\sve})$ [cf.~\Eq{eq:TnKDKK}],
$\sve_1$ is measured as the energy difference across the product $\varrho \Ac$ (not just $\Ac$) for $\D\Xb\Xh\Xt$ and $\K\D\D\D$ [cf.~\Eq{eq:ProjectorProduct4}].
By contrast,
$\sve_3$ is always associated with $\Dc$, up to a sign.
For $2$p PSFs, the argument $\sve$ is always associated with $\varrho\Ac$ or equivalently, $\Bc$, up to a sign.
Given that the $\ioc{\varrhored}{n}{\k}{\k}$'s are generally not diagonal, such a discrepancy in measuring $\sve_1$ introduces a numerical artifact to the distribution of spectral weights along the $\sve_1$ grid, for both the full $4$p PSFs and $S_{12;\sigma\sigma'} (\sve = \sve_1)$.
For $S_{34;\sigma\sigma'} (\sve = \sve_3)$, the $\sve_1$ dependence has been summed over, so the artifact is removed. 

As mentioned in Sec.~\ref{sec:ConnectedPart}, the connected part can have smaller values than the full $4$p correlator, and thus be more susceptible to numerical noise.
Hence, we need to remove the above-mentioned numerical artifact when subtracting the disconnected part.
For this, in evaluating \Eq{eq:Sc_w-dis-epsilon}, we substitute the $2$p PSFs obtained by summing over $(\sve_2, \sve_3)$, such as $S_{12;\sigma\sigma'}$, to the $2$p PSFs involving $\Oc^{\ovb{1}}$, and those obtained by summing over $(\sve_1, \sve_2)$, such as $S_{34;\sigma\sigma'}$, to the rest.
Moreover, to calculate the disconnected part for a given spin configuration ($\sigma \!=\! \sigma'$ or $\sigma \!\neq\! \sigma'$), we use the $2$p PSFs derived from $4$p PSFs for that configuration.

\section{Log-Gaussian broadening}
\label{app:logGaussian}

For standard NRG computations of 2p functions, 
it is customary to use a log-Gaussian function for broadening
discrete spectral data. For example, the broadening scheme of Ref.~\onlinecite{Lee2016}
broadens $2$p PSFs with the convolution of a symmetric log-Gaussian function $\delta_\SL$ and the derivative of the Fermi function $\delta_\Fermi$,
\begin{align}
\label{eq:logGaussFermibroadening}
\delta_{\SL+\Fermi} (\ve_i, \sve_i) 
&= 
\nint \md\ve' \, \delta_\Fermi (\ve_i, \ve') \, \delta_\SL (\ve',\sve_i), 
\\
\nonumber
\delta_{\SL} (\ve', \sve_i) 
&= 
\frac{\theta( \ve' \sve_i)}{\sqrt{\pi} b_\SL | \sve_i |} 
\exp 
\bigg[ - 
\Pig( \frac{\ln | \sve_i / \ve' |}{b_\SL} - \frac{b_\SL}{4} 
{\Pig)}^2 \,
\bigg] 
, 
\\
\nonumber
\delta_{\Fermi} (\ve_i, \ve') 
&= 
\frac{1}{2 b_\Fermi} 
\Pig( 1 + \cosh \frac{\ve_i - \ve'}{b_\Fermi} {\Pig)}^{-1} 
,
\end{align}
where  $\delta_{\SL}$ broadens discrete data, while $\delta_\Fermi$, with $b_\Fermi \sim T$, further smears out low-frequency artifacts arising at energies below the temperature $T$~\cite{Weichselbaum2007,Weichselbaum2012b}.
We use this broadening scheme (in conjunction with the self-energy trick \cite{Bulla1998}) for the $2$p correlators reattached to the vertex when 
computing connected 4p correlators (cf.~\Sec{sec:amputation}).

The symmetric log-Gaussian $\delta_\SL$ is, by design, symmetric under the exchange of arguments, $\delta_{\SL} (\ve', \sve_i) = \delta_{\SL} (\sve_i, \ve')$. This ensures that broadening a curve using $\delta_\SL$ (i) preserves peak heights but (ii) shifts peak positions (in a sense made precise in the Supplemental Material of Ref.~\onlinecite{Weichselbaum2007}). Property (i) is desirable, for example, when computing fermionic 2p spectral functions featuring a Kondo resonance of known height. However, property (ii) is undesirable 
when computing bosonic correlators,  such as dynamical susceptibilities---the peak positions of their imaginary parts are associated with characteristic energy scales such as the Kondo temperature, and broadening-induced
shifts in peak positions would induce errors in identifying such scales.
This problem can be avoided by using a \textit{centered} log-Gaussian~\cite{Bulla2008}, defined as
\begin{equation}
\delta_{\CL} (\ve', \sve_i) =
\frac{\theta( \ve' \sve_i)}{\sqrt{\pi} b_\CL | \sve_i |} 
\exp \Pig(  -  \frac{\ln^2 | \sve_i / \ve' |}{b_\CL^2} - \frac{b_\CL^2}{4} 
\Pig) .
\label{eq:centeredLogGauss}
\end{equation}
It preserves peak positions but yields $b_\CL$-dependent peak heights. We used $\delta_{\CL+\Fermi}$, with 
$\delta_\CL$ instead of $\delta_\SL$ in \Eq{eq:logGaussFermibroadening}, 
to broaden the PSFs of bosonic operators needed  in Sec.~IV A of Ref.~\onlinecite{Kugler2021}. 

Bosonic correlators also arise when computing the XAS and RIXS spectra of \Sec{sec:RIXS} of this paper. We broadened them as follows.
Energy dependencies not affected by $\Gammap$ were broadened
using $\delta_{\CL+\Fermi}$, with $b_\Fermi = T/5$.
This applies to the dependence of the RIXS kernel on $\ve_2$ regardless
of $\Gammap$.  For $\Gammap \! < \! T$, it also applies to the 
dependence of the XAS kernel on $\ve$, and of the RIXS kernel
on $\ve_1$ and $\ve_3$, because then their Lorentzian dependence on $\Gammap$
becomes irrelevant. By contrast, for $\Gammap \! > \! T$, we implemented
Lorentzian broadening of the $\ve$ or $\ve_1$ and $\ve_3$ dependencies
by replacing $\delta_\Fermi$ in \Eq{eq:logGaussFermibroadening} by a Lorentzian $\delta_\Lorentz$ 
of the form \eqref{eq:LorentzianKernel}, with $b_\Lorentz |\sve_i|$
there replaced by $\Gammap$. 
For the RIXS kernel, this yields the imaginary parts
of its Lorentzian denominators; their real parts were then obtained
using Kramers--Kronig transformations.
We used the broadening parameters $b_\CL \!=\! \ln \Lambda$,
except for the data shown in \Fig{fig:RIXS_AIM}, where a smaller value, 
$b_\CL \!=\! (1/2) \ln \Lambda$, was used to better resolve the separation of the Raman- and fluorescence-type peaks.

\begin{figure*}[tb]
\includegraphics[width=\linewidth]{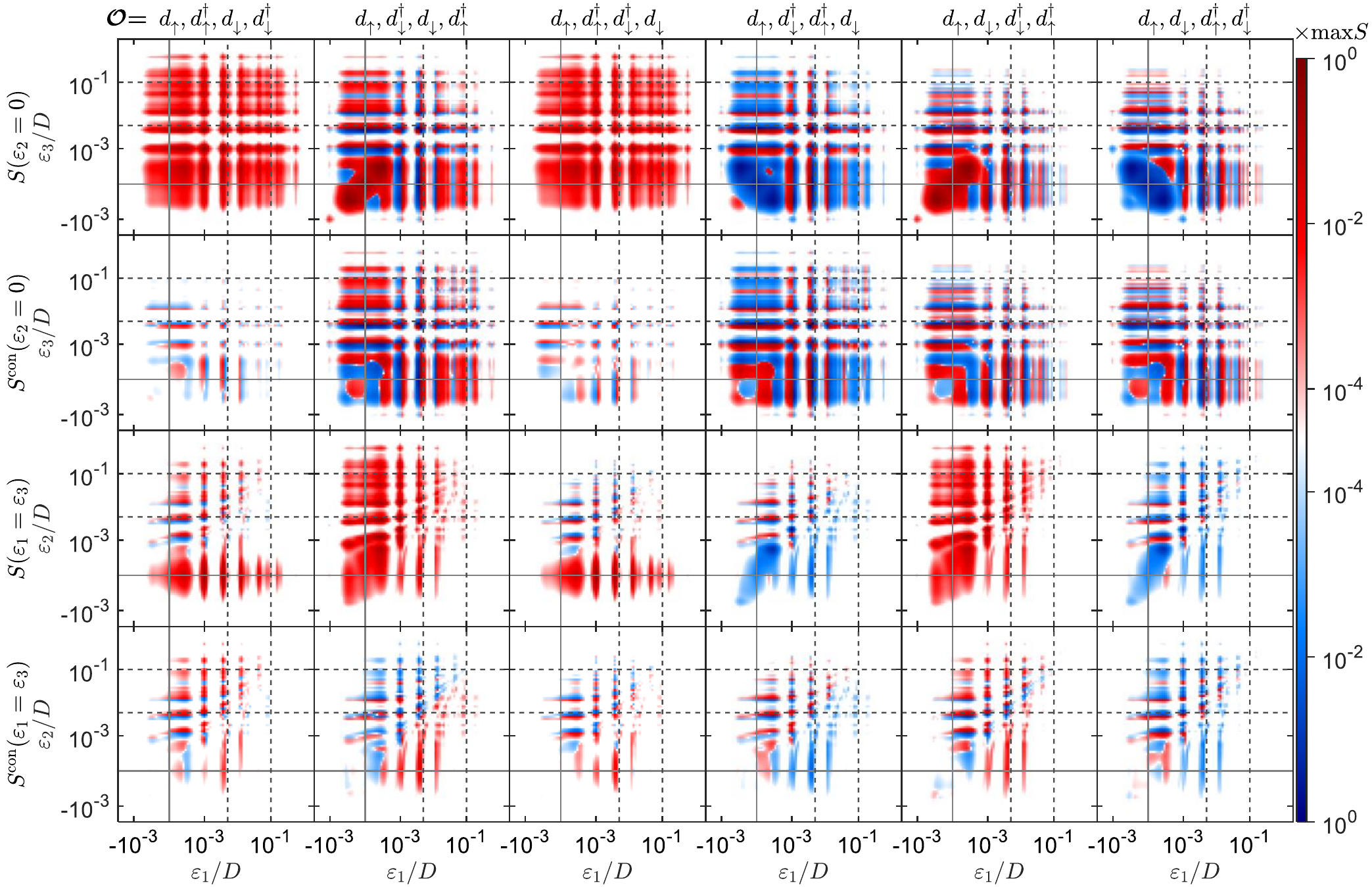}
\caption{%
All independent different-spin $4$p PSFs of the AIM, with the same parameters as used in Figs.~\ref{fig:Adisc4}(a), \ref{fig:Adisc4}(b), \ref{fig:IF-connected-GF-AIM}, and \ref{fig:ZF-KF-connected-GF-AIM}.
For all permutations $\vec{\Oc}_p$ of $\vec{\Oc}$ not listed here,
the corresponding PSFs $S_p$ are related to those shown  by spin and particle-hole symmetries.
The PSFs are obtained and narrowly broadened in the same way as in Figs.~\ref{fig:Adisc4} and \ref{fig:Adisc4_same_spin}.
The first and second rows show the PSFs as a function of $\ve_1$ and $\ve_3$ for fixed $\ve_2 \!=\! 0$;
the third and fourth rows as a function of $\ve_1$ and $\ve_2$ along the cross section of $\ve_1 \!=\! \ve_3$.
Odd rows show the full PSFs $S$; even rows show the connected PSFs $S^\connected$.
Each pair of the full and connected PSFs, with the same operators and frequency choices, are normalized by the maximum magnitude of the full PSF $S$ in the pair.
The dashed lines mark the Kondo temperature $\Tk/D \!\simeq\! 1/200$ and the position of Hubbard side peak $U/2 \!=\! 0.1D$.
}
\label{fig:Adisc4_all_permutation} 
\end{figure*}

\section{Further exemplary 4p PSFs}
\label{app:Example_4p_PSFs}

For completeness, in \Fig{fig:Adisc4_all_permutation}, we show the $4$p PSFs of the AIM at strong coupling for various permutations of operators other than those shown in \Fig{fig:Adisc4}.
Figures \ref{fig:Adisc4}(c) and \ref{fig:Adisc4}(d) appear again as the top two panels in the leftmost column of \Fig{fig:Adisc4_all_permutation}.
The PSFs shown here form a ``complete set,'' in the sense that those not shown are related to those shown by spin and particle-hole symmetries.
The former symmetry involves permutations exchanging $d_\uparrow^\pdag \leftrightarrow d_\downarrow^\pdag$ \textit{and} $d_\uparrow^\dagger \leftrightarrow d_\downarrow^\dagger$; 
the latter involve permutations exchanging $d_\uparrow^\pdag \leftrightarrow d_\uparrow^\dagger$ \textit{and} $d_\downarrow^\pdag \leftrightarrow d_\downarrow^\dagger$.
The spectral distributions of the full and connected PSFs change
significantly across the characteristic energy scales $\Tk$ and $U/2$, marked by dashed lines.

\section{Heuristic discussion of Anderson orthogonality}
\label{app:AndersonOrthogonality}

The power-law behavior observed for the XAS and RIXS spectra reported in 
Secs.~\ref{sec:XAS-RIXS-results-XES} and \ref{sec:XAS-RIXS-results-AAIM}
is due to Anderson orthogonality~\cite{Anderson1967,Anderson1967a}. We here review
how it comes about in the context of the MIM. When a free Fermi sea experiences
a sudden change in a local scattering potential, causing 
changes in the scattering phase shifts of all its single-particle wave functions, the overlap between the final and initial ground states
vanishes as $\langle g'|g\rangle \sim L^{- \Delta^2_\charge/2}$
in the limit of large system size $L$ ($\sim \Lambda^{N/2}$ for a length-$N$ Wilson chain \cite{Weichselbaum2011}). Here, $\Delta_\charge$
is the \textit{displaced charge}, i.e., the charge flowing inward from infinity into a large but finite region (of size $< L$) surrounding the scattering site in response to the change in scattering potential. In the present context, where the scattering potential 
is switched on by $\Tc = c^\dagger p$, the displaced charge is $\Delta_\charge \! = \! \Deltah \!-\!1$. Here, $\Deltah$ is the charge drawn in toward the scattering site by the core hole \cite{Weichselbaum2011,Muender2012}, related to 
the phase shift in \Eq{eq:XES_analytic} by the Friedel sum rule \cite{Friedel1956}, $\Deltah = \delta/\pi$; and $-1$ reflects the fact that 
the charge added to the Fermi sea by $c^\dagger$ 
flows outward to infinity in the long-time limit, hence contributing negatively to the displaced charge. (This argument is due to 
Hopfield \cite{Hopfield1969}. A thorough
discussion thereof is given in Ref.~\onlinecite{Muender2012}, which 
also contains a time-domain formulation of the argument presented next.)

The scaling behavior $I^\xas(\omegain) \sim \omegain^{-\alpha}$ of \Eq{eq:XES_analytic} can now be recovered by 
the following heuristic argument. 
In the limit $T,\Gammap \to 0$, \Eq{eq:XAS-textbook} takes the form
$I^\xas (\omegain) =  \sum_{\ub{2}} 
 \Tc^\pdag_{1_g \ub{2}}   \Tc^\dagger_{\ub{2} 1_g}
\delta(\omegain - \ve_{\ub{2}})$. It involves  matrix elements between the no-hole ground state
$|1_g\rangle$ and one-hole eigenstates $|\ub{2}\rangle$,
and transition energies $\ve_{\ub{2}} = E_{\ub{2}1_g} - \omega_\threshold$.
Since $\omega_\threshold = E_{2_g 1_g}$ is the subspace ground-state energy difference, $\ve_{\ub{2}} = E_{\ub{2}}- E_{2_g}$ is an excitation energy within the one-hole subspace. Viewing it as the low-energy level spacing of a finite box of size $L \sim 1/\ve_{\ub{2}}$ and evoking Anderson orthogonality, the corresponding matrix element scales as 
$ \Tc^\dagger_{\ub{2}1_g}  \sim (\ve_{\ub{2}})^{-\Delta^2_\charge/2}$. Expressing the sum
$\sum_{\ub{2}}$ as an integral, $\int \md \ve_{\ub{2}} / \ve_{\ub{2}}$ (with $1/\ve_{\ub{2}}$ representing the density of states
at the bottom of the spectrum of the finite box), we obtain 
\begin{flalign}
I^\xas  (\omegain) & 
\sim \! \int \! \frac{\md \ve_{\ub{2}}}{\ve_{\ub{2}}} (\ve_{\ub{2}})^{\Delta_\charge^2}  \,   \delta(\omegain \!-\!  \ve_{\ub{2}})  \sim  \omegain^{-1 + \Delta_\charge^2} \, .  \hspace{-1cm} &
\label{eq:XAS-heuristic-AO}
\end{flalign}
This reproduces \Eq{eq:XES_analytic}, since $1 \! -\! \Delta_\charge^2 = 
2 \Deltah \! - \! \Deltah^2 = \alpha$.

The above discussion gives some hints as to why the exponent $\alpha$
governing XAS spectra also governs RIXS spectra. 
The formula for RIXS spectra, \Eq{eq:RIXS_conventional}, 
likewise contains ground-to-excited states matrix elements, 
$ \Tc_{1_g \ub{2}} \sim (\ve_{\ub{2}})^{-(1-\alpha)^2/2}$. 
To fully rationalize the RIXS power-law behavior summarized in  \Eq{eq:RIXS_XES_powerlaws}, however, one would also need to know the
scaling behavior of excited-to-excited states matrix elements, $\Tc^\dagger_{\ub{2}\ub{3}}$, as a function of their energies $\ve_{\ub{2}}$ and $\ve_{\ub{3}}$.
Working out the corresponding details is left as an interesting task for future work. 

%

\end{document}